\documentclass[pdflatex,sn-nature]{sn-jnl}%

\usepackage{graphicx}
\usepackage{multirow}
\usepackage{amsmath,amssymb,amsfonts}
\usepackage{amsthm}
\usepackage[title]{appendix}
\usepackage{xcolor}
\usepackage{textcomp}
\usepackage{manyfoot}
\usepackage{booktabs}
\usepackage{algorithm}
\usepackage{algorithmicx}
\usepackage{algpseudocode}
\usepackage{listings}
\usepackage[normalem]{ulem}
\usepackage{hyperref}
\usepackage{tabularx}

\theoremstyle{thmstyleone}%

\theoremstyle{thmstyletwo}%

\theoremstyle{thmstylethree}%

\begin{document}

\title[Physics-guided surrogate learning enables zero-shot control of turbulent wings]{Physics-guided surrogate learning enables zero-shot control of turbulent wings}

\author*[1]{\fnm{Yuning} \sur{Wang}}\email{yuninw@umich.edu}

\author[2]{\fnm{Pol} \sur{Suárez}}\email{polsm@kth.se}

\author[3]{\fnm{Mathis} \sur{Bode}}\email{m.bode@fz-juelich.de}

\author*[1,2]{\fnm{Ricardo} \sur{Vinuesa}}\email{rvinuesa@umich.edu}

\affil[1]{\orgdiv{Department of Aerospace Engineering}, \orgname{University of Michigan}, \orgaddress{\city{Ann Arbor}, \postcode{48109}, \state{Michigan}, \country{USA}}}

\affil[2]{\orgdiv{FLOW, Engineering Mechanics}, \orgname{KTH Royal Institute of Technology}, \orgaddress{\city{Stockholm}, \postcode{SE-100 44}, \country{Sweden}}}

\affil[3]{\orgdiv{J{\"u}lich Supercomputing Centre}, \orgname{Forschungszentrum J{\"u}lich GmbH}, \orgaddress{\city{J{\"u}lich}, \postcode{52425}, \country{Germany}}}

\abstract{Turbulent boundary layers over aerodynamic surfaces are a major source of aircraft drag, yet their control remains challenging due to multiscale dynamics and spatial variability, particularly under adverse pressure gradients. Reinforcement learning has outperformed state-of-the-art strategies in canonical flows, but its application to realistic geometries is limited by computational cost and transferability. Here we show that these limitations can be overcome by exploiting local structures of wall-bounded turbulence. Policies are trained in turbulent channel flows matched to wing boundary-layer statistics and deployed directly onto a NACA4412 wing section at $Re_c = 2\times10^5$ without additional training on the wing, constituting zero-shot deployment. The transferred policies reduce total drag by 11.1\%, while state-of-the-art opposition control reduces it by 8.9\%. With the actuation cost taken into account, the corresponding net-energy savings are 10.9\% and 8.8\%, respectively. Surrogate training provides an estimated improvement of four orders of magnitude in exploration efficiency over direct on-wing training, enabling scalable flow control.}

\maketitle

\section*{Introduction}\label{sec:intro}

Turbulence is a primary source of skin-friction drag on aircraft wings~\cite{abbas_TBLCTRLReview_2017,fukagata_reviewCTRL_2024}. Despite its practical importance, its control remains a fundamental challenge, driven by the high dimensionality, chaotic dynamics, and multiscale nature of turbulent flows~\cite{cardesa_cascade_2017}, and further complicated by the geometric complexity of realistic configurations. Any practical flow-control strategy must balance four interdependent requirements: \textit{control performance}, the magnitude of drag reduction achieved; \textit{computational cost}, the resources required to train and deploy the controller; \textit{generalizability}, the ability to transfer across flow conditions and geometries; and \textit{turbulent-scale target}, linked to the level of complexity or richness of flow that the application demands. No existing framework achieves a satisfactory balance across all four. Active flow control introduces external actuation to modify the flow. When actuation is imposed \textit{a priori} and remains independent of the instantaneous flow state, the strategy is classified as open-loop control~\cite{seifert_1996,cattafesta_2011}. Such strategies are effective under fixed conditions, but they are sensitive to variations in angle of attack, Reynolds number, and freestream turbulence, limiting their practical utility~\cite{collis_2004}. Strategies that rely on system feedback are instead classified as closed-loop, and have emerged as a particularly promising direction in recent years~\cite{brunton_closedloop_2015,brunton_applying_2021}, driven by advances in machine learning, most notably deep reinforcement learning, which enables the autonomous discovery of nonlinear feedback laws directly from flow data~\cite{rabault_2019,vignon_2023}.

Turbulent-flow control strategies have historically been based on physical insight into the mechanisms sustaining wall-bounded turbulence~\cite{choi_active_1994,smits_highReynolds_2011}. Ref.~\citep{marusic_wallbounded_2010} showed that near-wall fluctuations can be predicted from a universal small-scale signal modulated by large-scale outer-layer motions. This finding provides a physical motivation for constructing surrogate environments that retain the near-wall dynamics relevant to control. A classical control strategy that exploits this near-wall structure is opposition control (OC)~\cite{choi_active_1994}, in which wall blowing and suction counteract the wall-normal velocity detected at an off-wall sensing plane, weakening the near-wall vortices responsible for momentum transfer and reducing skin-friction drag. Despite its simplicity, OC achieves meaningful drag reduction across different flow types~\cite{stroh_comparison_2015} and geometries~\cite{dacome_oppositionTBL_2024,yao_oc_PRF_2025}, a transferability attributed to the universality of the near-wall self-sustaining cycle of streaks and quasi-streamwise vortices~\cite{hamilton_regeneration_1995,jimenez_TBLvsTCF_2010}. From the perspective of the four pillars, OC scores favorably on computational cost, requiring no training.
However, its performance is inherently limited by the simplicity of the control law connecting the action and observation, essentially a simple linear proportional control. It cannot adapt its actuation to exploit flow-state-dependent drag-reduction opportunities, motivating the transition toward richer closed-loop strategies.

{Machine learning (ML) supports surrogate modelling and flow-field reconstruction, with applications constrained by data availability, computational cost and model validation~\cite{Drikakis2023AIFluidMechanics,Wang2024MachineLearningCFDSurvey}. Physics-informed neural networks can infer velocity and pressure fields~\cite{raissi_HFM_2020}, while convolutional networks reconstruct turbulent-flow images from low-resolution, noisy inputs~\cite{Sofos2024SparseNoisyFlow}. Reviews of super-resolution and generalization emphasize reconstruction accuracy, robustness and validation across flow conditions~\cite{Sofos2025SuperResolutionReview,Taira2025CriticalAssessment}. These developments address the computational and sensing demands of flow control. Deep reinforcement learning (DRL) complements them by learning nonlinear feedback policies through interaction with the flow.}

Closed-loop flow control can be formulated within DRL by modeling the system as a Markov decision process (MDP), in which a controller learns a policy mapping high-dimensional flow observations to control actions through data-driven optimization. Recent studies have reported that DRL-based policies can surpass conventional strategies in drag reduction across a range of configurations, from bluff-body wakes~\cite{rabault_2019,suarez_flow_2025} to wall-bounded turbulent flows~\cite{guastoni_DRL_2023,sonoda_RL_2023}, exploiting flow structures across a wide range of turbulent scales in ways beyond the reach of open-loop actuation and of closed-loop strategies derived from reduced-order or linear models. As a result, DRL can achieve substantial drag reduction and exploit turbulent-scale flow information effectively, but at the expense of the remaining two pillars: its application to realistic aerodynamic configurations is limited by the prohibitive cost of high-fidelity training and by poor generalization across geometries and actuator configurations.

{DRL training requires repeated interactions with the flow, with simulation costs increasing as higher Reynolds numbers require finer resolution. For three-dimensional cylinder flow, increasing the Reynolds number based on cylinder diameter from $Re_D=300$--$400$ to $3,900$ raises training costs by more than an order of magnitude~\cite{suarez_flow_2025,suarez_active_2025}. Generalization is also limited because observations, actions and characteristic time scales depend on the training configuration. Although transfer learning can extend applicability~\cite{varela_deep_2022,liu_design_2025}, policies may fail when deployed in a different geometry.}

The curse of dimensionality can be alleviated through multi-agent reinforcement learning (MARL), introduced in the context of flow control by Belus et al.~\cite{belus_exploiting_2019}, in which multiple agents operate on local domain partitions, i.e.\ pseudo-environments, under a shared policy, enabling efficient experience reuse and accelerated exploration~\cite{rabault_accelerating_2019}. MARL is particularly effective when the local invariances of the flow are known a priori and can be exploited. For example, turbulent channel flow has two homogeneous directions, which MARL can exploit to improve sampling efficiency~\cite{guastoni_DRL_2023,sonoda_RL_2023}. In contrast, it becomes less efficient in spatially developing flows such as flat plates with turbulent boundary layers, where streamwise inhomogeneity reduces the number of available pseudo-environments and a single shared policy may be insufficient. These limitations collectively motivate the search for training environments that are both computationally tractable and physically representative of the target flow.

Several proposed strategies have been good candidates to partially address the cost and generalizability pillars, including physics-informed observation design via explainable-learning tools~\cite{cremades_identifying_2024,sanchisagudo_xcal_2026}, coupling DRL with existing control schemes~\cite{font_deep_2025}, and enriching sparse observations through diffusion-based reconstruction~\cite{vishwasrao_diff-sport_2025}. 

A more fundamental approach is the replacement of high-fidelity simulations with a surrogate environment: a computationally inexpensive model that approximates the flow response to control inputs with sufficient fidelity to support policy learning. Surrogate-assisted DRL has shown promise in flow control, yet most studies have been limited to low-Reynolds-number laminar configurations~\cite{weiner_modelbased_2024,zhang_surrogate_2024}. Therefore, none of the existing approaches balances all four pillars: OC trades performance for simplicity, with limited control authority as its fixed feedback law cannot fully act on the targeted turbulent scales; conventional DRL frameworks sacrifice computational efficiency and generalizability; and surrogate-based methods have yet to demonstrate the turbulent-scale fidelity and transferability required under strong adverse pressure gradients.

Preliminary results on zero-shot deployment were presented as part of the HydroGym reinforcement-learning platform~\cite{hydrogym_2026}. Extending surrogate-based control to flows with strong, spatially varying adverse pressure gradients (APGs) requires further validation of the physical correspondence between the surrogate and the corresponding complex flows. The challenge is to retain the computational advantages of the surrogate training while achieving effective control with a positive net-energy saving under these conditions, a fact that is crucial for industrial applications.

\begin{figure}[H]
    \centering
    \includegraphics[
        width=\linewidth,
        trim=10 0 20 0,
        clip
    ]{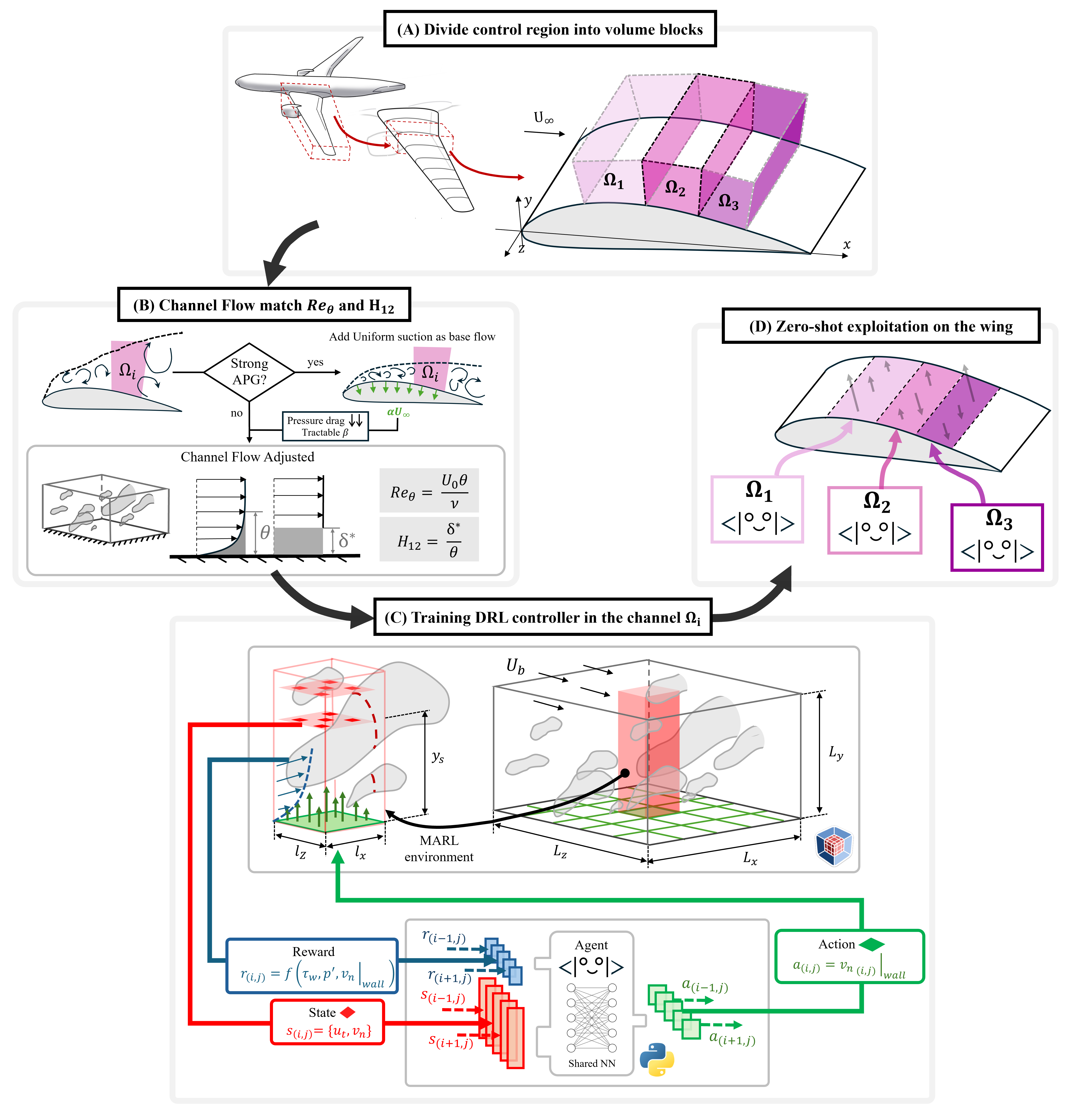}
    \caption{\textbf{Physics-guided zero-shot MARL framework for turbulent wing control.}
    (A) The suction side of a NACA4412 wing is decomposed into chordwise control blocks $\Omega_i$.
    (B) Each $\Omega_i$ is matched to a turbulent channel-flow surrogate by reproducing the local shape factor ($H_{12}$) and momentum-thickness Reynolds number ($Re_\theta$), capturing both flow conditions and pressure-gradient history of the local state.
    (C) MARL agents are trained exclusively within these surrogates using a shared neural network, with wall-parallel and wall-normal velocity fluctuations $\{u_t, v_n\}$ at sensing height $y_s$ as state, reward function based on the wall-shear stress $\tau_w$, $p'$, $v_{n,(i,j)}\vert_{{wall}}$, and local blowing/suction wall-normal velocity $v_{n,(i,j)}\vert_{{wall}}$ as action.
    (D) The trained policy is deployed directly to the corresponding wing blocks without additional on-wing training.}
    \label{fig:FIG01_schematic}
\end{figure}

To this end, we demonstrate that MARL control policies, trained exclusively in inexpensive TCF surrogates, can be deployed directly on a turbulent wing section without additional training, achieving reductions in both skin-friction and total drag with a positive net-energy saving.. By weakening the near-wall regeneration cycle, the controller attenuates near-wall turbulence, in the spirit of the control mechanism characterized by Ref.~\cite{choi_active_1994}. The accompanying changes in near-wall turbulence are examined in Supplementary Information~7. The wing surface is partitioned into chordwise blocks $\Omega_i$, each matched to a TCF surrogate via boundary-layer characteristics (e.g. shape factor, momentum thickness~\cite{hydrogym_2026}), exploiting the homogeneous directions of TCF to maximise MARL efficiency. By linking the surrogate design to the local physics of the wing boundary layer, the proposed framework balances all four pillars: surrogate training reduces the computational cost by four orders of magnitude; block-wise physical matching provides a mechanism for generalization; each surrogate preserves the range of turbulent scales accessible to DRL; and zero-shot transfer yields drag reduction that consistently outperforms OC. We validate the framework on the suction side of a NACA4412 wing section at $Re_c = 2\times 10^5$ and $AoA = 5^{\circ}$ ($Re_c = U_\infty c /\nu$ where $c$ is the chord length and $\nu$ is the kinematic viscosity), where a strong adverse pressure gradient (APG) downstream of chordwise coordinate $x_{ss}/c \approx 0.5$ (\emph{{\it ss}} denotes the suction side) provides a challenging and practically relevant test case, illustrated in Figure~\ref{fig:FIG01_schematic}.

\section*{Results}\label{sec:results}

This section presents results for active flow control on the suction side of a NACA4412 wing section at $Re_c = 2\times10^5$ and $AoA = 5^{\circ}$, validating the physics-guided zero-shot transfer framework introduced above. The presentation follows the three stages of the proposed methodology: characterization of the uncontrolled wing flow and definition of the chordwise control blocks $\Omega_i$ (Figure~\ref{fig:FIG01_schematic}(A,B)); MARL training within each TCF surrogate (Figure~\ref{fig:FIG01_schematic}(C)); and zero-shot deployment of the surrogate-trained controllers on the wing, benchmarked against opposition control (OC) as a classical closed-loop baseline (Figure~\ref{fig:FIG01_schematic}(D)).

The main results consider a control configuration in which a wall-normal suction of $0.2\%\,U_{\infty}$ is applied to the base flow to reduce pressure drag by attenuating the adverse pressure gradient along the suction side of the airfoil. This suction intensity was selected following the parametric study of Ref.~\cite{atzori_aero_2020}, in which it yielded the largest pressure-drag reduction among the configurations examined for the same wing section, Reynolds number and angle of attack. This suction modifies the local boundary layer state, altering both the momentum-thickness Reynolds number $Re_{\theta} = U_0\theta/\nu$ and the shape factor $H_{12} = \delta^*/\theta$ (where $\delta^*$ and $\theta$ are the displacement and momentum thickness, respectively).

The momentum and displacement thicknesses are defined as:
\begin{equation}
    \theta = \int^\psi_0 \frac{U}{U_0}\left(1 - \frac{U}{U_0}\right) \mathrm{d}y, \qquad
    \delta^* = \int^{\psi}_0 \left(1 - \frac{U}{U_0}\right) \mathrm{d}y,
\label{eq:metrics}
\end{equation}
with $U_0$ corresponding to the boundary-layer edge velocity $U_e$ for the wing TBL and to the centerline velocity $U_{cl}$ for the TCF. In the same way, the upper integration limit, $\psi$, corresponds to the $99\%$ boundary-layer thickness $\delta_{99}$ for the wing and to the channel height $h$ for the TCF. Note that the $U_e$ and $\delta_{99}$ are attained following the approach in Ref~\cite{vinuesa_determining_2016}.

The channel surrogates are therefore configured to match both quantities simultaneously, ensuring that the near-wall flow environment presented to the DRL controller approximately reproduces the local conditions on the wing under active suction. A secondary configuration was also considered for the presented study in which the surrogate matches only $Re_{\theta}$, is reported in Supplementary Information~3.

\subsection*{Exploration in local TCF surrogates}\label{subsec:explolocalTCF}

\begin{figure}[H]
  \centering
  \includegraphics[width=0.9\textwidth]{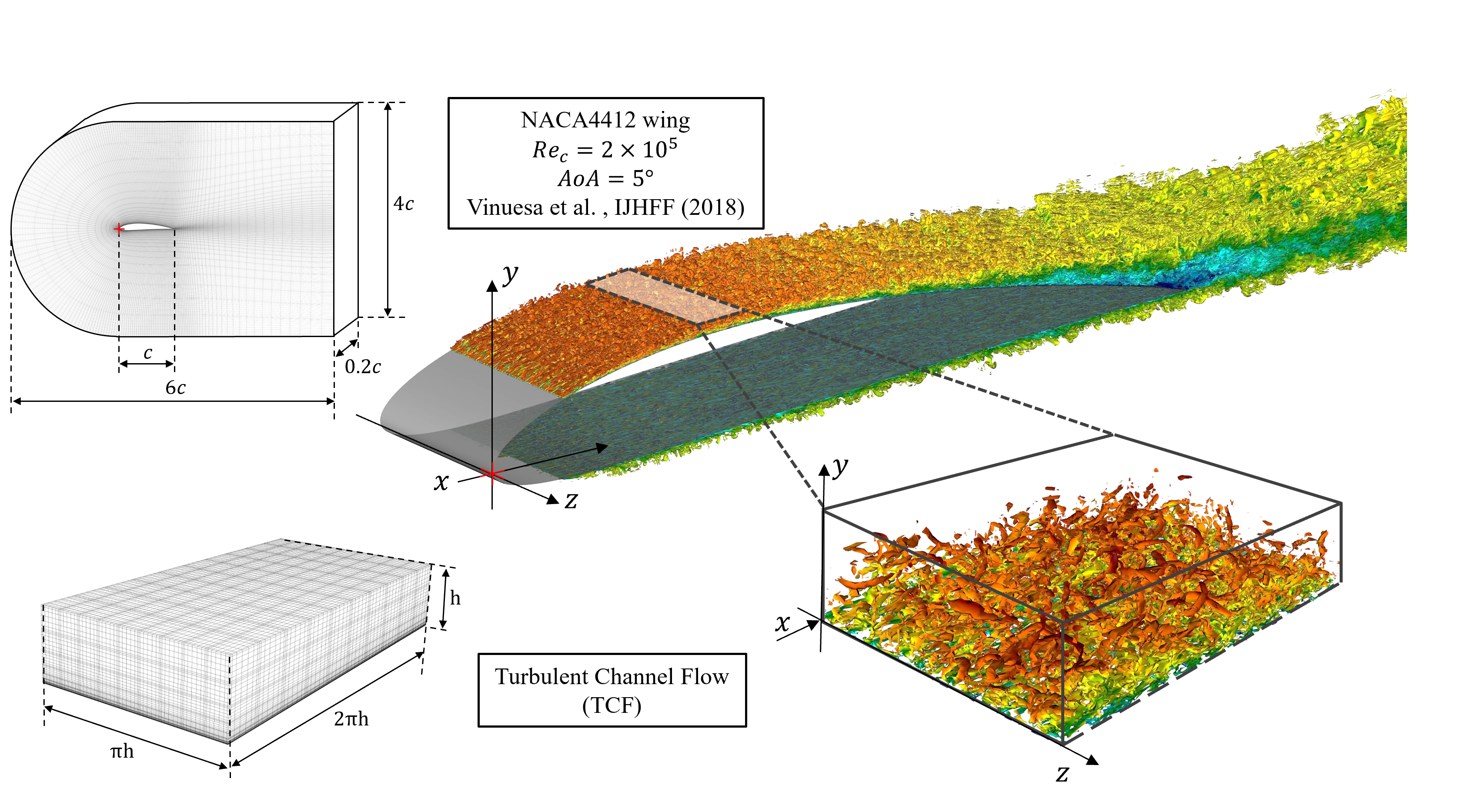}
  \caption{\textbf{Computational setup.}
NACA4412 wing at $Re_c=2\times10^5$ and $\alpha=5^{\circ}$ with the suction-side control region partitioned into four chordwise blocks $\Omega_1$--$\Omega_4$, together with representative turbulent structures from the wing and a matched turbulent channel-flow (TCF) surrogate.}
\label{fig:FIG02_3Dschemematching}
\end{figure}

\begin{figure}[H]
  \centering\includegraphics[width=0.95\textwidth]{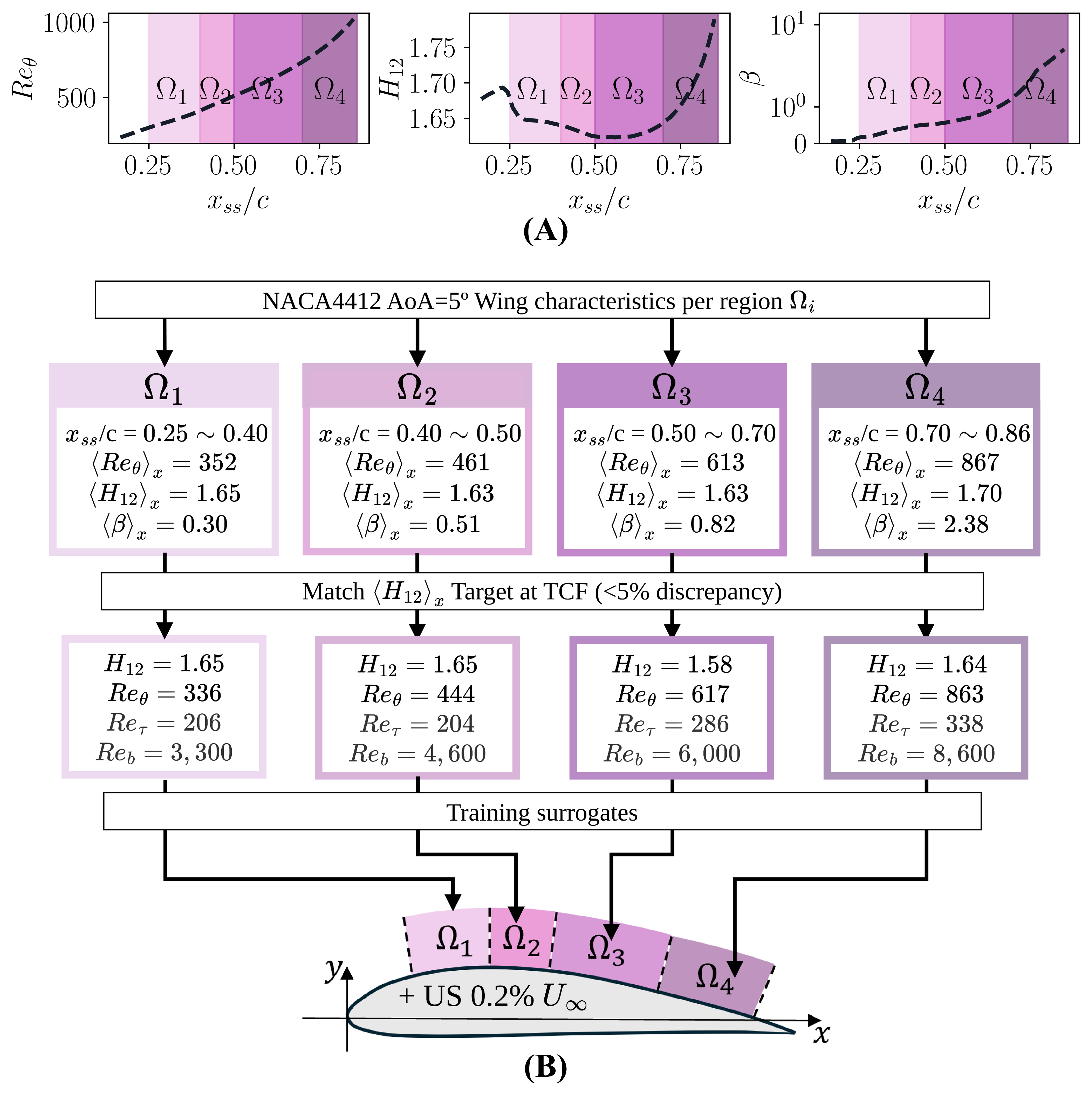}
  \caption{\textbf{Surrogate-matching strategy.}
(A) Chordwise evolution of $Re_{\theta}$, $H_{12}$, and $\beta$ on the suction side of the base flow; shaded regions indicate the blocks where statistics are averaged.
(B) Wing-to-TCF matching tree for DRL-US. Controllers trained on each surrogate are deployed directly to the corresponding wing blocks with uniform suction of $0.2\%\,U_{\infty}$.}
\label{fig:FIG03_treematching}
\end{figure}

Figure~\ref{fig:FIG02_3Dschemematching} illustrates the proposed surrogate approach: instantaneous turbulent structures on the suction side of the wing and in a matched TCF surrogate reveal the degree of similarity in near-wall dynamics that underpins the transfer strategy. The suction-side control region is partitioned into four chordwise blocks $\Omega_1$--$\Omega_4$, based on the chordwise evolution of the Clauser pressure-gradient parameter
\begin{equation}
\beta = \frac{\delta^*}{\tau_w}\frac{\mathrm{d}P}{\mathrm{d}x},
\end{equation}
\noindent where $\tau_w = \rho \nu ({\rm d U_t}/{\rm d} y_n)_{y_n =0}$ is the wall-shear stress, $\delta^*$ the displacement thickness, and $P$ the mean pressure. Defined here based on established knowledge of wing boundary-layer behavior, the partition keeps the APG variation within each block as smooth as possible, promoting the statistical homogeneity required to support periodic TCF surrogates. Values of $\beta \leq 1$ correspond to a mild APG regime, whereas $\beta \geq 5$ indicates a strong APG~\cite{harun_PGEffect_2013}; the four blocks span this range progressively from the leading edge of the control region toward the trailing edge. The block-averaged statistics of $Re_\theta$, $H_{12}$, and $\beta$ are reported in Figure~\ref{fig:FIG03_treematching}(A), alongside Figure~\ref{fig:FIG03_treematching}(B), which presents the matching tree of TCF parameters, including $H_{12}$ (within a 5\% discrepancy), $Re_b$, and $Re_\tau$. Here $Re_b = U_b h/\nu$ and $Re_\tau = u_\tau h/\nu$, where $U_b$ is the bulk velocity and $u_\tau = \sqrt{\tau_w / \rho}$ the friction velocity. We refer to {Method Section} for the details.

An initial reference case without base-flow modification (BASE, OC-noUS, DRL-noUS), presented in Supplementary Information~3, matches surrogates to the wing via $Re_{\theta}$ alone. In this case, the strong APG exhibited in $\Omega_4$ (i.e. $\langle\beta\rangle_x = 5.27$) drives the boundary layer close to separation, yielding local $Re_\tau$ values that cannot be faithfully reproduced in TCF, where the flow is attached. Consequently, MARL training in $\Omega_4$ becomes unfeasible, and only three surrogates were trained for $\Omega_1$--$\Omega_3$, while $\Omega_4$ is either left unactuated or assigned uniform blowing at $0.2\%\,U_\infty$~\cite{atzori_uniform_2021,wang_opposition_2025}. Leaving $\Omega_4$ unactuated yields the higher integral net-energy saving, $5.0\%$ against $4.8\%$ with downstream blowing, both above the $3.7\%$ of OC-noUS (Supplementary Table~3). This reference case highlights a limitation with respect to the four pillars introduced above, in particular generalization: without base-flow modification, the workflow cannot accommodate strong APG conditions, as the resulting boundary-layer states fall outside the range that can be reliably reproduced by TCF surrogates.

The main configuration showcased here (BASE-US, OC-US, DRL-US) introduces a uniform suction of $0.2\%\,U_{\infty}$ over the suction side within the control region. The suction attenuates the APG intensity, improves boundary-layer attachment, and reduces pressure drag by narrowing the wake.
It further brings the statistics of all four blocks into a regime that can be reliably matched by TCF surrogates, enabling independent MARL training in all blocks $\Omega_1$--$\Omega_4$.

Policies are trained over 100 episodes. In all cases, the selected DRL policies surpass the net-energy saving of opposition control in the deterministic evaluation (Supplementary Table~2), consistent with previous findings for MARL applied to TCF~\cite{guastoni_DRL_2023}. Since the episode rewards continue to fluctuate during training, a known characteristic of TD3 in which policy updates can temporarily degrade performance before recovering, the optimal policy is selected as the one achieving the highest reward over the full training history (see Supplementary Information~2 and Supplementary Figure~2).

\subsection*{Zero-shot deployment on the wing section}\label{subsec:zeroshotdeployment}

The DRL controllers, trained exclusively in TCF surrogates, are deployed directly to the suction-side turbulent boundary layer of the NACA4412 wing section at the aforementioned configuration without additional training (we refer to Methods for the numerical setup and Ref.~\cite{vinuesa_turbulent_2018}). Each controlled case is initialized from a single fully developed turbulent state, motivated by the high computational cost of wing simulations and the long integration times required to achieve statistical convergence~\cite{atzori_uniform_2021}; the resulting finite-time sampling uncertainty is quantified in Supplementary Information~4.

Figure~\ref{fig:FIG04_JARwing}(A) shows the chordwise distribution of the local net-energy saving, $\mathrm{NES}_{\mathrm{Wing},\tau}$, for DRL-US and OC-US, alongside the deterministic performance $\mathrm{NES}_{\mathrm{TCF},\tau}$ in the surrogate channels.
The NES, which is also the reward used to train the DRL policies, quantifies the drag-power saving that remains once the power spent on actuation is discounted,
\begin{equation}
\mathrm{NES} = \mathrm{PR} - \mathrm{PIR}, \qquad \mathrm{PR} = 1 - \mathcal{P}^{\rm ctrl}/\mathcal{P}^{\rm ref},
\label{eq:nes_brief}
\end{equation}
\noindent where the superscripts ``ctrl'' and ``ref'' denote the controlled and uncontrolled states, respectively. The quantity $\mathcal{P} = \tau_w U_{\rm drive}$ is the driving power per unit wall area~\citep{bewley_2001_dns}, with $U_{\rm drive}=U_b$ for the channels and $U_{\rm drive}=U_\infty$ for the wing section. Since $U_{\rm drive}$ is unchanged between each controlled case and its reference, PR equals the percentage reduction in wall-shear stress. PIR is the ZNMF actuation power, including pressure work and wall-normal kinetic-energy input, expressed as a percentage of the reference driving power. Positive local NES therefore indicates that the wall-shear-related power saving exceeds the modelled ZNMF actuation cost.

In each channel, the reference is the same surrogate without feedback actuation. On the wing, the local metric uses BASE-US, the flow with uniform suction but without feedback control, as its reference. It therefore measures the additional benefit of OC or DRL on the suction-modified boundary layer. The integral metric, $\mathrm{NES}_{\mathrm{Wing},D}$, instead uses the uncontrolled BASE case and accounts for both total drag and the power required for uniform suction. Its formulation is given in Methods, and the reference states are listed in Table~\ref{tab:wing_reference_table}. The integral results are discussed with Figure~\ref{fig:FIG04_JARwing}(B).

The block-averaged local NES on the wing, shown on the left of Figure~\ref{fig:FIG04_JARwing}(A), is compared with the corresponding channel results on the right. DRL achieves higher NES than OC in all four surrogates, and DRL-US retains this advantage in each corresponding wing block. The difference between the controllers is larger on the wing than in the channels. Averaged over the control region, the local NES is $21.5\%$ for DRL-US and $13.4\%$ for OC-US. These values are averages of locally normalized savings, rather than the integral energy-saving ratio for the complete wing section.

The chordwise distribution identifies where control is most effective. DRL-US reaches a local NES of nearly $28\%$ in $\Omega_2$, at $x_{ss}/c\approx0.40$--$0.50$. The relatively thin boundary layer and weak APG in this region may favour control based on observations at $y_n^+=15$, where the sensed fluctuations reflect the near-wall cycle of streaks and vortices.

Farther downstream, the APG strengthens through $\Omega_3$ and $\Omega_4$, while NES decreases and the actuation cost rises. The greater contribution of outer-layer motions may be only partly represented by observations from the near-wall sensing plane. Larger wall-pressure fluctuations may also increase the pressure-work cost of actuation. Uniform suction partly mitigates these adverse conditions and maintains boundary-layer attachment, reducing the block-averaged $\beta$ in $\Omega_4$ from $5.27$ to $2.38$ (Figure~\ref{fig:FIG03_treematching}(A)). Although the actuation cost reaches its highest value in $\Omega_4$, DRL-US still achieves approximately $40\%$ higher block-averaged local NES than OC-US there.

The balance between wall-shear reduction and actuation cost differs among the blocks. In $\Omega_2$ and $\Omega_4$, the ZNMF power-input ratios, $\mathrm{PIR}^{\mathrm{ZNMF}}_{\mathrm{Wing},\tau}$, are $0.9\%$ and $3.8\%$ for DRL-US, compared with at most $0.6\%$ for OC-US. The additional wall-shear reduction nevertheless compensates for these higher costs. In $\Omega_1$, DRL-US achieves both higher NES and lower actuation cost than OC-US.

The absolute NES values also differ between the wing and its surrogates, with higher values on the wing in $\Omega_2$ and lower values in $\Omega_3$--$\Omega_4$. OC exhibits similar channel--wing differences despite requiring no training. This suggests that differences between the underlying flows contribute to the performance differences, being consistent with the statistical comparisons in Supplementary Information~7. Note that the localized decrease in NES, most prominently near $x_{ ss}/c\approx0.70$, coincides with changes in the applied policy and actuation-update interval at block interfaces. The associated changes in the wall boundary condition may increase local pressure fluctuations and the pressure-work cost of actuation~\cite{stroh_comparison_2015}.

\begin{figure}[H]
    \centering
    \includegraphics[
        width=\linewidth,
        trim=0 0 0 0,
        clip
    ]{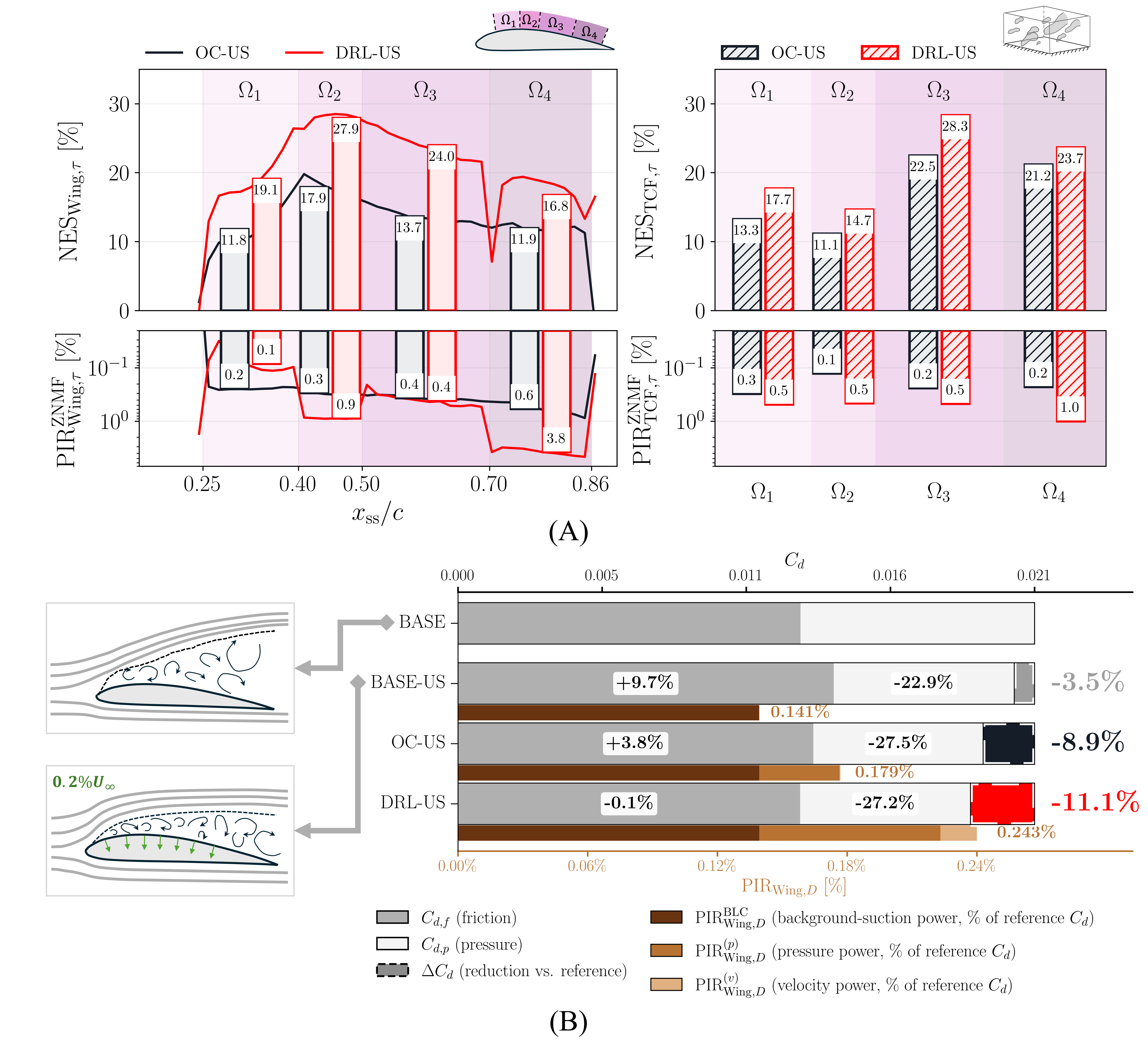}
    \caption{\textbf{Zero-shot control performance summary on the NACA4412 wing section.} (A) Left: chordwise distribution of the local net-energy saving $\mathrm{NES}_{\mathrm{Wing},\tau}$ (top) and ZNMF power-input ratio $\mathrm{PIR}^{\mathrm{ZNMF}}_{\mathrm{Wing},\tau}$ (bottom) for the DRL-US and OC-US, together with the block-averaged values over $\Omega_1$--$\Omega_4$ as bar charts. Right: block-averaged $\mathrm{NES}_{\mathrm{TCF},\tau}$ (top) and $\mathrm{PIR}^{\mathrm{ZNMF}}_{\mathrm{TCF},\tau}$ (bottom) obtained during MARL training in each TCF surrogate $\Omega_1$--$\Omega_4$. (B) Total drag-coefficient budget decomposed into friction ($C_{d,f}$) and pressure ($C_{d,p}$) contributions relative to the uncontrolled base flow, for the uncontrolled case with and without uniform suction (sketched on the left), and for OC-US and DRL-US. Labels on the right indicate the relative change in total drag with respect to BASE, $\Delta C_d/C_d^{\rm BASE} = -\mathrm{PR}_{\mathrm{Wing},D}$, and the dashed boxes mark the corresponding drag reduction. The bars below each case show the power-input ratios of the uniform suction, $\mathrm{PIR}^{\mathrm{BLC}}_{\mathrm{Wing},D}$, and of the ZNMF actuation, $\mathrm{PIR}^{(p)}_{\mathrm{Wing},D}$ and $\mathrm{PIR}^{(v)}_{\mathrm{Wing},D}$, with their sum annotated, on the lower axis in percent of the reference drag power.}
    \label{fig:FIG04_JARwing}
\end{figure}

Figure~\ref{fig:FIG04_JARwing}(B) moves from the local picture to the aerodynamic performance of the complete wing section, which is ultimately what matters for the control-performance pillar. The drag coefficient is defined as $C_d = f_d/(qc)$, where $c$ is the chord length, and $q = \tfrac{1}{2}\rho U_\infty^2$ is the free-stream dynamic pressure. It is decomposed into friction ($C_{d,f}$) and pressure ($C_{d,p}$) contributions obtained by integrating the corresponding wall stresses around the airfoil and projecting them onto the streamwise direction. All changes are measured with respect to the uncontrolled flow without suction (BASE). Uniform suction alone (BASE-US) illustrates the trade-off that the closed-loop controllers must resolve. The suction thins the boundary layer, which narrows the wake and reduces the pressure drag by $22.9\%$, but also increases the wall-shear stress, increasing the friction drag by $9.7\%$. The net result is a modest $3.5\%$ reduction in total drag, as most of the pressure-drag benefit is consumed by the friction penalty.

To distinguish the additional effect of feedback control from that of uniform suction, we evaluate the incremental reduction in each coefficient $X\in\{C_{d,f},C_{d,p},C_d\}$ as
\[
100\% \left (\frac{X^{\mathrm{BASE-US}}-X^{A\mathrm{-US}}}
{X^{\mathrm{BASE}}} \right),
\qquad A\in\{\mathrm{OC},\mathrm{DRL}\}. 
\]
Positive values indicate a reduction beyond that obtained with suction alone. Each coefficient is normalized by its corresponding BASE value. The OC reference uses $\alpha=0.5$ and $y_s^+=15$, which yield the highest net-energy saving among the configurations tested with background suction (Supplementary Information~5).

Adding OC and DRL reduces the friction drag by $6.0\%$ and $9.8\%$ of the BASE friction drag, respectively. The corresponding additional pressure-drag reductions are $4.6\%$ and $4.3\%$ of the BASE pressure drag. DRL therefore compensates almost completely for the friction penalty introduced by suction, while both controllers retain and further enhance its pressure-drag benefit. The incremental total-drag reductions are $5.4\%$ for OC and $7.6\%$ for DRL, normalized by the BASE total drag. Together with the suction contribution, these give total-drag reductions of $8.9\%$ and $11.1\%$, respectively.

Using the unrounded coefficients, DRL provides a $40.5\%$ larger incremental total-drag reduction beyond suction alone than OC. In a direct comparison of the controlled configurations, DRL-US has $2.4\%$ lower total drag than OC-US. This advantage is largely associated with lower friction drag. The approximate $95\%$ confidence interval for the total-drag reduction of DRL-US relative to OC-US is $2.0$--$2.8\%$ (see Supplementary Information~4).

Accounting for the modelled actuation costs, the integral net-energy saving is $3.4\%$ for suction alone, $8.8\%$ for OC-US, and $10.9\%$ for DRL-US. Subtracting the suction-only value gives incremental net-energy savings of $5.4$ and $7.5$ percentage points for OC and DRL, respectively. DRL-US thus improves the integral net-energy saving over OC-US by $2.1$ percentage points, despite its higher actuation cost. These comparisons use the common BASE driving power as the reference.

For reference, a $0.5\%$ reduction in cruise drag corresponds to approximately a $0.4\%$ reduction in aircraft fuel burn~\cite{quadrio_review_2011}, and each kilogram of fuel saved avoids about $3.16$~kg of CO$_2$~\cite{icao_carbon_2024}. Although the present results, obtained at a moderate Reynolds number with idealized actuation, are not intended as a direct estimate of such savings, they provide an indication of their order of magnitude.

\begin{figure}[H]
    \centering
    \includegraphics[
        width=\linewidth,
        trim=0 0 0 0,
        clip
    ]{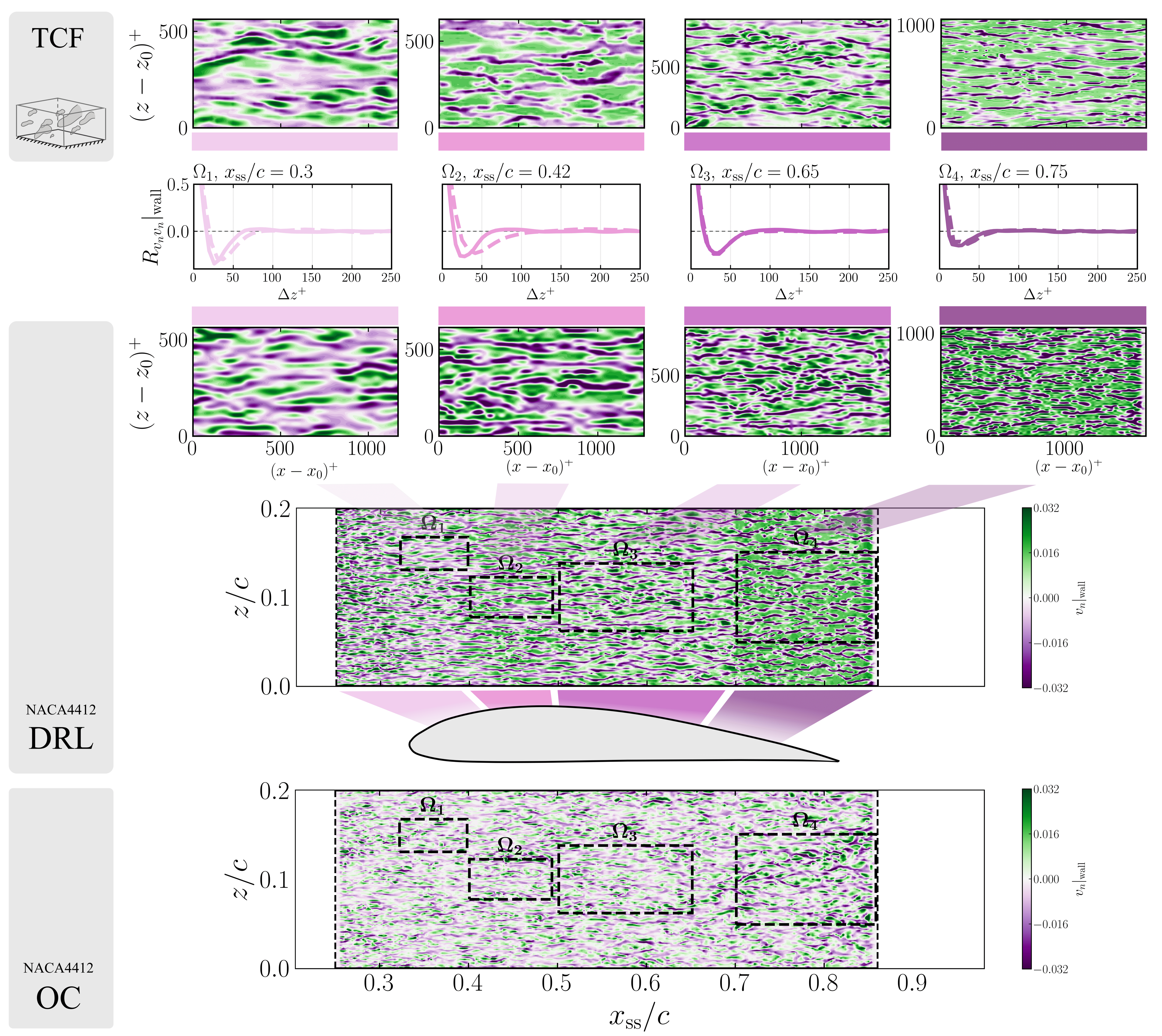}
    \caption{\textbf{Zero-shot transfer of the actuation structure from the channel surrogates to the NACA4412 wing.}
First row (TCF): instantaneous wall-normal actuation velocity $v_n\vert_{\mathrm{wall}}$ in the turbulent-channel-flow surrogate environments $\Omega_1$--$\Omega_4$.
Second row: spanwise two-point correlation of the wall actuation, $R_{v_n v_n}\vert_{\mathrm{wall}}$, as a function of the spanwise separation $\Delta z^+$, comparing each TCF surrogate, dashed, with the corresponding location on the wing ($x_{ss}/c = 0.30$, $0.42$, $0.65$ and $0.75$ for $\Omega_1$--$\Omega_4$), solid line.
Third and fourth rows (DRL): zero-shot on the suction side of the NACA4412 wing for DRL with uniform suction at $0.2\%\,U_\infty$ (DRL-US), showing close-up views of the windows marked by dashed boxes in the full suction-side field below. Wing quantities in wall units are scaled with the block-averaged viscous length $\langle l^* \rangle_{\Omega_i}$.
Bottom row (OC): corresponding deployment for opposition control with uniform suction (OC-US).}
    \label{fig:FIG05_actDISTRO}
\end{figure}

Next, we compare the spatial distribution of wall actuation in the channel surrogates and on the wing section. Figure~\ref{fig:FIG05_actDISTRO} shows that DRL produces streamwise-elongated regions of alternating blowing and suction in the channels (first row). These regions extend over several hundred viscous units and resemble the near-wall streaks of wall-bounded turbulence. This resemblance is consistent with the use of wall-tangential-velocity fluctuation at $y_n^+=15$ as one of the policy inputs. Similar patterns appear within each wing block $\Omega_i$ (third row) and across the controlled suction-side interval (fourth row), although their spatial distribution varies along the chord.

The second row of Figure~\ref{fig:FIG05_actDISTRO} compares the spanwise two-point correlations of wall actuation, $R_{v_nv_n}|_{\rm wall}$, providing a statistical measure of their spanwise distribution near the actuated surface. The representative wing stations are $x_{ss}/c=0.30$, $0.42$, $0.65$ and $0.75$ for $\Omega_1$--$\Omega_4$, respectively. In both geometries, the correlation decreases rapidly and reaches a negative minimum at separations of a few tens of viscous units, consistent with the alternating blowing and suction regions visible in the instantaneous fields. It subsequently approaches zero at separations of order $\Delta z^+\approx100$. These features are consistent with distribution at scales characteristic of near-wall streaks~\cite{jimenez_TBLvsTCF_2010,smits_highReynolds_2011}, although the negative minimum and subsequent decay do not directly define the full streak spacing. The channel and wing profiles show the closest agreement in $\Omega_3$ and $\Omega_4$. In $\Omega_1$ and $\Omega_2$, the channel correlations extend over somewhat larger spanwise separations.

The actuation patterns emerge from the interaction between the feedback law and the evolving flow. Although each agent uses local observations, the feedback modifies those observations through its impact on the near-wall turbulence. The correspondence between differences in the background fluctuations and those in the controlled fields suggests that the policies respond to the local flow distribution. Specifically, the control-induced displacement of inner peak of the wall-tangential-velocity fluctuation towards larger $y_n^+$, documented in Supplementary Information~7, which further illustrates this interaction. Additionally, we also find that our method is sensitive to both surrogate selection and block partitioning: deploying a mismatched policy degrades the performance (Supplementary Table~6) and the partitioning is not unique (Supplementary Figure~8). These findings motivate future work could use finer or adaptive partitions based on the boundary-layer development.

The comparison with OC-US provides additional context for the DRL actuation. In the bottom row of Figure~\ref{fig:FIG05_actDISTRO}, OC produces smaller patches of generally lower amplitude, with less pronounced streamwise elongation. Note that for the wing with uniform suction, we use the OC configuration with $\alpha=0.5$, which achieves the highest control-region average of local NES. Its lower actuation amplitude is therefore consistent with the lower power input observed in most blocks (Figure~\ref{fig:FIG04_JARwing}(A)).

These differences can be interpreted in terms of the observed velocity components and the feedback laws. OC sets the wall actuation in proportion to the negative wall-normal velocity at the sensing plane. DRL additionally observes the wall-tangential component and applies a learned nonlinear feedback law. The elongated DRL patterns are consistent with its access to the wall-tangential-velocity fluctuations associated with near-wall streaks. In turbulent channel flow, these fluctuations have been found to contain more information about future wall shear than the wall-normal component~\citep{Arranz_Lozano-Duran_2024}. Moreover, the single-component sensitivity study (see Supplementary Information~5) support the importance of both the observation and the nonlinear feedback. 

The role of feedback is also relevant when interpreting the organized actuation reported in earlier DRL studies. Using a reward based on wall-shear reduction, Ref.~\cite{sonoda_RL_2023} obtained policies that switch between strong blowing and suction and, in some cases, develop coherent wave-like patterns spanning the computational domain. Imposing the extracted coherent component alone as a predetermined open-loop input, however, produced little drag reduction, indicating that this component alone does not explain the performance of the complete feedback policy. In the present study, the actuation instead resembles the near-wall streaks observed at $y_n^+=15$. The reward formulation may contribute to this difference, since our NES objective explicitly accounts for pressure work and wall-normal kinetic-energy input. The influence of reward design on actuation is also illustrated by Ref.~\cite{tonti2026explainable}, which obtained different actuation patterns and energy costs by changing the flow information used to construct attribution-based rewards.

\subsection*{Computational savings}\label{subsec:compusavings}

\begin{figure}[H]
  \centering
  \includegraphics[width=0.85\textwidth]{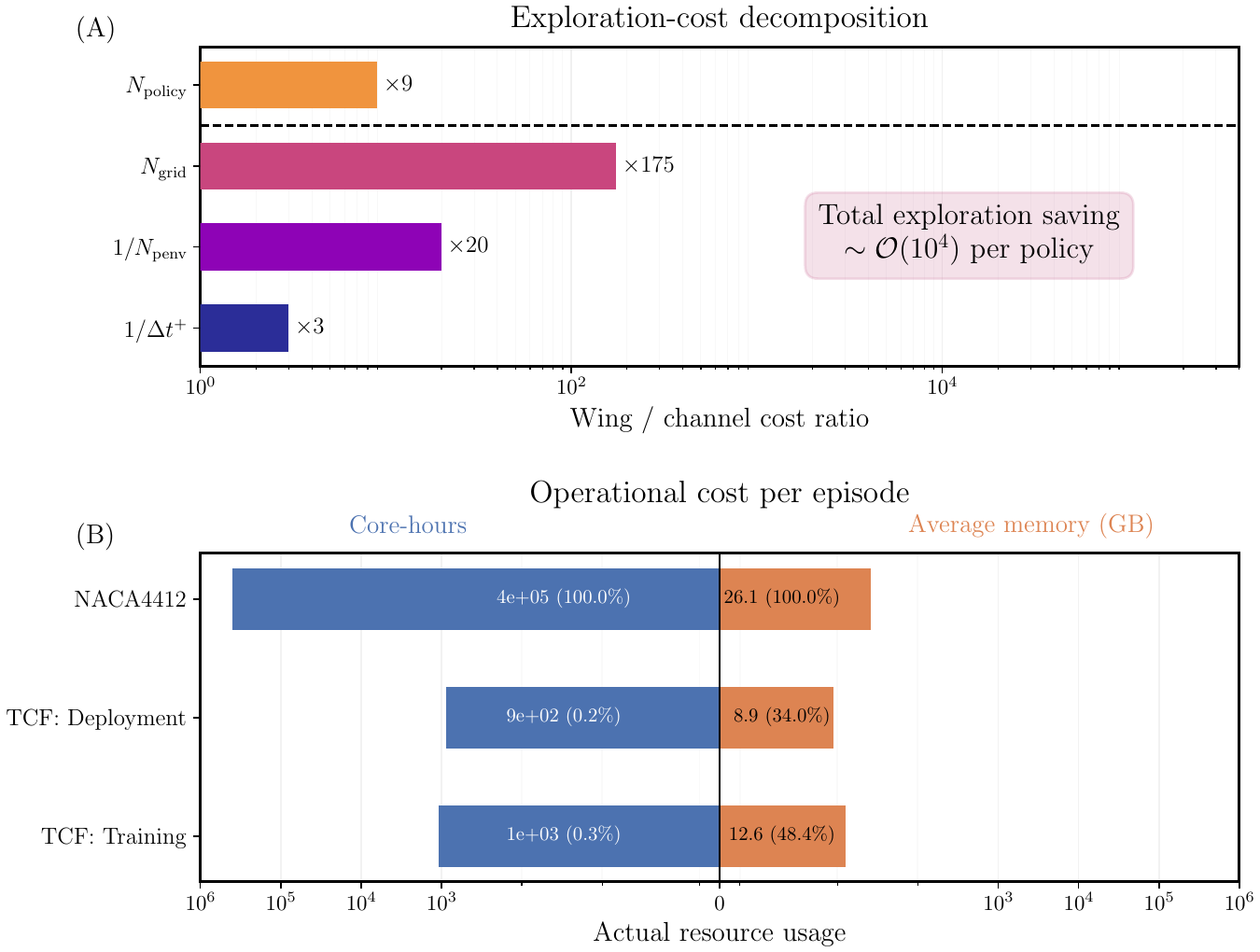}
    \caption{\textbf{Computational and operational cost of zero-shot DRL training and deployment.}
    (A) Computational savings for training DRL controllers, expressed as ratios between the NACA4412 wing and TCF configurations for the factors in equation~\ref{eq:efficiency}. Single-controller cost ratios are shown below the dashed line, and the policy-count ratio $N_{\rm policy}$ required to cover the full control region is shown above; data correspond to control block $\Omega_2$ and its surrogate channel flow.
    (B) Operational cost in terms of core-hours and average memory (GB) for one episode of zero-shot wing deployment, TCF exploitation, and TCF training using Intel Xeon Platinum 8168 CPUs. Percentages in parentheses denote the proportion relative to the cost of wing deployment.}
  \label{fig:FIG06_cost_saving}
\end{figure}

We estimate exploration efficiency in the TCF and wing environments as
\begin{equation}
\Gamma = \frac{N_{\rm grid}}{N_{\rm penv} \cdot \Delta t^+},
\label{eq:efficiency}
\end{equation}
{\noindent where $N_{\rm grid}$ is the number of grid points, $\Delta t^+=\Delta t/t^*$ the time step in viscous units, and $N_{\rm penv}$ the number of MARL pseudo-environments. These factors represent simulation cost and the benefit of parallel experience collection. Their ratios for wing block $\Omega_2$ and its surrogate yield an estimated $10{,}500\times$ efficiency gain per controller (Figure~\ref{fig:FIG06_cost_saving}(A)).}

{A lower bound on the cost of direct on-wing exploration can be inferred from the zero-shot deployment cost. The small turbulent scales near the wing leading edge require a viscous time step approximately three times smaller than in the TCF surrogate. Resolving the complete wing section also requires approximately 175 times more grid points, despite the surrogate's narrower spanwise extent. Streamwise inhomogeneity restricts wing pseudo-environments mainly to spanwise samples and requires approximately 9 times more independent policies to cover the control region.}

{Figure~\ref{fig:FIG06_cost_saving}(B) compares core-hours and memory for one episode of wing deployment, TCF exploitation and TCF training, using the same CPU architecture and near-optimal computing configurations~\cite{wang_opposition_2025}. Wing deployment already costs substantially more than TCF training and provides a lower bound on direct wing exploration costs. Although absolute costs depend on the solver and hardware, the disparity in resolved turbulent scales underlies the computational advantage of the surrogates.}

\section*{Discussion and conclusions}\label{sec:discuss}

We propose a physics-guided zero-shot DRL framework for drag-reduction control of turbulent boundary layers over wing sections. DRL controllers are trained in turbulent channel-flow surrogates matched to the local statistics of the target wing section, and then deployed without additional training. The suction side of a NACA4412 wing section at $Re_c = 2\times10^5$ and $AoA = 5^{\circ}$ is partitioned into chordwise blocks according to the local pressure-gradient intensity, and each block is paired with a TCF surrogate that reproduces its $H_{12}$ and $Re_\theta$ within $5\%$. A uniform suction of $0.2\%\,U_\infty$ alters the flow topology, resulting in a thinner boundary layer, a narrower wake and a weaker APG. Consequently, these changes bring the selected boundary-layer properties of all four blocks within the range represented by the channel surrogates. This includes the most downstream block, which approaches separation without suction and could not be matched using the present surrogates (Supplementary Information~3).

The selected MARL policies achieve higher NES than opposition control in the deterministic channel evaluations, and this ranking is preserved upon zero-shot deployment on the wing. DRL-US achieves a mean local NES of $21.5\%$ over the control region, compared with $13.4\%$ for OC-US, corresponding to an approximately $60\%$ relative improvement. The improvement in block-averaged local NES reaches approximately $75\%$ in $\Omega_3$. Note that these local values use BASE-US as the reference and account for the ZNMF actuation cost. The power required for background suction is included separately in the integral energy budget.

The advantage of DRL persists downstream, although the local NES of both controllers decreases towards the trailing edge, where the APG strengthens and the actuation cost increases. The growing contribution of outer-layer motions may also make observations from a single near-wall plane less representative of the dynamics relevant to control. This downstream deterioration motivates extending the sensing and surrogate descriptions to account more fully for the evolving boundary layer.

At the scale of the complete wing section, the integral drag budget distinguishes the contributions of uniform suction and feedback control. Suction alone reduces pressure drag but increases friction drag, giving a total-drag reduction of $3.5\%$ relative to the BASE case. Adding DRL largely offsets this friction penalty and provides a further reduction equivalent to $7.6\%$ of the BASE total drag, compared with $5.4\%$ for tuned OC. These increments describe the effect of adding feedback controllers upon the suction-modified boundary layer and include the interaction between the two forms of actuation. The incremental reduction is approximately $24.7\%$ larger with DRL, corresponding to a direct reduction of $2.4$ percentage points in total drag relative to OC-US. OC therefore remains an effective baseline. Under the adopted actuation-power model, DRL-US also increases the integral NES by $2.1$ percentage points relative to OC-US.

The actuation fields provide a complementary comparison between the training and deployment environments. DRL produces streamwise-elongated patterns with comparable near-wall spanwise scales in the channel surrogates and on the wing section. This correspondence supports the physical rationale for transfer, despite the remaining differences in the turbulence statistics (Supplementary Information~7). Organized actuation should nevertheless be interpreted in the context of the complete feedback policy. Ref.~\cite{sonoda_RL_2023} found that imposing the extracted coherent wave-like component alone as a predetermined open-loop input produced little drag reduction, showing that this component alone did not explain the performance of their learned controller.

Together, these results demonstrate the transfer of channel-trained policies to a spatially developing wing boundary layer without additional training, while achieving higher NES than the tuned OC benchmark. The estimated improvement of four orders of magnitude in exploration efficiency further supports the use of surrogate environments when direct training on the target geometry is computationally demanding.

Note that a practical limitation shared by OC and DRL is obtaining the required off-wall observations. Measuring the instantaneous velocity fluctuations at the sensing plane, particularly the wall-normal component, is challenging in experiments. Non-intrusive reconstruction from wall measurements offers a possible approach. Ref.~\citep{guastoni_CNN_2021} used convolutional-network models to reconstruct off-wall velocity fluctuations from wall-shear-stress components and wall pressure in turbulent channel flow. Applying such reconstruction to the present controlled wing boundary layer would require validation of both the estimated observations and the resulting closed-loop performance.

In addition to instantaneous observations, the framework requires prior target-flow statistics for surrogate selection, block partitioning and viscous scaling. Zero-shot therefore denotes deployment without additional policy learning on the wing section. These prior statistics are obtained from high-resolution LES in the present study. The Reynolds-averaged Navier--Stokes (RANS) comparisons in Supplementary Information~6 indicate a possible lower-cost alternative: RANS overpredicts the block-averaged $H_{12}$ by $4.9$--$8.5\%$ and underpredicts $Re_\theta$ by $7.3$--$10.5\%$. These estimates could inform surrogate selection, but deployment based entirely on RANS remains to be assessed, including the effects of errors in wall shear on the local viscous scales.

{Future work should assess whether additional boundary-layer statistics and finer or adaptive block partitions improve deployment under stronger APGs, building on the sensitivity tests in Supplementary Information~5. Sequential training could account for upstream--downstream coupling. These developments may reduce the need for background suction, although this requires testing beyond the configurations in Supplementary Information~3. Improved sensing and actuation are also needed.} The surrogate-to-target approach could extend to other complex flows for which direct DRL training is computationally intractable.

\section*{Methods}\label{sec11} 
\subsection*{Numerical simulation}
\subsubsection*{Nek5000 implementation}
The incompressible Navier--Stokes equations govern the fluid motion in the present study.
In non-dimensional form, the continuity and momentum equations are expressed as:
\begin{align}
    \frac{\partial {u}_i}{\partial x_i} &= 0, \\
    \frac{\partial {u}_i}{\partial t} + {u}_j \frac{\partial {u}_i}{\partial x_j} &=
    -\frac{\partial {p} }{\partial x_i} + \frac{1}{{Re}} \frac{\partial^2 {u}_i}{\partial x_j x_j},
\label{eq:incompressible_N-S}
\end{align}
where the Einstein summation convention is adopted, ${u}_i$ denotes the velocity components in Cartesian coordinates, and ${p}$ denotes the instantaneous kinematic pressure. 

All simulations are performed using the open-source, CPU-based solver Nek5000~\cite{nek5000}, which employs the spectral-element method (SEM) to achieve high numerical accuracy while maintaining computational efficiency. The computational domain is partitioned into hexahedral elements, within which the velocity and pressure fields are approximated by high-order Lagrange interpolants following the $\mathbb{P}_N\mathbb{P}_{N-2}$ formulation~\cite{maday_spectral_1989}. For polynomial order $N-1$, the velocity field is represented on $N^3$ grid points per element distributed according to the Gauss--Lobatto--Legendre (GLL) quadrature rule, while the pressure is defined on a staggered grid comprising ${(N-2)}^3$ points per element based on the Gauss--Legendre (GL) quadrature. Temporal integration employs a third-order explicit extrapolation scheme (EXT3) for the nonlinear convective terms and a third-order implicit backward differentiation scheme (BDF3) for the viscous contributions. To mitigate aliasing errors, overintegration is applied by oversampling the nonlinear terms with a factor of $3/2$ relative to the adopted polynomial order $N$ in each spatial direction.

The present simulation framework has been extensively employed in high-fidelity investigations of wing configurations with equivalent numerical setups at Reynolds numbers up to $Re_c = 1\times10^6$~\cite{vinuesa_turbulent_2018}, and a broad range of flow-control strategies has been successfully applied within this framework~\cite{atzori_uniform_2021,wang_opposition_2025}. This comprehensive prior validation motivates the adoption of the present solver.

All simulations are executed on CPU architectures, with parallelization achieved through a hybrid distributed-memory and shared-memory strategy based on OpenMPI. By exploiting the structured meshes and the tensor-product formulation inherent to the SEM, the solver exhibits strong scalability and robust parallel performance. Coupling with reinforcement learning agents is implemented through dynamic MPI-based communication, enabling stable and efficient interaction between the flow solver and the learning framework across all configurations considered.

\subsubsection*{High-resolution large-eddy simulation of the NACA4412 wing section}

{High-resolution large-eddy simulation (LES) resolves the turbulent boundary layers (TBLs) over the NACA4412 wing section at $Re_c = 2\times10^5$ and an angle of attack of $5^{\circ}$, with a grid density approaching that of direct numerical simulation (DNS).

The C-type mesh spans $L_x = 6c$, $L_y = 4c$ and $L_z = 0.2c$ and contains $1.27\times10^5$ spectral elements. The leading and trailing edges lie $2c$ and $3c$ from the inflow boundary, respectively. Previous studies support this spanwise extent for resolving the energetically relevant scales and largest coherent structures of the wing boundary layer~\cite{vinuesa_turbulent_2018,tanarro_effect_2020,atzori_uniform_2021}}

The near-wall spatial resolution, expressed in viscous units, satisfies $\Delta x^+_t < 18.0$, $\Delta y^+_n < (0.64,\,11.0)$, and $\Delta z^+ < 9.0$ in the wall-tangential, wall-normal, and spanwise directions, respectively. The viscous length scale is defined as $l^* = \nu/u_{\tau}$, where the friction velocity is given by $u_{\tau} = \sqrt{\tau_{w}/\rho}$ and the mean wall-shear stress is $\tau_{w} = \rho\nu(\mathrm{d}U_t/\mathrm{d}y_n)_{y_n=0}$. To achieve this resolution, a polynomial order of $N = 11$ is adopted, yielding a total grid-point count of approximately $N_{\rm grid} \approx 2.2\times10^{8}$. In the wake region, the mesh resolution satisfies $\Delta x/\eta < 9$, where $\eta = (\nu^3/\epsilon)^{1/4}$ is the Kolmogorov length scale and $\epsilon$ the local isotropic dissipation rate. 
The relaxation-term LES procedure was validated by Ref.~\cite{negi_unsteady_2018} against DNS for turbulent channel flow and a NACA4412 wing section at $Re_c=4\times10^5$.

The unresolved turbulent scales are represented through an SGS model based on a time-independent relaxation-term filter~\cite{negi_unsteady_2018}. This filtering operation is implemented implicitly via a volume-force formulation that dissipates energy at unresolved scales while preserving mass conservation. With the inclusion of the LES filter, the momentum equation~(\ref{eq:incompressible_N-S}) is augmented as
\begin{align}
    \frac{\partial {u}_i}{\partial t} + {u}_j \frac{\partial {u}_i}{\partial x_j} &=
    -\frac{\partial {p}}{\partial x_i} + \frac{1}{{Re}_c} \frac{\partial^2 {u}_i}{\partial x_j x_j} - \mathcal{H}({u}_i),
    \label{eq:LES_filter}
\end{align}
where $\mathcal{H}$ denotes the high-pass spectral filter that restricts the LES dynamics to a resolved subset of modes within each element. The implementation of this approach in Nek5000 has been extensively validated in prior studies~\cite{vinuesa_turbulent_2018}.

Dirichlet boundary conditions are prescribed at the inflow, upper, and lower boundaries using a far-field velocity distribution obtained from a supporting Reynolds-averaged Navier--Stokes (RANS) computation. The RANS solution is obtained using the $k$--$\omega$ shear-stress transport (SST) model~\cite{menter_kmSST_1994} within a circular domain of radius $200c$. At the outflow boundary, the formulation proposed by~\cite{dong_robustoutlet_2014} is employed to prevent the unphysical influx of kinetic energy. Boundary-layer transition is triggered at $x/c = 0.1$ on both the suction and pressure sides through a localized wall-normal body force~\cite{schlatter_tripping_2012}, which generates intense, time-dependent streaks that subsequently break down to promote transition to turbulence~\cite{vinuesa_wingTBL_2017}.

One flow-over time ($T_{\rm FO}$) is defined as the duration required for a fluid particle convecting at the free-stream velocity $U_{\infty}$ to traverse one chord length $c$. The procedure adopted to generate the uncontrolled reference simulations is described in~\cite{wang_opposition_2025}. Controlled simulations are initialized from fully developed turbulent fields obtained from the corresponding base-flow cases, and statistical convergence is achieved after $\simeq 10\,T_{\rm FO}$~\cite{atzori_uniform_2021, wang_opposition_2025}. The computational cost of simulating $10\,T_{\rm FO}$ of uncontrolled flow amounts to approximately $1\times10^{6}$ core-hours on a BullSequana~X1000 system equipped with Intel Xeon Platinum~8168 CPUs operating at 2.70~GHz with a thermal design power of 205~W.

\subsubsection*{Direct numerical simulation of turbulent channel flow}
{DNS is employed for the TCF configurations that serve as surrogate training environments. The TCF surrogates use an open-channel configuration with an upper symmetry boundary to avoid complications associated with wall actuation~\cite{guastoni_DRL_2023}. Scaling equation~(\ref{eq:incompressible_N-S}) by the channel height $h$ and bulk velocity $U_b$ gives $Re_b = U_b h/\nu$, where $\nu$ is the kinematic viscosity.

Each surrogate matches the streamwise-averaged properties $\langle\cdot\rangle_x$ of the wing base flow within a control block $\Omega_i$. The approximation treats each block as locally fully developed and neglects streamwise growth, thereby justifying the use of TCF, which is by definition fully developed. The matching targets are $Re_\theta = U_0\theta/\nu$ and $H_{12} = \delta^*/\theta$, defined in Eq.~\ref{eq:metrics}.}

Matching $Re_\theta$ is achieved by adjusting $Re_b$ alone, whereas matching $H_{12}$ additionally requires a body-force damping (BDF) term defined as $g(v) = -\gamma v$ within the near-wall region $y < y^+_b$, where $\gamma$ is the forcing intensity and $y^+_b$ is the upper limit of the damped region. This volume force acts as an additional term in the wall-normal momentum equation, attenuating the near-wall momentum to modify $H_{12}$. In practice, $Re_b$ is first adjusted to reproduce the target $Re_\theta$, after which the body-force damping is introduced to achieve the desired shape factor.
The resulting configurations are summarized in Table~\ref{tab:tcf_match_wing}, where all cases exhibit deviations within $5\%$ of the respective target metrics, confirming excellent agreement with the wing base flow.

{The domain dimensions are $(L_x,\,L_y,\,L_z) = (2\pi h,\,h,\,\pi h)$, with spatial resolutions listed in Table~\ref{tab:tcf_configs}. The near-wall spacing in viscous units closely matches the wing simulations to maintain comparable effective actuator sizes. The TCF resolution does not strictly satisfy classical DNS criteria because scales smaller than those resolved on the wing are intentionally omitted. No-slip and symmetry conditions are imposed at the lower and upper boundaries, respectively, with periodic streamwise and spanwise boundaries. A constant bulk velocity drives the flow.}

All simulations are advanced for up to $400$ bulk time units ($T_{b} = h/U_b$) to ensure the turbulent flow attains a statistically stationary state, where $T_b < 200$ were excluded from calculating the turbulence statistics to avoid the initial transient.
Configurations employing body-force damping are initialized from fully developed TCF fields previously advanced for $T_b = 400$, after which an additional $150$ $T_b$ are computed to ensure statistical convergence.
Here, the last $100~T_b$ is used for statistics calculation, whereas the prior fully-developed TCF has to match the $Re_\theta$.
To promote DRL exploration, random perturbations are applied to the initial velocity fields to yield distinct fully developed turbulent states. For each configuration, six independent simulations are conducted with fixed random seeds to control the amplitude of the initial perturbations.

\begin{table}[ht]
  \centering
  \resizebox{\textwidth}{!}{
  \begin{tabular}{cccccccccc}
\toprule
  Case
  &  Block
  & {$x_{ss}/c$}
  & {$\langle Re_\theta \rangle_x$}
  & {$\langle H_{12} \rangle_x$}
  & {$\langle \beta \rangle_x$}
  & {Target}
  & {Value}
  & {Deviation ($\leq 5\%$)}
  & Name  \\
  \midrule
  Base Flow & $\Omega_1$ & $0.25$--$0.40$ & 365  & 1.70 & 0.39  & $Re_\theta$ & 359 & 1.6\% & TCF-1 \\
  -- & $\Omega_2$ & $0.40$--$0.50$ & 522  & 1.69 & 0.73  & --  & 518 & 0.6\% & TCF-2 \\
  -- & $\Omega_3$ & $0.50$--$0.70$ & 739  & 1.72 & 1.42  & -- & 732 & 0.9\% & TCF-3 \\
  -- & $\Omega_4$ & $0.70$--$0.86$ & 1074 & 1.87 & 5.27  & -- & -- & -- & -- \\
  \botrule
  BASE-US $0.2\%\,U_\infty$ & $\Omega_1$ & $0.25$--$0.40$ & 352  & 1.65 & 0.30  & $H_{12}$ & 1.65 & 0.0\% & TCF-BDF-1 \\
  -- & $\Omega_2$ & $0.40$--$0.50$ & 461  & 1.63 & 0.51  & -- & 1.65 & 1.0\% & TCF-BDF-2 \\
  -- & $\Omega_3$ & $0.50$--$0.70$ & 613  & 1.63 & 0.82  & -- & 1.58 & 2.4\% & TCF-BDF-3 \\
  -- & $\Omega_4$ & $0.70$--$0.86$ & 867  & 1.70 & 2.38  & -- & 1.64 & 3.0\% & TCF-BDF-4 \\
  \botrule
  \end{tabular}
}
\caption{\textbf{Flow characteristics of the surrogate channel flows.} Block-averaged statistics of the wing base flows and matching parameters of the surrogate channel flows that reproduce the target metrics within 5\% discrepancy.}
\label{tab:tcf_match_wing}
\end{table}

\begin{table}[ht]
  \centering
  \begin{tabular}{ccccccccccc}
  \toprule
  Name
  & {$Re_b$}
  & {$Re_\tau$}
  & {$N_x$}
  & {$N_y$}
  & {$N_z$}
  & {$\Delta x^+$}
  & {$\Delta y^+_{w}$}
  & {$\Delta z^+$}
  & {$\gamma$}
  & {$y^+_{b}$}\\
  \midrule

  TCF-1  & 3,300 & 206 & 80 & 64 & 72 & 18.8 & 0.5 & 10.1 & -- & -- \\
  TCF-2  & 4,600 & 280 & 80 & 64 & 72 & 25.6 & 0.7 & 15.4 & -- & -- \\
  TCF-3  & 7,000 & 406 & 120 & 96 & 108 & 20.6 & 0.5 & 14.4 & -- & -- \\
  TCF-BDF-1  & 3,300 & 185 & 80 & 64 & 72 & 16.8 & 0.4 & 9.0 & 5.0 & 20.0 \\
  TCF-BDF-2  & 4,300 & 204 & 80 & 64 & 72 & 18.7 & 0.5 & 11.2 & 15.0 & 30.0 \\
  TCF-BDF-3  & 6,000 & 286 & 120 & 96 & 108 & 14.5 & 0.4 & 10.1 & 15.0 & 60.0 \\
  TCF-BDF-4  & 8,600 & 338 & 144 & 108 & 132 & 16.4 & 0.3 & 11.7 & 25.0 & 100.0 \\
\botrule
  \end{tabular}
\caption{\textbf{Channel flow configurations.} The domain size is $(L_x,\,L_y,\,L_z) = (2\pi h,\,h,\,\pi h)$ for all configurations. $\Delta y^+_w$ denotes the wall-normal distance of the first grid point from the wall.}
\label{tab:tcf_configs}
\end{table}

\subsubsection*{Baseline control schemes}
Opposition control (OC)~\cite{choi_active_1994} is employed as the closed-loop reference strategy for benchmarking the drag-reduction performance of the learned controllers. OC imposes a wall-normal blowing or suction velocity at the wall that is opposite in sign to the off-wall fluctuation of the wall-normal velocity component, thereby attenuating near-wall momentum. This is expressed as:
\begin{equation}
 v(x,0,z,t) = -\alpha\left[ v(x,y_s,z,t) - \langle v(x,y_s,z,t)\rangle \right],
\label{eq:oc}
\end{equation}
where $\alpha\in(0,1]$ denotes the control amplitude and $y_s$ is the sensing plane location. An off-wall sensing plane at $y^+_s = 15$ and an amplitude of $\alpha = 1$ are prescribed for the uncontrolled base flow, whereas for OC-US the amplitude is reduced to $\alpha = 0.5$, which yields the highest NES among the tested configurations (Supplementary Information~5), consistent with the generally optimal configuration identified in both channel and wing flows~\cite{hammond_observed_1998,wang_opposition_2025}. The local friction velocity $u_\tau$ is used to evaluate wall units when OC is applied to the wing; further implementation details are provided in Ref.~\cite{wang_opposition_2025}.

In addition, uniform blowing (UB) and uniform suction (US) are implemented as Dirichlet boundary conditions at the wall, with the Cartesian velocity components imposed such that the magnitude of the wall-normal velocity equals the prescribed control intensity.

\subsection*{Net-energy-saving (NES) control policies}

\begin{figure}
    \centering
    \includegraphics[width=\linewidth]{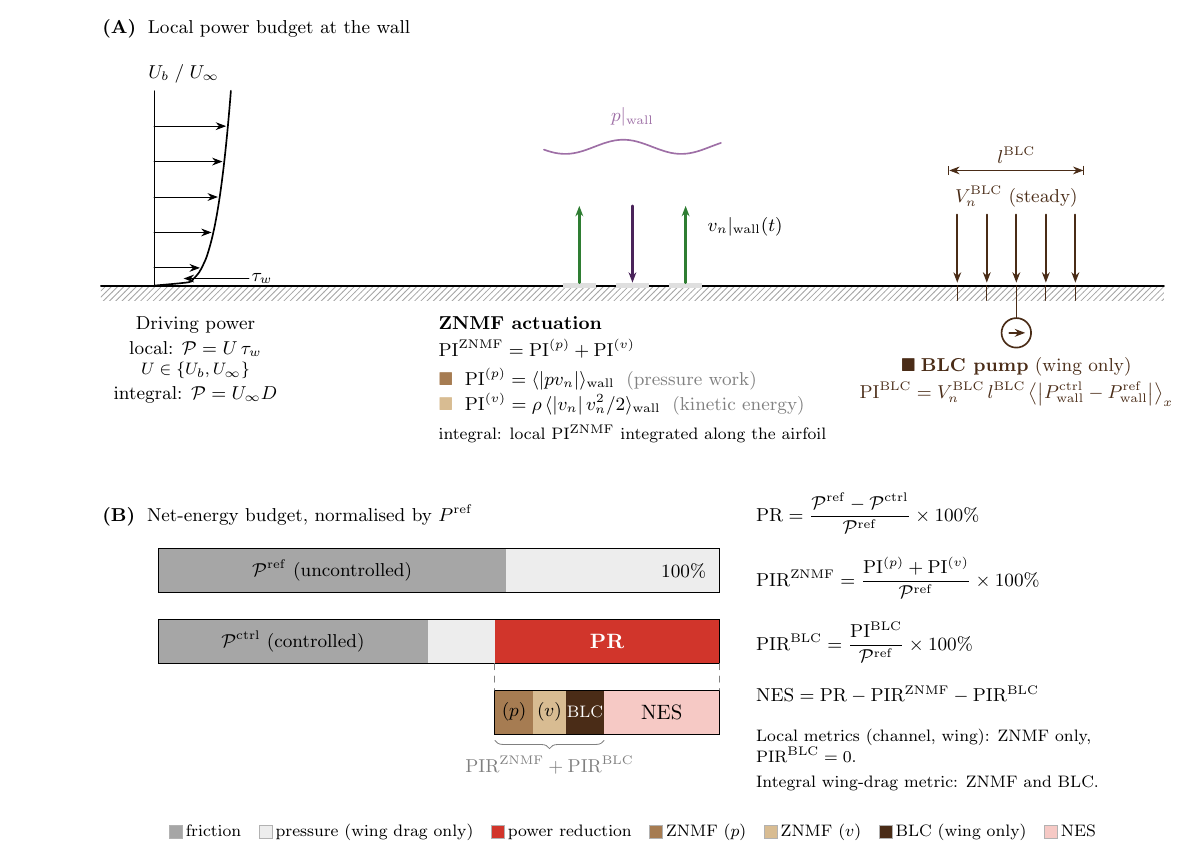}
    \caption{Sketch of the energy budget underlying the net-energy saving (NES). (A) Driving-power density at the wall, $\mathcal{P}=U\tau_w$, with $U\in\{U_b,U_\infty\}$, together with the pressure-work and kinetic-energy contributions to ZNMF actuation power, $\mathrm{PI}^{(p)}$ and $\mathrm{PI}^{(v)}$, and the pumping power associated with uniform suction, $\mathrm{PI}^{\mathrm{BLC}}$, on the wing section. (B) Power budget normalized by the reference driving power $\mathcal{P}^{\mathrm{ref}}$. Subtracting the applicable actuation-power ratios, $\mathrm{PIR}^{\mathrm{ZNMF}}$ and $\mathrm{PIR}^{\mathrm{BLC}}$, from the power-reduction ratio PR gives NES. Bar lengths are illustrative.}
    \label{fig:nes_sketch}
\end{figure}

Drag reduction alone does not determine the net energetic benefit of control, because the actuation power can offset the power saved by reducing drag. For wall-normal blowing and suction, the actuation cost depends on the wall-pressure fluctuations and the imposed velocity through pressure-work and kinetic-energy contributions. We therefore use net-energy saving (NES) as the training reward for the main DRL policies and as the metric for evaluating control performance. Figure~\ref{fig:nes_sketch} illustrates the contributions to the energy budget.

The subscript $\mathcal{C}\in\{\mathrm{TCF},\mathrm{Wing}\}$ identifies the flow configuration (i.e.\ turbulent channel flow or wing section), and $\mathcal{Q}\in\{\tau,D\}$ identifies the wall-shear-based or integral-drag metric. The integral-drag metric is used only for the wing section. Superscripts $\mathrm{ctrl}$ and $\mathrm{ref}$ denote the controlled and reference states, respectively. The actuation types are identified by $\mathrm{ZNMF}$ and $\mathrm{BLC}$, while $(p)$ and $(v)$ identify the pressure-work and kinetic-energy contributions. The operator $\langle\cdot\rangle$ denotes averaging over time and the spanwise direction. For the channels, $\langle\cdot\rangle_{\mathrm{wall}}$ denotes averaging over time and the entire actuated wall. The operator $\langle\cdot\rangle_x$ denotes averaging over the chordwise extent of the BLC region. Integral forces and powers for the wing section are reported per unit span.

The driving power required to overcome drag at a fixed characteristic velocity is
\begin{align}
    \mathcal{P}_{\mathrm{TCF},\tau} &= U_b\, \tau_w, \label{eq:p_tcf}\\
    \mathcal{P}_{\mathrm{Wing},\tau}(\xi) &= U_\infty\, \tau_w(\xi), \label{eq:p_wing_tau}\\
    \mathcal{P}_{\mathrm{Wing},D} &= U_\infty\, D, \label{eq:p_wing_d}
\end{align}
\noindent where $\xi$ is the curvilinear coordinate along the airfoil perimeter and $D$ is the total drag per unit span. Equations~\eqref{eq:p_tcf} and~\eqref{eq:p_wing_tau} define local power densities per unit wall area, whereas equation~\eqref{eq:p_wing_d} gives the integral wing-section power per unit span. The power-reduction ratio is
\begin{equation}
    \mathrm{PR}_{\mathcal{C},\mathcal{Q}} = \frac{\mathcal{P}^{\mathrm{ref}}_{\mathcal{C},\mathcal{Q}} - \mathcal{P}^{\mathrm{ctrl}}_{\mathcal{C},\mathcal{Q}}}{\mathcal{P}^{\mathrm{ref}}_{\mathcal{C},\mathcal{Q}}} \times 100\%.
    \label{eq:pr}
\end{equation}

Following Ref.~\citep{bewley_2001_dns}, we account for the pressure-work and kinetic-energy contributions to the ZNMF actuation power:
\begin{equation}
    \mathrm{PI}^{\mathrm{ZNMF}}_{\mathcal{C},\tau} = \mathrm{PI}^{(p)}_{\mathcal{C},\tau} + \mathrm{PI}^{(v)}_{\mathcal{C},\tau},
    \label{eq:pi_znmf_local}
\end{equation}
\noindent with
\begin{align}
    \mathrm{PI}^{(p)}_{\mathcal{C},\tau} &= \left\langle |p v_n| \right\rangle \big\vert_{\mathrm{wall}}, \label{eq:pi_p}\\
    \mathrm{PI}^{(v)}_{\mathcal{C},\tau} &= \rho \left\langle |v_n|\, v_n^2/2 \right\rangle \big\vert_{\mathrm{wall}}, \label{eq:pi_v}
\end{align}
\noindent where $p\vert_{\mathrm{wall}}$ is the wall-pressure fluctuation, and $v_n\vert_{\mathrm{wall}}$ is the velocity imposed by the ZNMF actuation, and $\rho$ is the fluid density. The absolute values count both signs of the pressure work and kinetic-energy flux as actuation costs, following the Type-I convention. These local contributions are power densities per unit wall area. In the channels, they are additionally averaged over the streamwise direction, giving the wall average $\langle\cdot\rangle_{\mathrm{wall}}$. On the wing, they are evaluated at each position $\xi$. The ZNMF power input for the integral wing-drag budget is then obtained by integrating along the airfoil perimeter $\Xi$:
\begin{equation}
    \mathrm{PI}^{\mathrm{ZNMF}}_{\mathrm{Wing},D} = \int_{\Xi} \mathrm{PI}^{\mathrm{ZNMF}}_{\mathrm{Wing},\tau}(\xi)\, \mathrm{d}\xi .
    \label{eq:pi_znmf_integral}
\end{equation}

On the wing, uniform suction or blowing is additionally applied, known as boundary-layer control (BLC). Unlike ZNMF actuation, BLC imposes a mean, rather than fluctuating, wall-normal velocity, so its power input is accounted for separately. Ref.~\citep{fahland_2021_blc_drag} estimates the BLC cost from the air-supply and pump powers. We retain only the pump power, since the air-supply power requires assumptions on the actuator hardware~\citep{fahland_2024_drag} that are beyond the scope of this numerical study:
\begin{equation}
    \mathrm{PI}^{\mathrm{BLC}}_{\mathrm{Wing},D} = V^{\mathrm{BLC}}_{n}\, l^{\mathrm{BLC}} \left\langle \left| P^{\mathrm{ctrl}}_{\mathrm{wall}} - P^{\mathrm{ref}}_{\mathrm{wall}} \right| \right\rangle_x ,
    \label{eq:pi_blc}
\end{equation}
\noindent where $V^{\mathrm{BLC}}_n$ is the mean wall-normal velocity, $l^{\mathrm{BLC}}$ the chordwise extent of the BLC region, and $P^{\mathrm{ctrl}}_{\mathrm{wall}}$ and $P^{\mathrm{ref}}_{\mathrm{wall}}$ the corresponding mean wall pressures. Using the chordwise-averaged pressure difference is a more favourable estimate than that of Refs.~\citep{beck_2018_drag,fahland_2021_blc_drag}, which use only the minimum pressure difference within the control region.

The power-input ratio is defined relative to the reference driving power,
\begin{equation}
    \mathrm{PIR}^{a}_{\mathcal{C},\mathcal{Q}} = \frac{\mathrm{PI}^{a}_{\mathcal{C},\mathcal{Q}}}{\mathcal{P}^{\mathrm{ref}}_{\mathcal{C},\mathcal{Q}}} \times 100\%, \qquad a \in \{\mathrm{ZNMF}, \mathrm{BLC}\}.
    \label{eq:pir}
\end{equation}
\noindent For BLC, using the pressure coefficient $c_p = (P - P_\infty)/(\tfrac{1}{2}\rho U_\infty^2)$ and the drag coefficient $C_d = D/(\tfrac{1}{2}\rho U_\infty^2 c)$, where $P_\infty$ is the mean free-stream pressure and $c$ is the chord, this becomes
\begin{equation}
    \mathrm{PIR}^{\mathrm{BLC}}_{\mathrm{Wing},D} = \frac{V^{\mathrm{BLC}}_n}{U_\infty}\,\frac{l^{\mathrm{BLC}}}{c}\,
    \frac{\left\langle \left| c_p^{\mathrm{ctrl}} - c_p^{\mathrm{ref}} \right| \right\rangle_x}{C_d^{\mathrm{ref}}} \times 100\%.
    \label{eq:pir_blc}
\end{equation}
\noindent Finally, the net-energy saving is
\begin{equation}
    \mathrm{NES}_{\mathcal{C},\mathcal{Q}} = \mathrm{PR}_{\mathcal{C},\mathcal{Q}} - \sum_{a} \mathrm{PIR}^{a}_{\mathcal{C},\mathcal{Q}},
    \label{eq:nes}
\end{equation}
\noindent where the sum includes only the actuation types present in each case: ZNMF for the TCF and the local wing metric, and both ZNMF and BLC for the wing drag. As illustrated in Figure~\ref{fig:nes_sketch}(b), a positive NES therefore requires the power saved through the drag reduction to exceed the total power spent on the actuation.

\subsection*{Multi-agent reinforcement learning (MARL) framework}
\subsubsection*{MARL framework implementation}

{The MARL agents share a policy $\pi(a|s)$ that maps local observations $s$ to actions $a$ and maximizes a common global reward $r$. Policy sharing reduces the control-space dimensionality and allows agents to reuse observations under locally invariant conditions~\cite{belus_exploiting_2019}.

Each agent observes a local partition, or pseudo-environment (Figure~\ref{fig:FIG01_schematic}). One pseudo-environment is assigned to each active wall grid point. The TCF and wing grid spacings are matched in viscous units to maintain comparable actuator sizes. Each agent produces one action, $a\in\mathbb{R}^{1\times1}$, from $N_{\rm feat}$ observations, $s\in\mathbb{R}^{N_{\rm feat}\times1}$.

Of the $N$ velocity nodes per spectral-element edge, the first and last are shared with adjacent elements. Only unique wall nodes define pseudo-environments, avoiding duplicate replay-buffer samples. Shared nodes receive identical actuation. The number of pseudo-environments is
\begin{equation}
 N_{\rm penv} = N_{{\rm CTRL},x} \times N_{{\rm CTRL},z}
              = N'_{\rm elem,x} \times N'_{\rm elem,z} \times (N-1)^2,
\label{eq:n_penv}
\end{equation}
where $N'_{\rm elem,x}$ and $N'_{\rm elem,z}$ are the numbers of spectral elements within the control region in the streamwise and spanwise directions.}

The reward function is defined as the net-energy saving of the channel flow,
\[
r = \mathrm{NES}_{\mathrm{TCF},\tau} = \mathrm{PR}_{\mathrm{TCF},\tau} - \mathrm{PIR}^{\rm ZNMF}_{\mathrm{TCF},\tau},
\]
\noindent introduced in Eq.~\eqref{eq:nes}, which rewards the reduction in wall-shear stress while penalizing the power spent on the ZNMF actuation. The power input is evaluated with the Type-I definition, which was found to yield better-performing policies than the alternative Type-II definition (Supplementary Information~1). In channel flow, the wall-shear stress and the actuation power are statistically uniform along both the streamwise and spanwise directions.
This property offers the global invariance that ensures the controller learns to exploit local flow invariance, which is the core principle underlying the MARL approach.

The observable state available to each agent consists of the inner-scaled fluctuation components of the streamwise ($u'^+$) and wall-normal ($v'^+$) velocity at an off-wall sensing plane located at $y_s^+ = 15$, yielding $N_{\rm feat} = 2$ input features. The spatial mean is subtracted from each feature prior to its use: in channel flow, averaging is performed over both the streamwise and spanwise directions, whereas on the wing, averaging is performed exclusively in the spanwise direction at each chordwise location $x_{ss}/c$, owing to the absence of streamwise homogeneity.

The control action $a$ corresponds to the time-varying wall-normal velocity imposed at the wall as a Dirichlet boundary condition, with the action space bounded by the local friction velocity magnitude, $|a|\leq u_\tau$. This boundary treatment models localized jet actuation inducing blowing or suction at the solid surface. To prevent unrealistically large instantaneous actuation arising from the incompressibility constraint, a zero-net-mass-flux (ZNMF) condition is enforced across the control region by subtracting the spatial mean of the wall actuation prior to its application.

Model-free, off-policy algorithms such as deep deterministic policy gradient (DDPG)~\cite{lillicrap_ddpg_2015} and its extension, twin delayed DDPG (TD3)~\cite{fujimoto_td3_2018}, have been widely adopted in turbulent flow-control studies~\cite{sonoda_RL_2023,guastoni_DRL_2023,cavallazzi_deep_2025,beneitez_improving_2025,zhou_reinforcement-learning-based_2025}. TD3 mitigates the overestimation bias inherent in DDPG by employing two independent critic networks, and is adopted here as the learning algorithm owing to its robustness and demonstrated effectiveness in recent flow-control studies~\cite{zhou_reinforcement-learning-based_2025,beneitez_improving_2025}. The TD3 architecture comprises an actor network and two critic networks: the actor approximates the policy mapping from states to actions, while the critics estimate the action--value function $Q(s,a)$, with the the minimum of the two target-critic estimates used to construct the Bellman target~\cite{fujimoto_td3_2018}. Both networks are implemented as fully connected multi-layer perceptrons (MLPs). The actor comprises a single hidden layer with 8 neurons, whereas each critic consists of three hidden layers with 16, 64, and 64 neurons, respectively.

The DRL framework is implemented in Python using Stable-Baselines3~\cite{stable-baselines3} as the reinforcement-learning library, coupled to a custom vectorized PettingZoo/Gym environment~\cite{terry_pettingzoo_2020}. Communication between the Python-based DRL framework and the \textsc{Fortran}~77 CFD solver is achieved through a dedicated message-passing interface (MPI) coupling layer, which offers superior efficiency and scalability compared with file-based input/output approaches~\cite{suarez_active_2025}. The solver and learning framework exchange observations, actions and rewards through MPI, as in Ref.~\cite{guastoni_DRL_2023}. In the present implementation, the two programs are launched concurrently, whereas Ref.~\citep{guastoni_DRL_2023} launch the solver as a child process.
The current design choice facilitates coupling with computationally intensive wing simulations.
Note that the adoption of \textsc{Fortran}~77 precludes the use of more advanced coupling interfaces such as those proposed in Ref.~\cite{font_deep_2025}, making MPI an optimal compromise in terms of robustness, efficiency, and scalability.

\subsubsection*{Exploration in the channel-flow surrogates}
Training is conducted using six distinct initial conditions over 100 episodes. At the beginning of each episode, an initial condition is selected via a random number generator with a prescribed seed to ensure reproducibility. Each episode spans a simulation time of $t^+ = 1{,}500$, scaled by the viscous time unit $t^* = \nu/u^2_\tau$, and the policy is updated every $\Delta t^+ = 180$ using 64 mini-batch gradient steps. The control policy is evaluated at intervals of $\Delta t^+ = 0.6$, resulting in 2{,}500 actuation updates per episode. Mini-batches of size $\mathcal{B}$ are sampled from a replay buffer of capacity $\mathcal{D}$ containing state--action--reward tuples; both $\mathcal{B}$ and $\mathcal{D}$ scale with the grid resolution consistently with the methodology outlined in Ref.~\cite{guastoni_DRL_2023}. {The the adopted values are summarized in Supplementary Information~8; all remaining parameters are set to the default values of the Stable-Baselines3 implementation~\cite{stable-baselines3}.}

\subsubsection*{Zero-shot exploitation on the wing section}

Following exclusive training in the TCF surrogate environments, the optimized policies are deployed directly to their corresponding control blocks on the NACA4412 wing section without any additional learning, constituting zero-shot exploitation. The controlled region on the suction side is defined as $x_{ss}/c = 0.25$--$0.86$, consistent with prior investigations~\cite{atzori_uniform_2021,wang_opposition_2025}. Table~\ref{tab:wing_reference_table} summarizes the control configurations applied on the wing sections in the present study.

\begin{table}[ht]
  \centering
  \small
  \setlength{\tabcolsep}{4pt}
  \renewcommand{\arraystretch}{1.15}

  \begin{tabularx}{\textwidth}{@{}l X c cc@{}}
    \toprule
    \multirow{2}{*}{Case}
    & \multirow{2}{*}{Control configuration}
    & \multirow{2}{*}{\shortstack{Background\\flow}}
    & \multicolumn{2}{c}{NES reference} \\
    \cmidrule(lr){4-5}
    & & & Local & Integral \\
    \midrule

    BASE
    & Uncontrolled flow
    & -- & -- & -- \\

    OC-noUS
    & OC with $\alpha=1.0$ and $y_n^+=15$
    & BASE & BASE & BASE \\

    DRL-noUS
    & DRL on $\Omega_1$--$\Omega_3$
    & BASE & BASE & BASE \\

    DRL-noUS-UB
    & DRL on $\Omega_1$--$\Omega_3$, with UB at
      $0.2\%\,U_\infty$ on $\Omega_4$
    & BASE & BASE & BASE \\

    \midrule

    BASE-US
    & US at $0.2\%\,U_\infty$
    & BASE & BASE & BASE \\

    OC-US
    & OC with $\alpha=0.5$ and $y_n^+=15$
    & BASE-US & BASE-US & BASE \\

    DRL-US
    & DRL on $\Omega_1$--$\Omega_4$
    & BASE-US & BASE-US & BASE \\

    \bottomrule
  \end{tabularx}

  \caption{\textbf{Wing control configurations and reference states
  for net-energy saving.}
  The listed control is applied to the indicated background flow.
  US and UB denote uniform suction and uniform blowing, respectively.
  OC and US act over $0.25\leq x_{ss}/c\leq0.86$.
  The last two columns identify the reference states used to evaluate
  the local wall-shear-based NES, $\mathrm{NES}_{\mathrm{Wing},\tau}$,
  and the integral drag-based NES, $\mathrm{NES}_{\mathrm{Wing},D}$,
  respectively.}
  \label{tab:wing_reference_table}
\end{table}
Several critical modifications to the channel-flow implementation are required to accommodate the spatial development inherent to TBLs. Chief among these is the fact that, unlike in fully developed channel flow, the friction velocity varies continuously along the chordwise direction, i.e.\ $u_\tau = u_\tau(x_{ss}/c)$. This spatial variation is reconstructed using a ninth-order polynomial fit of the $u_\tau$ distribution from the database of Ref.~\cite{vinuesa_turbulent_2018}, yielding an accurate representation of the reference friction-velocity profile along the suction side.

As a direct consequence, four key control parameters: (1) the bounds of the actuation amplitude, (2) the scaling of the observable state, (3) the wall-normal position of the sensing plane at $y^+ = 15$, and (4) the control update interval $\Delta t^+ = 0.6$ all become explicit functions of $x_{ss}/c$. The formulations of (1), (2), and (3) become straightforward once $u_\tau = u_\tau(x_{ss}/c)$ is established. Enforcing a consistent update interval at each streamwise location, however, presents a considerably greater challenge. To address this, the update interval is prescribed to be spatially uniform within each control block $\Omega_i$, with the viscous time scale computed from the block-averaged friction velocity as $t^*_{\Omega_i} = \nu/\langle u_\tau\rangle^2_{\Omega_i}$. This block-wise treatment results in distinct actuation update frequencies across control blocks, so that the information exchange frequency between the DRL framework and the CFD solver is set to the greatest common divisor (GCD) of the individual block update frequencies, thereby preserving local viscous scaling throughout the deployment phase. This spatially resolved treatment of viscous scales substantially increases the computational cost relative to a uniform-scaling approach, underscoring the importance of the surrogate-based training strategy for making DRL-based control of wing TBLs tractable.

According to Eq.~(\ref{eq:n_penv}), $N_{{\rm CTRL},x} = 1$ when exploration is conducted directly on the wing within the MARL formulation. However, under the assumption of locally parallel flow within each control block and deterministic policies, the effective number of pseudo-environments during deployment is instead determined by the local mesh resolution. For the NACA4412 configuration, the control region comprises $N_{{\rm elem},x} = 34$ and $N_{{\rm elem},z} = 27$ spectral elements, yielding $N_{\rm penv} = 111{,}078$ pseudo-environments during zero-shot exploitation.

The state vector for the wing TBLs consists of the inner-scaled wall-tangential ($u'^+_t$) and wall-normal ($v'^+_n$) velocity fluctuations sampled at an off-wall sensing plane at $y_s^+ = 15$, determined locally from $u_\tau(x_{ss}/c)$. The spatial mean of each feature is removed by averaging in the spanwise direction at each chordwise location $x_{ss}/c$, owing to the absence of streamwise homogeneity in spatially developing TBLs. The control action is bounded using the local friction velocity $u_\tau(x_{ss}/c)$, and the ZNMF constraint is enforced by subtracting the spanwise-averaged wall actuation within each control block.

As the numerical solver operates in Cartesian coordinates whereas the actuation is defined in the wall-normal direction, the horizontal and vertical velocity components are modified such that their projection along the wall-normal direction recovers the prescribed control intensity. Conversely, the observable states are defined in the wall-normal reference frame, so that both Cartesian velocity components contribute to the wall-tangential and wall-normal velocities according to the local surface curvature. 

\clearpage

\section*{Data availability}
The processed data underlying the main results, together with the associated analysis scripts, will be made publicly available through the University of Michigan's Deep Blue Data repository at \url{https://doi.org/10.7302/5yeg-rv23} under a Creative Commons Attribution 4.0 International (CC BY 4.0) license. The full CFD simulation datasets are available from the corresponding author upon reasonable request owing to their large size and the solver-specific tools required for post-processing.

\section*{Code availability}
The complete production and analysis code underlying the results reported in this study will be made publicly available upon acceptance at \url{https://github.com/VinuesaLAB-AI/ZeroShot_MARL_Nek5000.git}. The release will include the trained policy checkpoints and instructions for installing the software dependencies and running the control cases.

\section*{Acknowledgments}
\label{sec:acknowledges}
R.V. acknowledges financial support from ERC grant no.2021-CoG-101043998, DEEPCONTROL. Views and opinions expressed are however those of the author(s) only and do not necessarily reflect those of the European Union or the European Research Council. Neither the European Union nor the granting authority can be held responsible for them. The authors gratefully acknowledge the Gauss Centre for Supercomputing e.V. (\url{www.gauss-centre.eu}) for funding this project by providing computing time through the John von Neumann Institute for Computing (NIC) on the GCS Supercomputer JUWELS at J{\"u}lich Supercomputing Centre (JSC).

\section*{Author Contributions}
Wang, Y.: Methodology, Software, Validation, Investigation, Data Curation, Writing---Original Draft, Writing---Review \& Editing, Visualization.
Suárez, P.: Methodology, Validation, Investigation, Writing---Original Draft, Writing---Review \& Editing, Visualization.
Bode, M.: Methodology, Software, Validation, Writing---Review \& Editing.
Vinuesa, R.: Ideation, Methodology, Writing---Review \& Editing, Resources, Funding Acquisition.

\section*{Competing interests}
The authors declare no competing interests. 

\bibliography{sn-bibliography}%

\end{document}

\setcounter{page}{\numexpr\MainArticlePages+1\relax}

\title{
    \vspace{-1cm}
    \centering
    \textbf{Supplementary Information} \\[1.2ex]
    \large{Physics-guided surrogate learning enables zero-shot control of turbulent wings}
}

\author*[1]{\fnm{Yuning} \sur{Wang}}\email{yuninw@umich.edu}

\author[2]{\fnm{Pol} \sur{Suárez}}\email{polsm@kth.se}

\author[3]{\fnm{Mathis} \sur{Bode}}\email{m.bode@fz-juelich.de}

\author*[1,2]{\fnm{Ricardo} \sur{Vinuesa}}\email{rvinuesa@umich.edu}

\affil[1]{\orgdiv{Department of Aerospace Engineering}, \orgname{University of Michigan}, \orgaddress{\city{Ann Arbor}, \postcode{48109}, \state{Michigan}, \country{USA}}}

\affil[2]{\orgdiv{FLOW, Engineering Mechanics}, \orgname{KTH Royal Institute of Technology}, \orgaddress{\city{Stockholm}, \postcode{SE-100 44}, \country{Sweden}}}

\affil[3]{\orgdiv{J{\"u}lich Supercomputing Centre}, \orgname{Forschungszentrum J{\"u}lich GmbH}, \orgaddress{\city{J{\"u}lich}, \postcode{52425}, \country{Germany}}}

\maketitle
\newpage

\section*{Supplementary Information 1: Net-energy-saving (NES) control policies}

\paragraph{Power-input formula for wall-bounded flows: A revisit}
The power required for wall actuation includes contributions from pressure work and kinetic-energy flux~\citep{bewley_2001_dns,Hasegawa_Quadrio_Frohnapfel_2014,kametani_effect_2015}. These contributions have been considered in studies of opposition control~\citep{stroh_comparison_2015,wang_opposition_2025,yao_oc_PRF_2025} and DRL control~\citep{sonoda_RL_2023,zhou_reinforcement-learning-based_2025,beneitez_improving_2025,tonti2026explainable}. We use the power, PR, PIR, and NES definitions given in Methods. Here, Type-I denotes the absolute-value convention adopted there: both signs of the pressure work and wall-normal kinetic-energy flux contribute to the actuation cost~\citep{bewley_2001_dns,stroh_comparison_2015}.

For comparison, the alternative Type-II convention retains only positive contributions:
\begin{align}
    {\rm PI}^{(p)}_{\mathrm{TCF},\tau} &= \langle pv_n S_1\rangle_{\mathrm{wall}}, \\
    {\rm PI}^{(v)}_{\mathrm{TCF},\tau} &= \rho \langle v_n^3 S_2 / 2\rangle_{\mathrm{wall}},
    \label{eq:nes_def_2}
\end{align}
where $S_1=1$ when $pv_n>0$ and zero otherwise, and $S_2=1$ when $v_n^3>0$ and zero otherwise. The wall averaging and normalization follow Methods.

We compared policies trained with the two NES conventions in TCF-2. The remaining training settings followed Methods, with the same randomization settings in both experiments. The best-performing policies were then evaluated deterministically using each convention.

Supplementary Table~\ref{tab:tcf_nes_two_def} reports the results. The Type-I-trained policy achieves NES values of $28.42\%$ and $25.20\%$ under Type-I and Type-II evaluation, respectively. The corresponding values for the Type-II-trained policy are $25.55\%$ and $23.37\%$. The Type-I-trained policy also gives a higher PR, $30.19\%$ compared with $26.61\%$. OC has the same reported NES under both evaluations. Previous DRL studies at similar Reynolds numbers also report actuation costs alongside the drag reduction~\citep{sonoda_RL_2023,zhou_reinforcement-learning-based_2025,beneitez_improving_2025,tonti2026explainable}.

The Type-I reward yields higher NES under both evaluation conventions for the policies considered here. We therefore use the Type-I convention for the NES-based training and evaluations in the channel and wing configurations. 

\begin{table}
\centering
\resizebox{\textwidth}{!}{%
\begin{tabular}{ccccccc}
\toprule
Learning type & Evaluation type & Configuration & $\mathrm{PR}_{\mathrm{TCF},\tau}$ $[\%]$ & ${\rm PIR}_{\mathrm{TCF},\tau}^{(p)}$ $[\%]$ & ${\rm PIR}_{\mathrm{TCF},\tau}^{(v)}$ $[\%]$ & $\mathrm{NES}_{\mathrm{TCF},\tau}$ $[\%]$ \\
\midrule
-  & I & OC & 22.75\ & 0.40\ & 0.22\ & 22.12\ \\
I  & I & DRL & 30.19\ & 1.55\ & 0.22\ & \textbf{28.42}\ \\
II & I & DRL & 26.61\ & 0.96\ & 0.10\ & 25.55\ \\
\midrule 
-  & II & OC & 22.75\ & 0.40\ & 0.22\ & 22.12\ \\
I  & II & DRL & 30.19\ & 4.93\ & 0.04\ & \textbf{25.20}\ \\
II & II & DRL & 26.61\ & 3.19\ & 0.05\ & 23.37\ \\
\bottomrule
\end{tabular}%
}
\caption{\textbf{Effect of the power-input convention on training and evaluation in TCF-2.} PR, pressure-work and kinetic-energy input ratios, and NES are reported for OC and the selected DRL policies. Learning type identifies the reward convention. Evaluation type identifies the convention used to report the costs and NES. Type-I follows Methods and Type-II is given in equation~\ref{eq:nes_def_2}. Means use six initial conditions after excluding $t^+<500$. Bold entries identify the highest reported NES under each evaluation convention.}
\label{tab:tcf_nes_two_def}
\end{table}

\begin{figure}
    \centering
    \includegraphics[width=0.8\linewidth]{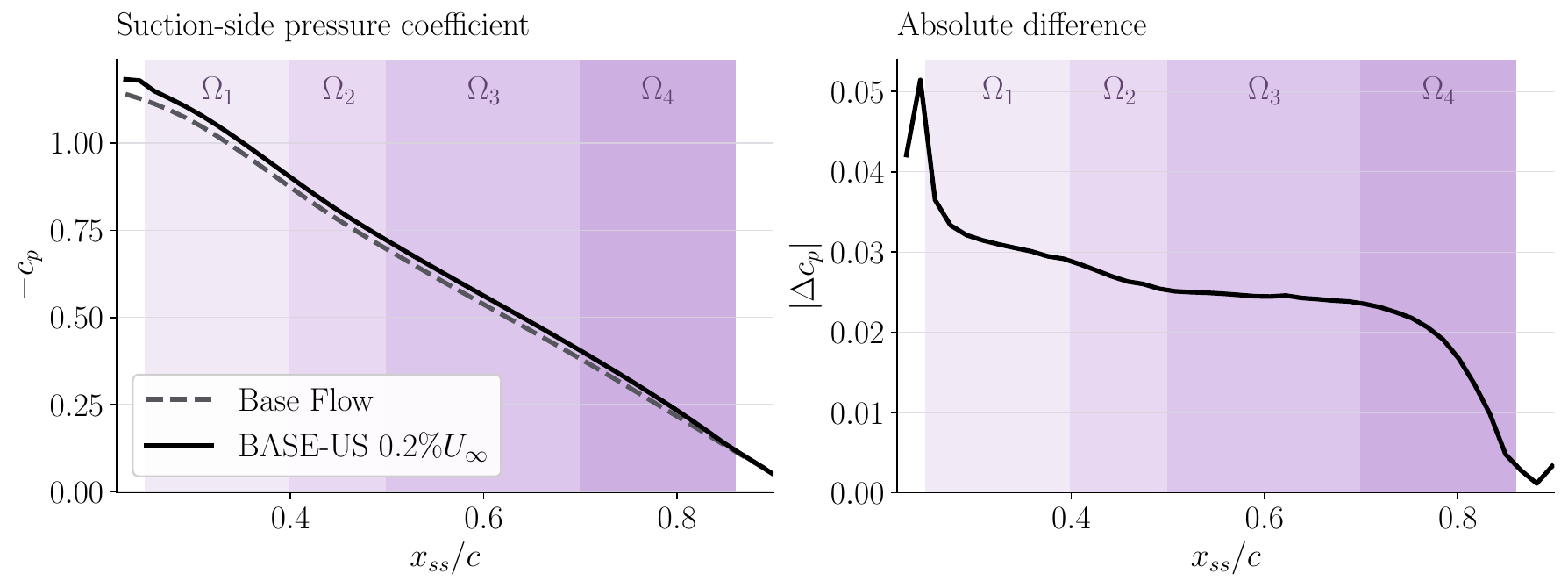}
    \caption{\textbf{Pressure distributions used to estimate the uniform-suction power input.} Left: suction-side $-c_p$ for the uncontrolled Base Flow (grey dashed) and uniform suction at $0.2\%\,U_\infty$ (black solid). Right: the absolute difference $|c_p^{\mathrm{US}}-c_p^{\mathrm{Base}}|$. Shading identifies the four control blocks within $0.25\leq x_{\mathrm{ss}}/c\leq0.86$. The pressure difference enters the BLC power estimate defined in Methods.}
    \label{fig:cp_difference}
\end{figure}

\paragraph{Power inputs for integral-drag reduction on turbulent wing sections}
Uniform suction or blowing imposes a mean wall-normal velocity, so its BLC power input is accounted for separately from the fluctuating ZNMF actuation. We use the integral power budget and BLC estimate defined in Methods. Supplementary Figure~\ref{fig:cp_difference} shows the suction-side pressure distributions used to estimate the uniform-suction cost. Over the actuation interval, the chordwise mean absolute pressure-coefficient difference is approximately $\langle |c_p^{\mathrm{ctrl}}-c_p^{\mathrm{ref}}|\rangle_x=0.025$.

\section*{Supplementary Information 2: Deterministic policies in surrogate channels}
\begin{figure}[ht]
    \centering
    \includegraphics[width=\linewidth]{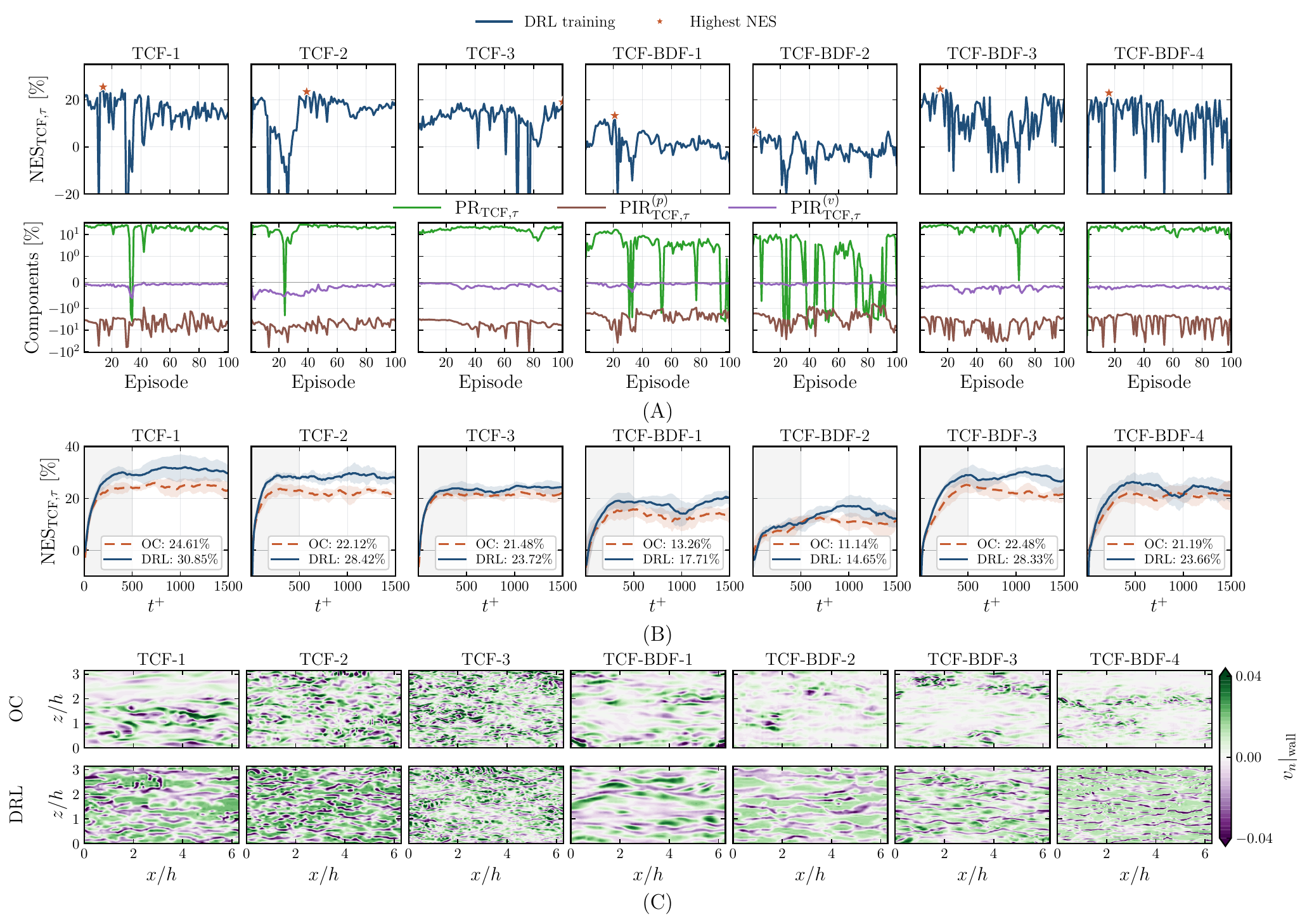}
    \caption{\textbf{Training, deterministic performance, and wall actuation in the channel surrogates.} (A) Episode-mean NES (top) and its wall-shear, pressure-work, and kinetic-energy contributions (bottom). Curves are smoothed using sliding-window averaging with a window size equal to one episode. Stars mark maximum rewards. Component axes are symmetric-logarithmic, with costs negative.  (B) Realization-mean deterministic NES for OC and DRL, with coloured bands of $\pm$ one standard deviation. Grey shading marks $t^+<500$. Legend values are means over the retained data. (C) Instantaneous OC (top) and DRL (bottom) wall actuation against $x/h$ and $z/h$. Colour limits are $\pm0.04$ for $v_n |_{\rm wall}$. Positive and negative values denote blowing and suction, respectively.}
    \label{fig:tcf_combined}
\end{figure}

\nopagebreak[4]
Supplementary Figure~\ref{fig:tcf_combined}(A) shows the training histories for the seven channel configurations. After the initial episodes, TCF-1 and TCF-2 attain predominantly positive rewards, whereas TCF-3 exhibits intermittent large negative rewards. Positive rewards are also obtained repeatedly in TCF-BDF-3 and TCF-BDF-4. In TCF-BDF-1 and TCF-BDF-2, the mean rewards over episodes 81--100 are approximately $-0.8\%$ and $-1.8\%$, respectively. This reflects a known characteristic of TD3 in which policy updates can temporarily degrade performance before recovering. Thus, policies with positive NES are obtained during training, although the episode rewards continue to fluctuate. The optimal policy is therefore selected as the one achieving the highest reward over the full training history. The evolution of components reveals that the pressure-work contribution generally exceeds the wall-normal kinetic-energy input. During episodes with large negative rewards, the actuation cost exceeds the power saved through wall-shear reduction. This is connected to the exploration noise with an amplitude of $10\%$ of the control intensity is included during training. 

Supplementary Figure~\ref{fig:tcf_combined}(B) compares the instantaneous NES obtained during deterministic evaluation. The results are averaged over six independent TCF realizations after excluding the initial transient ($t^+<500$). DRL yields a higher mean NES than OC in all seven cases (Supplementary Table~\ref{tab:tcf_deterministic_energy_performance}). The additional wall-shear reduction compensates for its higher actuation cost.

\begin{table}
\centering
\resizebox{\textwidth}{!}{%
\begin{tabular}{lccccc}
\toprule
Case & Configuration & $\mathrm{PR}_{\mathrm{TCF},\tau}$ $[\%]$ & $\mathrm{PIR}^{(p)}_{\mathrm{TCF},\tau}$ $[\%]$ & $\mathrm{PIR}^{(v)}_{\mathrm{TCF},\tau}$ $[\%]$ & $\mathrm{NES}_{\mathrm{TCF},\tau}$ $[\%]$\\
\midrule
TCF-1 & OC & 25.00\ & 0.29\ & 0.10\ & 24.61\ \\
 & DRL & 31.40\ & 0.49\ & 0.07\ & \textbf{30.85}\ \\
TCF-2 & OC & 22.75\ & 0.40\ & 0.22\ & 22.12\ \\
 & DRL & 30.19\ & 1.55\ & 0.22\ & \textbf{28.42}\ \\
TCF-3 & OC & 22.12\ & 0.44\ & 0.20\ & 21.48\ \\
 & DRL & 24.81\ & 0.96\ & 0.13\ & \textbf{23.72}\ \\
\midrule
TCF-BDF-1 & OC & 13.57\ & 0.23\ & 0.08\ & 13.26\ \\
 & DRL & 18.20\ & 0.43\ & 0.05\ & \textbf{17.71}\ \\
TCF-BDF-2 & OC & 11.27\ & 0.10\ & 0.03\ & 11.14\ \\
 & DRL & 15.12\ & 0.42\ & 0.04\ & \textbf{14.65}\ \\
TCF-BDF-3 & OC & 22.72\ & 0.18\ & 0.06\ & 22.48\ \\
 & DRL & 28.80\ & 0.41\ & 0.06\ & \textbf{28.33}\ \\
TCF-BDF-4 & OC & 21.42\ & 0.16\ & 0.06\ & 21.19\ \\
 & DRL & 24.66\ & 0.86\ & 0.14\ & \textbf{23.66}\ \\
\bottomrule
\end{tabular}%
}
\caption{\textbf{Deterministic energy performance in the seven channel surrogates.} PR, pressure-work and kinetic-energy input ratios, and NES use the Type-I convention defined in Methods. Means use six initial conditions after excluding $t^+<500$. Bold entries identify the higher NES in each channel.}
\label{tab:tcf_deterministic_energy_performance}
\end{table}

The instantaneous wall-actuation fields in Supplementary Figure~\ref{fig:tcf_combined}(C) illustrate the spatial distribution of blowing and suction. For OC, the imposed velocity is opposite to the negative wall-normal velocity fluctuation at the sensing plane. The DRL policies produce a different spatial distribution of actuation from OC.

\section*{Supplementary Information 3: Discussion on DRL control configuration on the uncontrolled base flow}
\begin{figure}
    \centering
    \includegraphics[width=\linewidth]{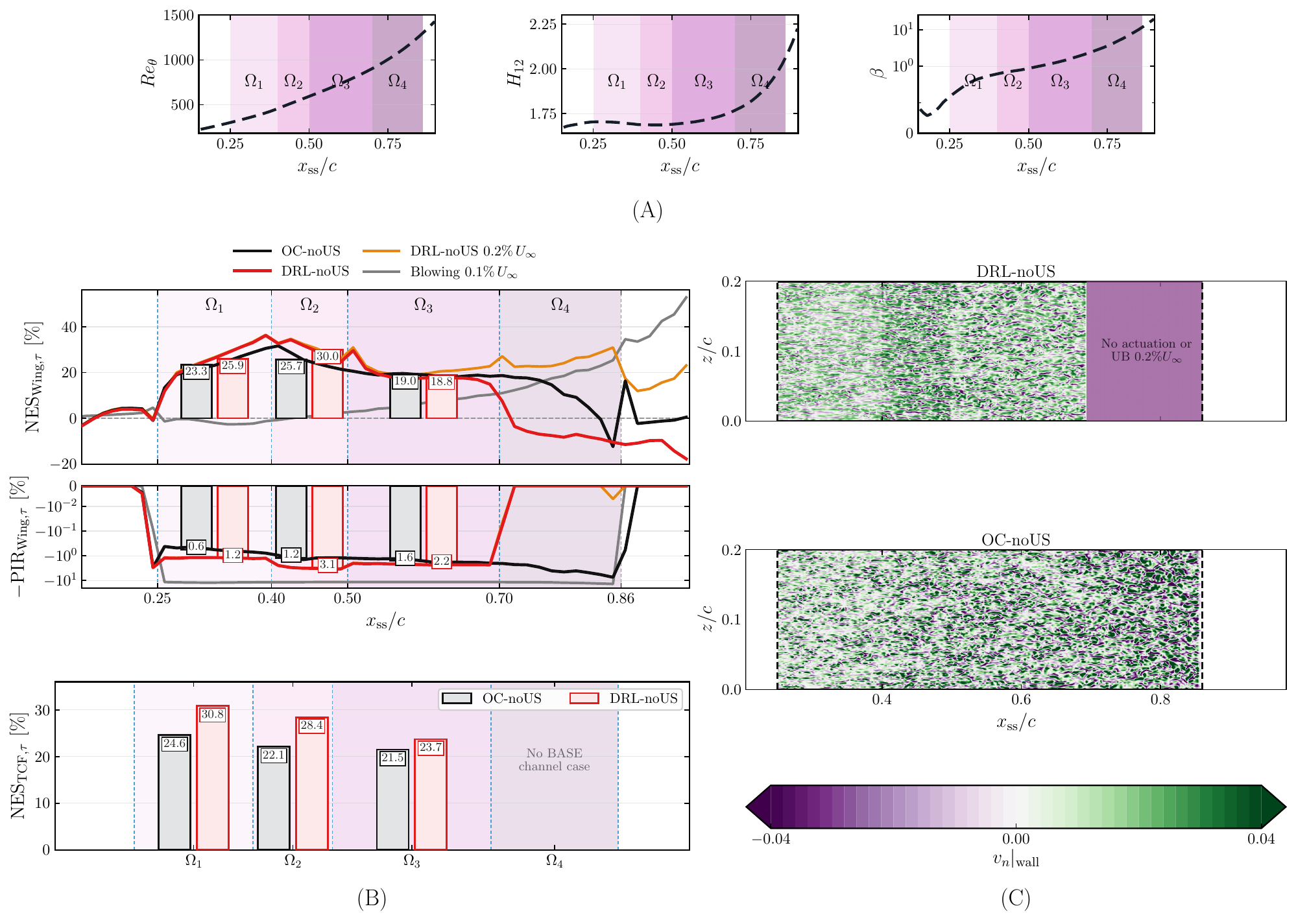}
    \caption{\textbf{Zero-shot wing control without background suction.} (A) Uncontrolled suction-side $Re_\theta$, $H_{12}$, and $\beta$, with the control blocks shaded. (B) Local wing NES and the negative ZNMF power-input ratio, referenced to the uncontrolled Base Flow. Bars compare block means for OC and DRL without downstream blowing over $\Omega_1$--$\Omega_3$. The lower bar chart gives the corresponding deterministic channel results. Additional curves show hybrid DRL with downstream blowing at $0.2\%\,U_\infty$ and uniform blowing alone at $0.1\%\,U_\infty$. The local metric excludes BLC pump power. (C) Instantaneous hybrid DRL (top) and OC (bottom) wall actuation, with common colour limits of $\pm0.04$ in the exported simulation velocity units. Positive and negative values denote blowing and suction. The hybrid's downstream blowing block $\Omega_4$ is masked.}
    \label{fig:wing_noUS_deterministic}
\end{figure}

Supplementary Figure~\ref{fig:wing_noUS_deterministic} compares two DRL configurations without background suction. Both apply channel-trained policies over $\Omega_1$--$\Omega_3$ ($0.25<x_{ss}/c<0.70$). The first leaves the downstream region unactuated. The second, referred to here as the hybrid configuration, adds uniform blowing at $0.2\%\,U_\infty$ in $\Omega_4$ ($0.70<x_{ss}/c<0.86$). The uncontrolled pressure-gradient parameter increases from approximately $2.5$ to $11.6$ across this downstream block (panel A). 

Without downstream blowing, DRL yields a higher local NES than OC in $\Omega_1$ and $\Omega_2$, but nearly the same NES in $\Omega_3$ (panel B). In the third block, the additional actuation cost offsets most of the additional wall-shear reduction obtained with DRL. The difference between DRL and OC is therefore smaller than in the corresponding deterministic channel comparison.

The two DRL configurations have similar local NES profiles in the first two blocks. In $\Omega_3$, the block mean increases from approximately $18.8\%$ without downstream blowing to $21.6\%$ in the hybrid case, with little change in the local ZNMF penalty. The largest difference occurs in $\Omega_4$: the no-blowing configuration has a mean local NES of approximately $-5.9\%$, whereas the hybrid reaches $25.5\%$. OC remains active through $\Omega_4$ and gives a smaller positive block mean than the hybrid. Uniform blowing alone also becomes more favourable downstream, although its upstream performance are weak. Moreover, Panel (C) shows spatially organized blowing and suction in the hybrid DRL case, with $\Omega_4$ masked to display only its learned actuation.

To assess the effect on the complete wing section, the next section compares integral lift and drag, aerodynamic efficiency, and the global energy budget, including the cost of downstream blowing.

\section*{Supplementary Information 4: Control compact on aerodynamic efficiency and its statistical convergence}

\paragraph{Integral net-energy saving and aerodynamic efficiency}
Supplementary Table~\ref{tab:wing_drag_energy_summary} extends the integral power budget in the main text to the configurations with and without background suction. DRL gives higher total-drag PR and NES than OC in both comparisons. Without background suction, the DRL configuration with no actuation in $\Omega_4$ achieves a higher global NES than the hybrid with downstream blowing at $0.2\%\,U_\infty$. The larger downstream wall-shear reduction in the hybrid configuration therefore does not yield a higher integral NES. 

\begin{table}[ht]
\centering
\resizebox{\textwidth}{!}{%
\begin{tabular}{lrrrr}
\toprule
Case & $\mathrm{PR}_{\mathrm{Wing},D}$ & $\mathrm{PIR}^{(\mathrm{ZNMF})}_{\mathrm{Wing},D}$ & $\mathrm{PIR}^{(\mathrm{BLC})}_{\mathrm{Wing},D}$ & $\mathrm{NES}_{\mathrm{Wing},D}$ \\
\midrule
OC-noUS & $3.85\%$ & $0.12\%$ & $0.00\%$ & $3.73\%$ \\
DRL-noUS & $5.18\%$ & $0.14\%$ & $0.00\%$ & $5.04\%$ \\
DRL-noUS-UB  & $4.95\%$ & $0.14\%$ & $0.01\%$ & $4.80\%$ \\
\midrule
BASE-US  & $3.53\%$ & $0.00\%$ & $0.14\%$ & $3.39\%$ \\
OC-US    & $8.93\%$ & $0.04\%$ & $0.14\%$ & $8.75\%$ \\
DRL-US  & $11.11\%$ & $0.10\%$ & $0.14\%$ & $10.87\%$ \\
\bottomrule
\end{tabular}%
}
\caption{\textbf{Integral wing-drag power budget.} All entries are percentages of the uncontrolled BASE driving power. ZNMF costs are integrated over the active interval using the Type-I convention in Methods. The BLC estimate uses the uniform-suction pressure profile for all US cases. For the hybrid DRL-noUS-UB case, ZNMF acts over $0.25\leq x_{\mathrm{ss}}/c\leq0.70$ and the downstream-blowing cost uses its DRL pressure profile over $0.70\leq x_{\mathrm{ss}}/c\leq0.86$. NES subtracts both applicable costs from the total-drag PR.}
\label{tab:wing_drag_energy_summary}
\end{table}

Supplementary Table~\ref{tab:aero_coeff_all_cases} reports the integral aerodynamic coefficients and their changes relative to the uncontrolled base flow. Consistent with the drag calculation in Methods, we use $C_l=L/(\tfrac{1}{2}\rho U_\infty^2 c)$ for lift per unit span $L$, and $L/D=C_l/C_d$. DRL produces lower total drag than OC. Both controllers also increase lift, except for the hybrid DRL configuration. However, enhanced lift is not unconditionally beneficial from an aerodynamic performance perspective: for a finite wing operating at a fixed angle of attack, increased lift is accompanied by a corresponding rise in induced drag. Drag reduction, by contrast, constitutes an unambiguous aerodynamic benefit regardless of the underlying assumptions, which motivates the emphasis placed on drag modification throughout the main text. The following analysis assesses the finite-time sampling uncertainty of these coefficients.

\begin{table}[ht]
\centering
\resizebox{\textwidth}{!}{
\begin{tabular}{lccccc}
\toprule
Case & $C_l$ & $C_{d,f}$ & $C_{d,p}$ & $C_d=C_{d,f}+C_{d,p}$ & $L/D$ \\
\midrule
Base Flow & $0.8611$ & $0.0127$ & $0.0087$ & $0.0214$ & $40.3$ \\
OC-noUS & $0.8734\;(+1.4\%)$ & $0.0119\;(-6.4\%)$ & $0.0087\;(0.0\%)$ & $0.0206\;(-3.8\%)$ & $42.5\;(+5.5\%)$ \\
DRL-noUB & $0.8775\;(+1.9\%)$ & $0.0118\;(-6.9\%)$ & $0.0085\;(-2.6\%)$ & $0.0203\;(-5.2\%)$ & $43.3\;(+7.5\%)$ \\
DRL-noUS-UB & $0.8568\;(-0.5\%)$ & $0.0117\;(-8.3\%)$ & $0.0087\;(-0.1\%)$ & $0.0203\;(-5.0\%)$ & $42.1\;(+4.7\%)$ \\
\midrule
BASE-US & $0.9180\;(+6.6\%)$ & $0.0139\;(+9.7\%)$ & $0.0067\;(-22.9\%)$ & $0.0206\;(-3.5\%)$ & $44.5\;(+10.5\%)$ \\
OC-US   & $0.9349\;(+8.6\%)$ & $0.0132\;(+3.8\%)$ & $0.0063\;(-27.5\%)$ & $0.0195\;(-8.9\%)$ & $48.0\;(+19.2\%)$ \\
DRL-US  & $0.9418\;(+9.4\%)$ & $0.0127\;(-0.1\%)$ & $0.0063\;(-27.2\%)$ & $0.0190\;(-11.1\%)$ & $49.5\;(+23.0\%)$ \\
\bottomrule
\end{tabular}
}
\caption{\textbf{Integral aerodynamic coefficients and lift-to-drag ratio.} Values are arithmetic means of complete, non-overlapping averaging segments starting at elapsed $t\geq6$ flow-over time units. Parentheses give signed percentage changes relative to the uncontrolled Base Flow. $L/D$ is the mean of the segmentwise ratios. DRL-noUB denotes the no-suction DRL configuration without downstream blowing, called DRL-noUS elsewhere.}
\label{tab:aero_coeff_all_cases}
\end{table}

\begin{figure}[ht]
    \centering
    \includegraphics[width=\linewidth]{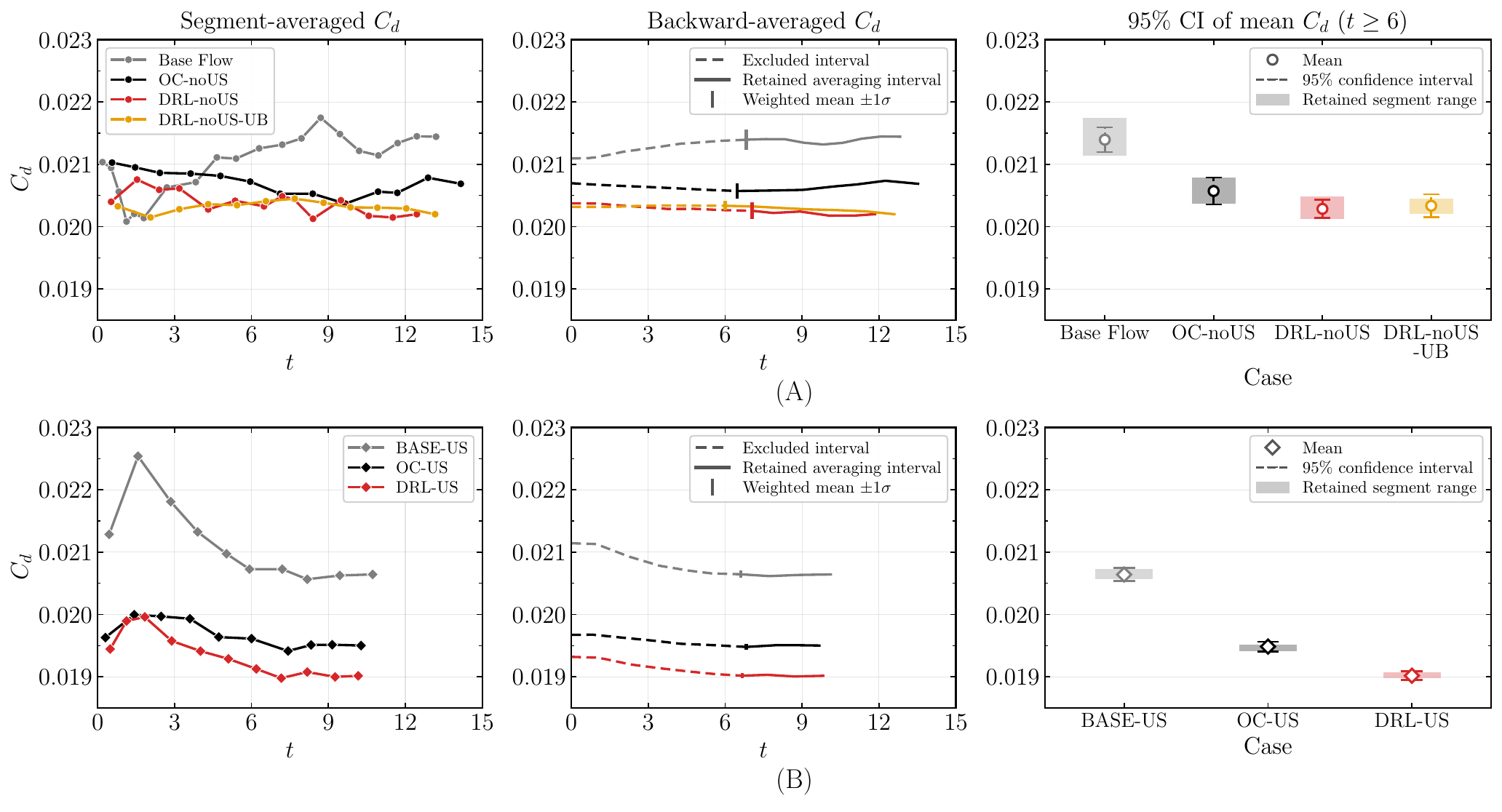}
    \caption{\textbf{Total-drag convergence and finite-time sampling uncertainty.} (A) Cases without background suction. (B) Cases with uniform suction at $0.2\%\,U_\infty$. Columns show (left) segment-averaged $C_d$, (middle) duration-weighted backward means, and (right) retained arithmetic segment means. In the middle column, dashed and solid curves identify excluded and retained regions. Vertical bars show the retained weighted mean plus and minus one weighted population standard deviation. In the right column, open markers show arithmetic means, capped dashed bars give approximate 95\% confidence intervals allowing for estimated positive serial correlation, and shaded rectangles show the observed segment ranges. Only complete segments starting at elapsed $t\geq6$ are retained. Time is measured from each case's first available segment.}
    \label{fig:wing_clcd_CI}
\end{figure}

\paragraph{Statistical convergence and uncertainty estimation}
We assess statistical convergence using aerodynamic coefficients obtained from successive time-averaged fields. Each field corresponds to a simulation interval, referred to as an averaging segment. Supplementary Figure~\ref{fig:wing_clcd_CI} shows the segment-averaged total-drag coefficient, its backward-cumulative average, and the estimated mean with an approximate 95\% confidence interval. Panels (A) and (B) show cases without and with background suction, respectively.
Note that the cases without background suction were run more than $10$ flow-over time units to improve the convergence as their segment-averaged $C_d$ are more fluctuating than the suction-background cases~\citep{atzori_aero_2020}. 

Following the convergence analysis of Ref.~\citep{atzori_aero_2020}, we calculated duration-weighted backward averages,
\[
B_j(X)=
\frac{\sum_{i=j}^{N}\Delta t_i X_i}
{\sum_{i=j}^{N}\Delta t_i},
\]
\noindent where $X_i$ is the coefficient averaged over segment $i$, $\Delta t_i$ is its flow-over-time duration, and $N$ is the number of available segments. Segments are ordered chronologically, thus $B_j$ includes segment $j$ and all later segments. Varying $j$ tests the sensitivity to excluding earlier data. Dashed and solid curves distinguish excluded and retained averaging regions. The vertical bars at the start of the retained region show the duration-weighted mean plus or minus one weighted population standard deviation,
\[
\sigma_{w,j}=
\left[
\frac{\sum_{i=j}^{N}\Delta t_i\{X_i-B_j(X)\}^{2}}
{\sum_{i=j}^{N}\Delta t_i}
\right]^{1/2}.
\]
This standard deviation describes the variation among segment averages. Note that we used a prescribed cutoff of $t_{\mathrm{cut}}=6.0$ flow-over time units, with elapsed time measured from the start of each case's first available averaging segment. This choice yields the minimal weighted standard deviation for each case. 

The confidence-interval (CI) analysis includes only complete segments beginning at or after $t_{\mathrm{cut}}$, excluding those that cross the cutoff. Reindexing the retained segments as $i=1,\ldots,n$, we calculate the arithmetic mean and sample variance as
\[
\bar X=\frac{1}{n}\sum_{i=1}^{n}X_i,
\qquad
s_X^2=\frac{1}{n-1}\sum_{i=1}^{n}(X_i-\bar X)^2.
\]
Note that we use arithmetic means in order to follow the standard CI calculation. Here, the arithmetic segment mean differs from the duration-weighted backward mean when segment durations are unequal. Under stationary sampling, each segment estimates the same underlying mean, but unequal durations can give different segment variances. The CI calculation is therefore approximate. 

However, the difference between the mean drag coefficients is insensitive to the choice of weights: the total-drag advantage of DRL-US over OC-US, $\bar C_{d,\mathrm{OC}}-\bar C_{d,\mathrm{DRL}}$, is $4.68\times10^{-4}$ with equal segment weights and $4.61\times10^{-4}$ with duration weights. The difference is approximately 1.4\% of the equal-segment estimate and does not change the trends. This comparison assesses the sensitivity of the estimated mean difference to the weighting procedure.

Subsequently, we estimate serial dependence from the segment autocorrelation,

\[
\widehat{\rho}_k=
\frac{\sum_{i=1}^{n-k}(X_i-\bar X)(X_{i+k}-\bar X)}
{\sum_{i=1}^{n}(X_i-\bar X)^2},
\qquad \widehat{\rho}_0=1.
\]
The lag $k$ counts averaging segments. Unequal durations or gaps prevent a unique conversion to physical time. We inspect lags up to
\[
K=\min\!\left(\left\lfloor n/3\right\rfloor,5\right)
\]
and sum consecutive positive correlations from lag 1 until the first nonpositive or non-finite value, or lag $K$. If $m$ positive lags are retained, the correlation factor and effective segment count are
\[
g_X=\max\left(1+2\sum_{k=1}^{m}\widehat{\rho}_k,1\right),
\qquad
n_{\mathrm{eff},X}=\min\left\{n,\max\left(2,\frac{n}{g_X}\right)\right\}.
\]
The sum is zero when no positive lag is retained. The effective count can be noninteger and is restricted to $[2,n]$. The lower bound is imposed for numerical stability and was inactive in the primary analysis. It does not imply that two independent observations are available.

The estimated standard error and effective degrees of freedom are
\[
SE_X=\frac{s_X}{\sqrt{n_{\mathrm{eff},X}}},
\qquad
\zeta_X=\max(n_{\mathrm{eff},X}-1,1).
\]
Writing $q_{\mathrm{St}}(p;\zeta)$ for the $p$-quantile of a
Student-$t$ distribution with $\zeta$ degrees of freedom, the
approximate two-sided 95\% interval is
\[
\bar X\pm h_X,
\qquad
h_X=q_{\mathrm{St}}(0.975;\zeta_X)SE_X.
\]
The right-hand panels of Supplementary Figure~\ref{fig:wing_clcd_CI} distinguish these intervals from the observed minimum--maximum range of the retained segment averages.

\begin{figure}[ht]
    \centering
    \includegraphics[width=\linewidth]{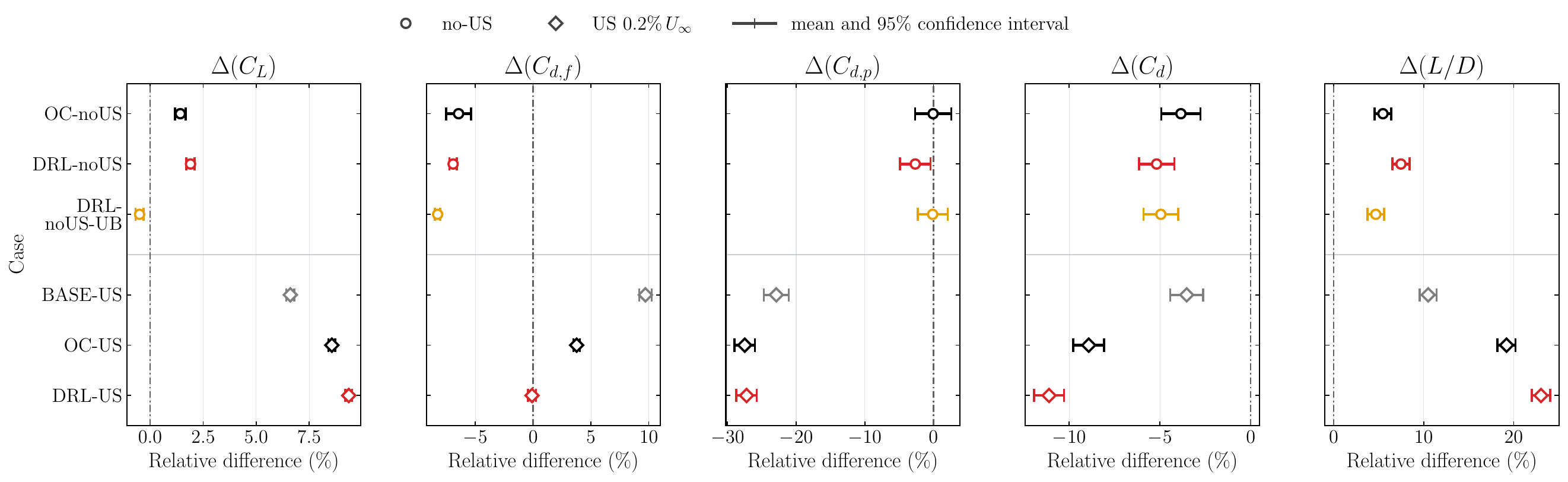}
    \caption{\textbf{Aerodynamic changes relative to the common uncontrolled Base Flow.} Panels show signed changes in lift ($C_l$, labelled $C_L$ in the plot), friction drag, pressure drag, total drag, and lift-to-drag ratio, using $\Delta(X)$ defined in the text. Circles and diamonds denote cases without and with background suction at $0.2\%\,U_\infty$, respectively. Open markers give the estimated changes. Horizontal bars give approximate individual 95\% confidence intervals incorporating sampling uncertainty in both case and reference means. Dashed vertical lines mark zero change. The analysis retains complete segments starting at elapsed $t\geq6$ flow-over time units.}
    \label{fig:wing_clcd_relative_CI}
\end{figure}

Supplementary Figure~\ref{fig:wing_clcd_relative_CI} compares lift, friction drag, pressure drag, total drag, and lift-to-drag ratio with the same uncontrolled Base Flow, including for cases with background suction. The signed relative change is
\[
\Delta(X)=100\left(\frac{\bar X_{\mathrm{case}}}
{\bar X_{\mathrm{Base}}}-1\right).
\]
Assuming negligible covariance between the case and reference mean
estimates, first-order delta propagation gives
\[
SE_{\Delta(X)}^2=\chi_{\mathrm{case}}+\chi_{\mathrm{Base}},
\]
\noindent where the two variance contributions are
\[
\chi_{\mathrm{case}}=
\left(\frac{100}{\bar X_{\mathrm{Base}}}\right)^2
SE_{\mathrm{case}}^2,
\qquad
\chi_{\mathrm{Base}}=
\left(\frac{100\bar X_{\mathrm{case}}}
{\bar X_{\mathrm{Base}}^2}\right)^2
SE_{\mathrm{Base}}^2.
\]
Using the effective degrees of freedom obtained separately for each
mean, the Welch--Satterthwaite approximation is
\[
\zeta_{\Delta(X)}=
\frac{(\chi_{\mathrm{case}}+\chi_{\mathrm{Base}})^2}
{\chi_{\mathrm{case}}^2/\zeta_{\mathrm{case}}
+\chi_{\mathrm{Base}}^2/\zeta_{\mathrm{Base}}}.
\]
The resulting approximate 95\% interval is
\[
\Delta(X)\pm
q_{\mathrm{St}}\!\left(0.975;\zeta_{\Delta(X)}\right)
SE_{\Delta(X)}.
\]
The relative change is expressed as a percentage, and its standard error and CI half-width are in percentage points. For $L/D$, we analyse the arithmetic mean of the segmentwise ratios $C_{l,i}/C_{d,i}$, rather than the ratio of the mean lift and drag coefficients. The results indicate that the obtained control performance are statistically robust in the current realizations. Nevertheless, it is worth noting that these intervals do not provide simultaneous coverage for all quantities or describe variability between independently initialized realizations.

\section*{Supplementary Information 5: Sensitivity analysis of control configurations}
\paragraph{Optimal configuration for the opposition control applied to the wing section with $0.2\%U_\infty$ uniform suction}
The OC configuration identified for the wing section without background suction uses $y_n^+=15$ and $\alpha=1.0$~\citep{wang_opposition_2025}. To assess whether this choice remains suitable with uniform suction, we varied amplitude at $y_n^+=15$ ($\alpha=0.25,0.5,1.0$) and sensing height at $\alpha=1.0$ ($y_n^+=7.5,15.0,22.5$). These two one-parameter sweeps comprise five distinct configurations. Each simulation covers approximately two flow-over time units.

\begin{figure}[ht]
    \centering
    \includegraphics[width=0.75\linewidth]{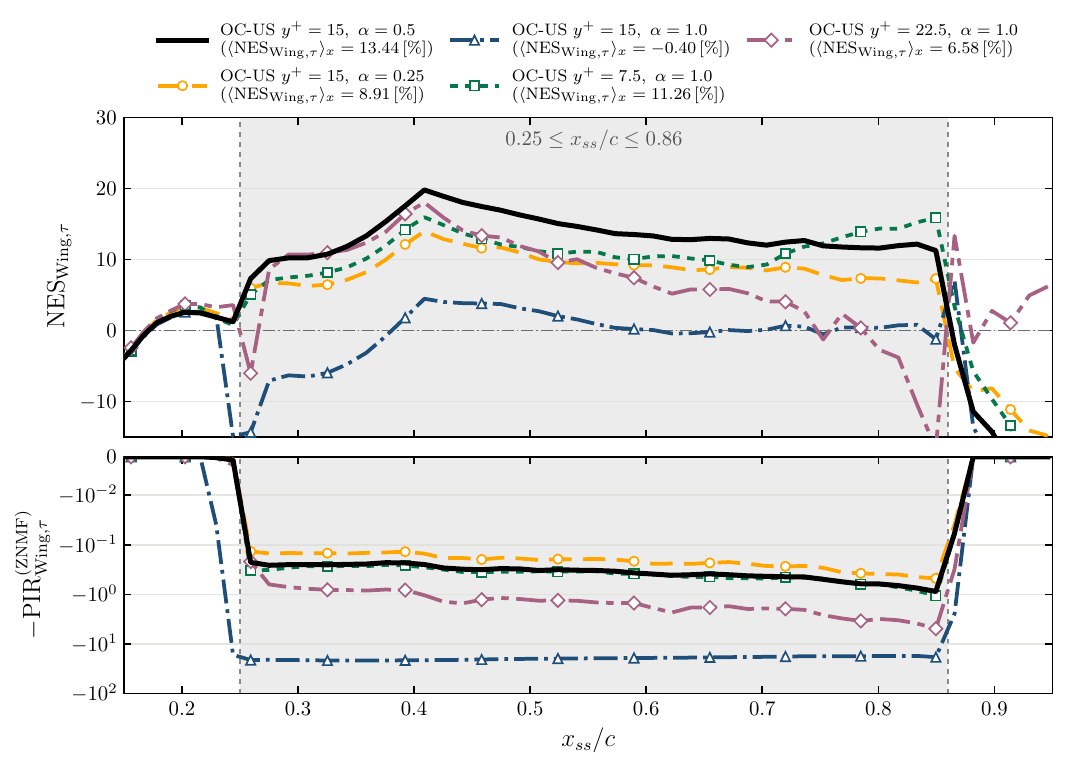}
    \caption{\textbf{OC parameter sensitivity with background suction at $0.2\%\,U_\infty$.} Local NES (top) and negative ZNMF power-input ratio (bottom), referenced to BASE-US. The five tested pairs $(y_n^+,\alpha)$ are $(15,0.5)$, $(15,0.25)$, $(15,1)$, $(7.5,1)$, and $(22.5,1)$. Shading identifies $0.25\leq x_{\mathrm{ss}}/c\leq0.86$. Legend values are arithmetic station means of NES over this interval. The power-input panel uses a symmetric-logarithmic scale.}
    \label{fig:wing_oc_parametric}
\end{figure}

Supplementary Figure~\ref{fig:wing_oc_parametric} compares the local NES profiles and their means over the control region. The configuration $y_n^+=15.0$, $\alpha=1.0$ gives the lowest mean NES among those tested, $-0.40\%$. Reducing the amplitude to $\alpha=0.5$ at the same height gives the highest mean, $13.44\%$. This configuration yields the highest NES among the five tested pairs and is used as the OC reference for the uniform-suction case.

\paragraph{Impact of the off-wall sensing plane}
Both OC and DRL use observations from an off-wall sensing plane. The reference height, $y_n^+=15$, follows previous OC~\citep{choi_active_1994,hammond_observed_1998,yao_oc_PRF_2025} and DRL studies~\citep{sonoda_RL_2023,guastoni_DRL_2023,zhou_reinforcement-learning-based_2025}. Ref.~\citep{chung_effectivenessTCF_2011} examined OC sensing heights up to $y_n^+=30$ and found the greatest wall-shear reduction at $y_n^+=15$. Ref.~\citep{guastoni_yplusdrl_2023} reported DRL advantages over OC at several sensing heights using a different MARL formulation. Here we assess the effect of sensing height on both wall-shear reduction and actuation cost.

We trained and evaluated additional TCF-1 policies at $y_n^+=5$ and $25$, retaining the other hyperparameters and randomization settings of the $y_n^+=15$ case. Supplementary Table~\ref{tab:tcf_sensing_plane} compares the selected deterministic policies. Both DRL and OC attain their highest NES at $y_n^+=15$ among these three heights. At $y_n^+=5$ and $25$, DRL gives NES values of $19.03\%$ and $11.22\%$, compared with $10.62\%$ and $-49.78\%$ for OC. The OC wall-shear trends are consistent with Ref.~\citep{chung_effectivenessTCF_2011}.

DRL gives greater wall-shear reduction and NES than OC at all three tested heights. Among the three heights considered in TCF-1, $y_n^+=15$ is therefore the most favourable for NES. The dependence on sensing height may differ for other flows and control strategies.

\begin{table}
\centering
\resizebox{\textwidth}{!}{%
\begin{tabular}{lccccc}
\toprule
 Configuration & $y^+_n$ & $\mathrm{PR}_{\mathrm{TCF},\tau}$ $[\%]$ & $\mathrm{PIR}^{(p)}_{\mathrm{TCF},\tau}$ $[\%]$ & $\mathrm{PIR}^{(v)}_{\mathrm{TCF},\tau}$ $[\%]$ & $\mathrm{NES}_{\mathrm{TCF},\tau}$ $[\%]$\\
\midrule
OC & 5 & 10.76 & 0.14 & $\sim 0.0$ & 10.62 \\
DRL        & 5 & 19.32 & 0.29 & $\sim 0.0$ & \textbf{19.03} \\
\midrule
OC & 15 & 25.00 & 0.29 & 0.10 & 24.61 \\
DRL        & 15 & 31.40 & 0.49 & 0.07 & \textbf{30.85} \\
\midrule
OC & 25 & -38.57 & 5.09 & 6.11 & -49.78 \\
DRL        & 25 & 11.31 & 0.09 & $\sim 0.0$ & \textbf{11.22} \\
\bottomrule
\end{tabular}%
}
\caption{\textbf{Sensing-height sensitivity in TCF-1.} PR, pressure-work and kinetic-energy input ratios, and NES for OC and DRL trained and evaluated at the indicated $y_n^+$. Means use six initial conditions after excluding $t^+<500$. Bold entries identify the higher NES at each height. Kinetic-energy input ratios below $10^{-4}\%$ are shown as $\sim0$.}
\label{tab:tcf_sensing_plane}
\end{table}

\paragraph{Impact of observation components}

The reference DRL policy uses inner-scaled tangential and normal velocity fluctuations~\citep{sonoda_RL_2023,guastoni_DRL_2023,beneitez_improving_2025,tonti2026explainable}, whereas OC uses only the normal component. We trained single-component policies in TCF-1 using PR as the reward to compare their drag-reduction performance with OC. All other settings are as described in Methods, with the same randomization settings for both policies.

The tangential-only and normal-only policies achieve PR values of $27.84\%$ and $25.30\%$, respectively, compared with $25.00\%$ for OC. The conditional mean actuation of the normal-only policy in Supplementary Figure~\ref{fig:tcf_component_action_map}(A) depends approximately linearly on the observed velocity. Its slope is close to that of OC, with small nonlinear deviations, consistent with the feedback behaviour discussed in Ref.~\citep{sonoda_RL_2023}. This comparison concerns drag reduction, since these policies were trained using PR.

\begin{figure}[ht]
    \centering
    \includegraphics[width=\linewidth]{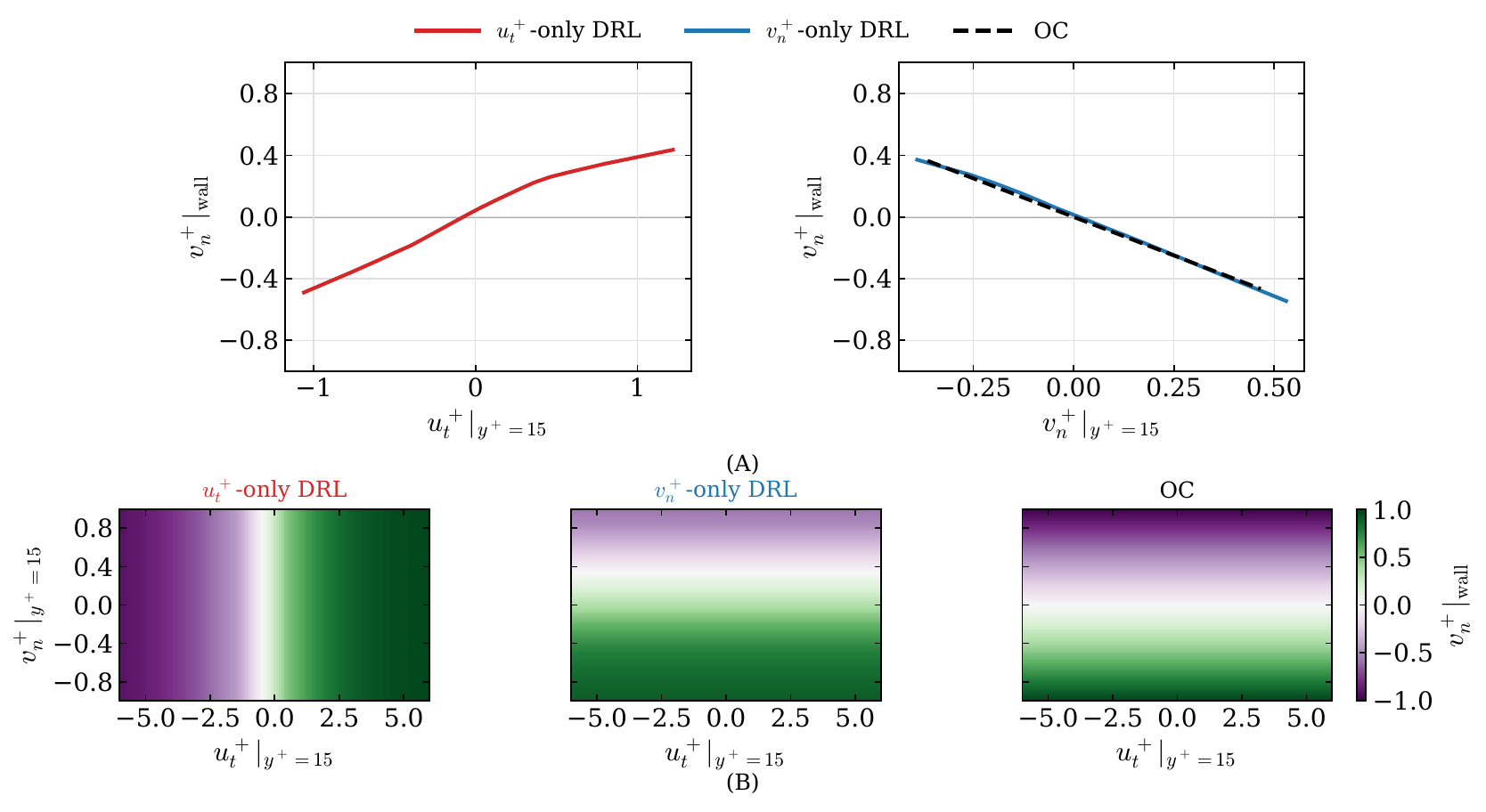}
    \caption{\textbf{Single-component feedback in TCF-1 at $y_n^+=15$.} (A) Trajectory-conditioned mean wall actuation after the GLL-weighted ZNMF correction, plotted against the observed velocity component: tangential-only DRL (red), normal-only DRL (blue), and OC (black dashed). (B) Direct outputs of the same controllers on a synthetic $(u_t^+,v_n^+)$ input mesh, with common wall-action limits $[-1,1]$. These maps evaluate the feedback laws before the instantaneous spatial ZNMF correction. All velocities use the common reference friction-velocity scale. The DRL policies in this ablation were trained with PR as the reward.}
    \label{fig:tcf_component_action_map}
\end{figure}

The tangential-only policy thus achieves greater drag reduction in this comparison. Ref.~\citep{zhou_reinforcement-learning-based_2025} also found poorer performance with only wall-normal velocity, and Ref.~\citep{Arranz_Lozano-Duran_2024} identified streamwise fluctuations as more informative about future wall shear. These observations concern the information used to determine the wall actuation.

We also tested OC based on $u_t|_{y_n^+=15}$ and on $(u_t/2-v_n)|_{y_n^+=15}$, as considered in Ref.~\citep{sonoda_RL_2023}. The respective amplitudes were $\alpha=0.07$ and $0.1$, the largest used without numerical instability in these tests. Both configurations increased drag power. The drag response therefore depends on both the observed component and the feedback law. The nonlinear tangential-only policy reduces drag, whereas the two linear laws tested here increase it.

\paragraph{Impact of mismatching the local boundary-layer properties}
To test sensitivity to differences between the training and evaluation flows, we transferred the policy trained in TCF-BDF-1 ($Re_\theta=336$, $H_{12}=1.65$) to TCF-BDF-3 ($Re_\theta=617$, $H_{12}=1.58$). The latter represents a farther-downstream wing block.

In TCF-BDF-3, the transferred policy achieves $23.92\%$ NES, compared with $28.33\%$ for the policy trained in that channel (Supplementary Table~\ref{tab:tcf_mismatch}). Both exceed OC ($22.48\%$). For this pair of channels, matching the training environment to the evaluation flow yields a higher NES. Since both $Re_\theta$ and $H_{12}$ differ, their individual effects cannot be distinguished from this test.

\begin{table}
\centering
\resizebox{\textwidth}{!}{%
\begin{tabular}{ccccccc}
\toprule
 Evaluation & Learning & Configuration & $\mathrm{PR}_{\mathrm{TCF},\tau}$ $[\%]$ & $\mathrm{PIR}^{(p)}_{\mathrm{TCF},\tau}$ $[\%]$ & $\mathrm{PIR}^{(v)}_{\mathrm{TCF},\tau}$ $[\%]$ & $\mathrm{NES}_{\mathrm{TCF},\tau}$ $[\%]$\\
\midrule
 TCF-BDF-3 & --         & OC & 22.72 & 0.18 & 0.06 & 22.48 \\
 TCF-BDF-3 & TCF-BDF-3  & DRL & 28.80 & 0.41 & 0.06 & \textbf{28.33} \\
 TCF-BDF-3 & TCF-BDF-1  & DRL & 24.38 & 0.40 & 0.05 & 23.92 \\
\bottomrule
\end{tabular}%
}
\caption{\textbf{Policy transfer between two channel surrogates.} All policies are evaluated in TCF-BDF-3. Learning identifies the training environment. PR, pressure-work and kinetic-energy input ratios, and NES are deterministic means over six initial conditions after excluding $t^+<500$. Bold identifies the highest NES. The TCF-BDF-1 policy is transferred without retraining.}
\label{tab:tcf_mismatch}
\end{table}

\paragraph{Impact of partitioning the chordwise control blocks}
\nopagebreak[4]
The reference configuration divides $0.25\leq x_{\mathrm{ss}}/c\leq0.86$ into four control blocks. We also tested two blocks, spanning $0.25$--$0.65$ and $0.65$--$0.86$. Supplementary Table~\ref{tab:tcf_match_wing_2blocks} gives their mean boundary-layer properties. The shape factors of TCF-BDF-1 and TCF-BDF-4 differ from the corresponding block means by less than $5\%$. This agreement is specific to $H_{12}$ and does not extend to all the listed properties. Their policies were deployed without further training. The wing simulation covers approximately two flow-over time units, with spanwise averaging as in the wing-flow analysis of Ref.~\citep{wang_separation_2025}. The comparisons below are based on this finite averaging interval.

\begin{table}[ht]
  \centering
  \resizebox{\textwidth}{!}{
  \begin{tabular}{cccccccccc}
  \toprule
  Case
  &  Block
  & {$x_{\rm ss}/c$}
  & {$\langle Re_\theta \rangle_x$}
  & {$\langle H_{12} \rangle_x$}
  & {$\langle \beta \rangle_x$}
  & {Target}
  & {Value}
  & {Deviation ($\leq 5\%$)}
  & Name  \\
  \midrule
  BASE-US $0.2\%\,U_\infty$ & $\Omega_1$ & $0.25$--$0.65$ & 467  & 1.64 & 0.51  & $H_{12}$ & 1.65 & 0.6\% & TCF-BDF-1 \\
  -- & $\Omega_2$ & $0.65$--$0.86$ & 828  & 1.69 & 2.08  & -- & 1.64 & 3.0\% & TCF-BDF-4 \\
  \botrule
  \end{tabular}
}
\caption{\textbf{Targets for the two-block wing-control configuration.} Wing properties are averaged over each listed interval. The Target and Value columns identify the matched quantity, $H_{12}$, and its value in the selected channel surrogate. The reported deviations compare the surrogate and wing shape factors. The $5\%$ criterion does not refer to all listed boundary-layer properties.}
\label{tab:tcf_match_wing_2blocks}
\end{table}

\begin{figure}[ht]
    \centering
    \includegraphics[width=0.75\linewidth]{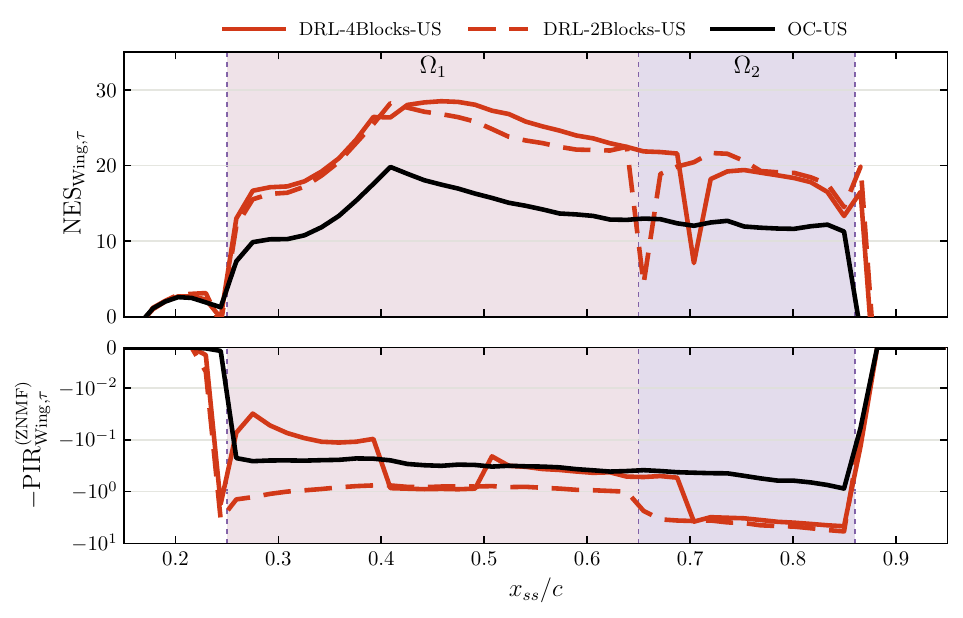}
    \caption{\textbf{Sensitivity to the chordwise partition with background suction at $0.2\%\,U_\infty$.} Local NES (top) and negative ZNMF power-input ratio (bottom), referenced to BASE-US. Solid red, dashed red, and black curves denote four-block DRL, two-block DRL, and OC, respectively. Shading and vertical lines mark the two-block intervals, $0.25$--$0.65$ and $0.65$--$0.86$. Four-block DRL is the configuration used in the main results. The lower panel uses a symmetric-logarithmic scale.}
    \label{fig:wing_nes_2blocks}
\end{figure}

Supplementary Figure~\ref{fig:wing_nes_2blocks} compares the two-block and four-block DRL configurations with OC. The DRL NES profiles nearly coincide over $0.25\lesssim x_{\mathrm{ss}}/c\lesssim0.40$, corresponding to the original $\Omega_1$. A local reduction in NES occurs at the interface between the two blocks, where the applied policy changes. The downstream power-input distribution also changes. Averaged across $0.25\leq x_{\mathrm{ss}}/c\leq0.86$, the local NES is $21.5\%$ for four blocks and $20.8\%$ for two blocks.

The two partitions thus yield similar averages of the local NES over the available records. These results show that an alternative partition yields similar mean local performance over the available averaging interval. We therefore do not claim that the four-block partition is unique or optimal.

\section*{Supplementary Information 6: Obtaining boundary-layer characteristics from lower-fidelity simulations for creating surrogate channels}
\nopagebreak[4]
Zero-shot deployment requires prior target-flow statistics to define the surrogate channels. In the present study, these statistics are obtained from high-resolution LES. We examine whether lower-cost simulations can estimate the boundary-layer properties needed for surrogate selection.

We use two-dimensional Reynolds-averaged Navier--Stokes simulations in OpenFOAM with the $k$--$\omega$ SST model. Previous comparisons provide a basis for assessing this model against high-resolution wing LES~\citep{atzori_aero_2020,mallor_bayesian_2024}.

The square domain has dimensions $L_x=L_y=100c$, with the airfoil centred at an angle of attack of $5^{\circ}$. Structured quadrilateral cells resolve the near-airfoil and near-wake regions. The remaining mesh is unstructured and triangular. The mesh contains slightly more than $250,000$ cells, approximately half of them in the boundary-layer and near-wake regions. The first off-wall point lies below one viscous unit. Further details of the numerical setup are given in Ref.~\citep{atzori_aero_2020}.

Supplementary Table~\ref{tab:rans_LES_clcd} compares integral lift, friction drag, pressure drag, total drag, and lift-to-drag ratio for the uncontrolled and uniform-suction cases. RANS underpredicts total drag by approximately 10\% in both cases. Comparisons of RANS with high-resolution LES at moderate Reynolds numbers are also reported in Ref.~\citep{atzori_aero_2020}. The present results quantify the discrepancy but do not identify its source. RANS and LES nevertheless predict the same direction of the suction-induced changes in lift and the drag components.

\begin{table}
  \centering
  \resizebox{\textwidth}{!}{%
  \begin{tabular}{llccccc}
  \toprule
  Case & Method & $C_l$ & $C_{d,f}$ & $C_{d,p}$ & $C_d$ & $L/D$ \\
  \midrule
  Base & LES & 0.861 & 0.0127 & 0.0087 & 0.0214 & 40.2 \\
   & RANS & 0.848 \, (-1.5\%) & 0.0113 \,(-11.0\%)  & 0.0079 \, (-9.2\%) & 0.0192 \, (-10.3\%) & 44.1 \, (+9.7\%) \\
  \midrule
  US $0.2\%\,U_\infty$ & LES & 0.925 & 0.0140 & 0.0066 & 0.0206 & 44.9 \\
   & RANS & 0.906 \, (-2.1\%) & 0.0128 \, (-8.6\%) & 0.0056 \, (-15.2\%) & 0.0183 \, (-11.2\%)  & 49.5 \, (-11.2\%) \\
  \bottomrule
  \end{tabular}%
  }
  \caption{\textbf{RANS and LES integral aerodynamic quantities.} Lift, friction drag, pressure drag, total drag, and lift-to-drag ratio are compared for the uncontrolled and uniform-suction cases. Parentheses report signed RANS deviations from the corresponding LES reference, using $100(\mathrm{RANS}-\mathrm{LES})/\mathrm{LES}$.}
  \label{tab:rans_LES_clcd}
\end{table}

\begin{figure}[ht]
    \centering
    \includegraphics[width=\linewidth]{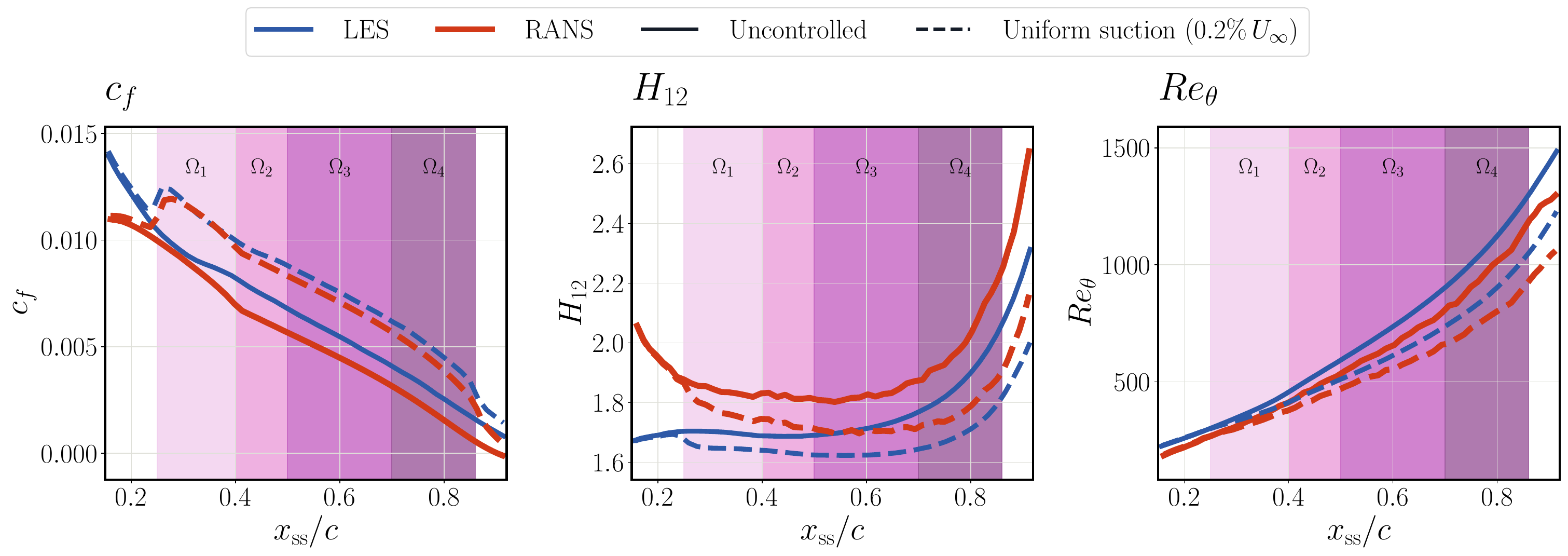}
    \caption{\textbf{RANS and LES suction-side boundary-layer properties.} Skin-friction coefficient, shape factor, and momentum-thickness Reynolds number for NACA4412 at $Re_c=200{,}000$ and an angle of attack of $5^\circ$. Blue and red denote LES and RANS. Solid and dashed curves denote the uncontrolled and uniform-suction ($0.2\%\,U_\infty$) cases. The wall-shear stress is normalized by the free-stream dynamic pressure. Shading marks $\Omega_1$--$\Omega_4$.}
    \label{fig:rans_LES_characteristics}
\end{figure}

Supplementary Figure~\ref{fig:rans_LES_characteristics} compares the suction-side boundary-layer properties. The wall-shear stress is normalized by the free-stream dynamic pressure, giving $c_f=2\tau_w/(\rho U_\infty^2)$. RANS captures the streamwise trend in $c_f$ but underpredicts the downstream wall shear, particularly without suction. It generally overpredicts $H_{12}$ and underpredicts $Re_\theta$, although the streamwise trends are similar to those obtained with LES.

Supplementary Table~\ref{tab:rans_les_blockwise} quantifies these differences over the control blocks. RANS overpredicts the block-mean $H_{12}$ by $4.9$--$8.5\%$ and underpredicts $Re_\theta$ by $7.3$--$10.5\%$. RANS can therefore provide approximate values for selecting the surrogate channels. The present policies, however, were matched using LES statistics. Deployment based entirely on RANS requires further assessment, particularly because errors in wall shear also affect the local viscous scales.

\begin{table}
  \centering
  \resizebox{\textwidth}{!}{%
  \begin{tabular}{lccrrcrrc}
  \toprule
  Case & Block & $x_{\rm ss}/c$ & \multicolumn{3}{c}{$\langle Re_\theta\rangle_x$} & \multicolumn{3}{c}{$\langle H_{12}\rangle_x$} \\
  & & & LES & RANS & Deviation [\%] & LES & RANS & Deviation [\%] \\
  \midrule
  Base flow & $\Omega_1$ & $0.25$--$0.40$ & 373 & 334 & -10.4\% & 1.699 & 1.843 & +8.5\% \\
   & $\Omega_2$ & $0.40$--$0.50$ & 525 & 478 & -8.9\% & 1.688 & 1.820 & +7.8\% \\
   & $\Omega_3$ & $0.50$--$0.70$ & 736 & 666 & -9.4\% & 1.717 & 1.827 & +6.4\% \\
  \midrule
  US $0.2\%\,U_\infty$ & $\Omega_1$ & $0.25$--$0.40$ & 353 & 320 & -9.4\% & 1.648 & 1.779 & +7.9\% \\
   & $\Omega_2$ & $0.40$--$0.50$ & 462 & 428 & -7.3\% & 1.632 & 1.726 & +5.8\% \\
   & $\Omega_3$ & $0.50$--$0.70$ & 613 & 562 & -8.4\% & 1.627 & 1.706 & +4.9\% \\
   & $\Omega_4$ & $0.70$--$0.86$ & 867 & 777 & -10.5\% & 1.699 & 1.782 & +4.9\% \\
  \bottomrule
  \end{tabular}%
  }
  \caption{\textbf{Block-mean boundary-layer properties from RANS and LES.} $Re_\theta$ and $H_{12}$ are averaged over the listed suction-side intervals. Signed deviations from the corresponding LES values are computed as $100(\mathrm{RANS}-\mathrm{LES})/\mathrm{LES}$ and expressed in percent.}
  \label{tab:rans_les_blockwise}
\end{table}

\section*{Supplementary Information 7: Turbulence statistics and spectral analysis for the surrogate channels and wing sections}
\begin{figure}[ht]
    \centering
    \includegraphics[width=0.85\textwidth]{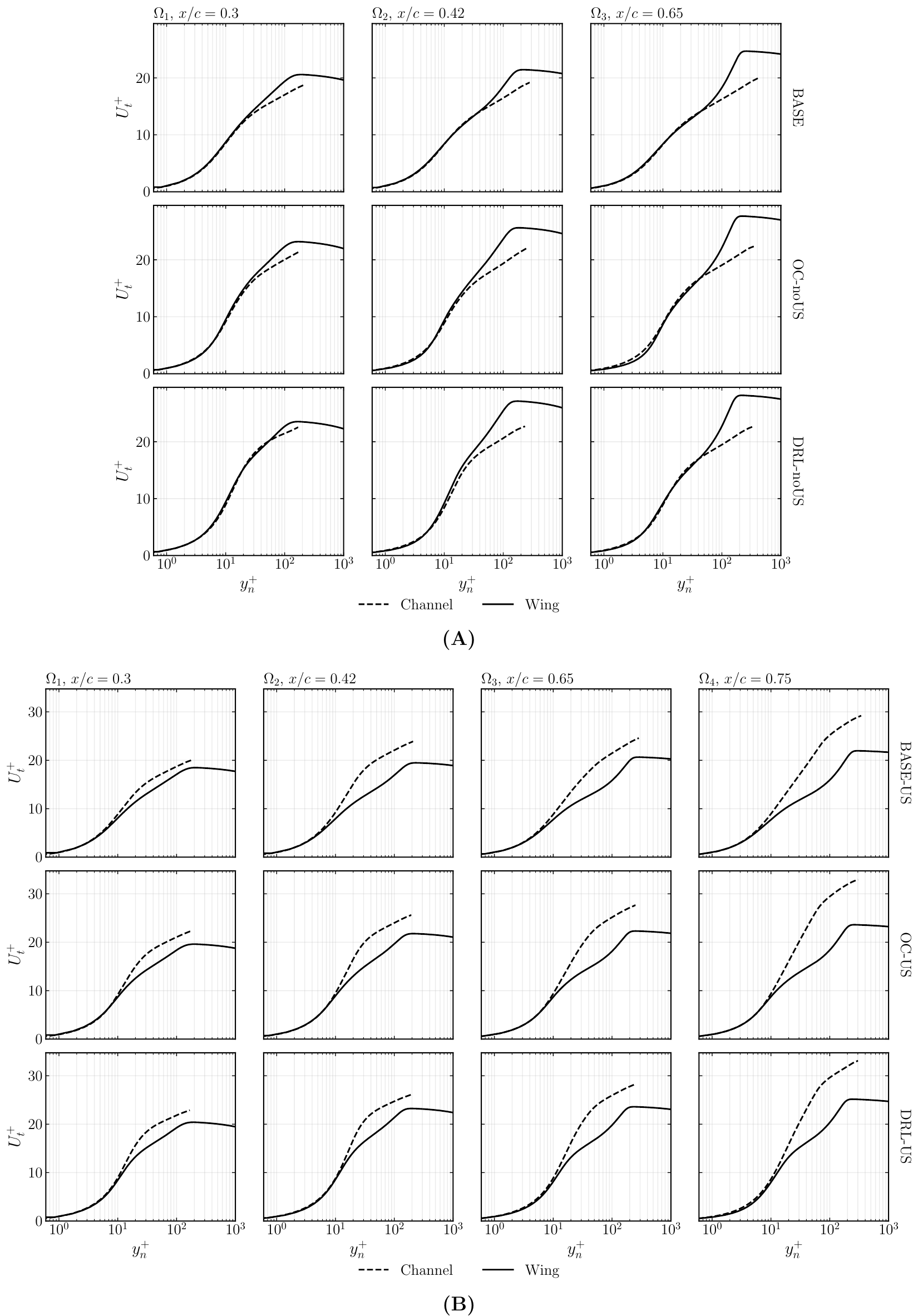}
    \caption{\textbf{Mean tangential velocity in local viscous units.} (A) Wing-section cases without background suction (noUS). (B) Wing-section cases with uniform suction at $0.2\%\,U_\infty$ (US). Rows show the background without feedback, OC, and DRL, from top to bottom. Columns compare the indicated nominal wing stations with their channel surrogates, covering $\Omega_1$--$\Omega_3$ in (A) and $\Omega_1$--$\Omega_4$ in (B). Dashed and solid curves denote channel and wing-section results, respectively. Velocity and wall-normal distance are expressed in viscous units based on the local friction velocity of each case. DRL in (A) is applied without downstream blowing.}
    \label{fig:u_mean}
\end{figure}

\begin{figure}[ht]
    \centering
    \includegraphics[width=0.85\textwidth]{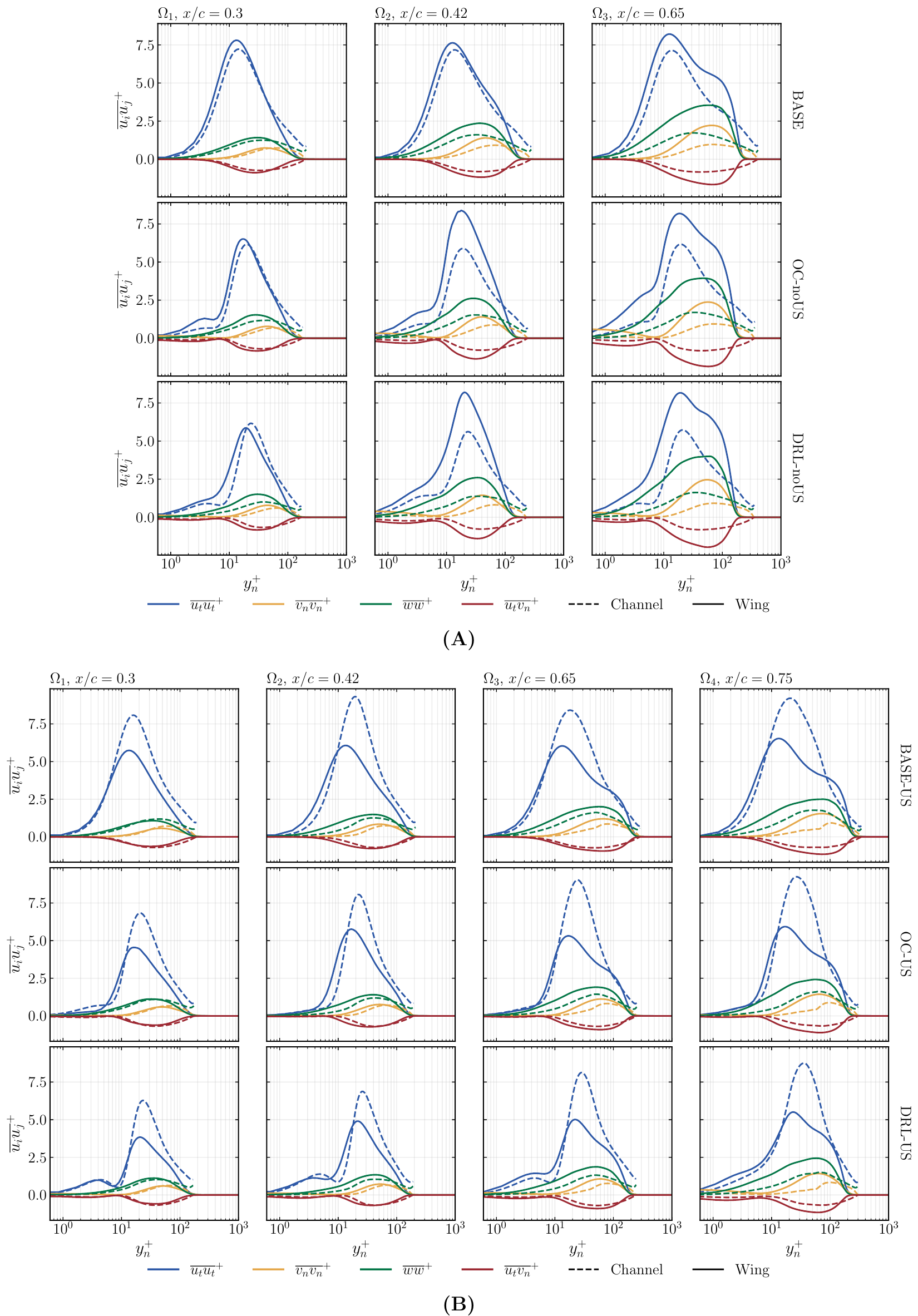}
    \caption{\textbf{Reynolds stresses in local viscous units.} (A) Wing-section cases without background suction (noUS). (B) Wing-section cases with uniform suction at $0.2\%\,U_\infty$ (US). Rows show the background without feedback, OC, and DRL, from top to bottom. Columns compare the indicated nominal wing stations with their channel surrogates, covering $\Omega_1$--$\Omega_3$ in (A) and $\Omega_1$--$\Omega_4$ in (B). Dashed and solid curves denote channel and wing-section results, respectively. Blue, orange, green, and red denote $\overline{u_t^2}^{+}$, $\overline{v_n^2}^{+}$, $\overline{w^2}^{+}$, and $\overline{u_tv_n}^{+}$, respectively. Stresses and wall-normal distance use the local viscous scales of each case. DRL in (A) is applied without downstream blowing.}
    \label{fig:uiuj}
\end{figure}

\paragraph{Turbulence statistics}
Supplementary Figures~\ref{fig:u_mean} and \ref{fig:uiuj} compare the mean tangential velocity and Reynolds-stress profiles on the wing with those of the corresponding channel surrogates. The nominal wing stations are $x_{\mathrm{ss}}/c=0.30$, $0.42$, $0.65$, and $0.75$, corresponding to $\Omega_1$--$\Omega_4$. The comparison without background suction includes $\Omega_1$--$\Omega_3$ and considers DRL without downstream blowing. With uniform suction, all four blocks are included. The wing velocities are resolved in the local tangential, normal, and spanwise directions. Profiles are taken at the nearest available station and expressed in viscous units based on the local friction velocity of each case.

Without background suction, the uncontrolled mean-velocity profiles agree closely near the wall (Supplementary Figure~\ref{fig:u_mean}). Larger differences are observed in the outer region, where the wing profiles exhibit a more pronounced wake contribution, whereas the channel profiles approach the centreline. A similar distinction is observed in the Reynolds stresses (Supplementary Figure~\ref{fig:uiuj}). The near-wall maximum of $\overline{u_t^2}^{+}$ occurs at approximately $y_n^+=12$--$14$ in both geometries, with values of $7.6$--$8.2$ on the wing and $7.1$--$7.2$ in TCF. The remaining stress components also show close agreement at $x_{\mathrm{ss}}/c=0.30$. Farther downstream, the wing exhibits higher $\overline{u_t^2}^{+}$ in the outer region, together with larger wall-normal and spanwise variances and a larger magnitude of the Reynolds shear stress. At $x_{\mathrm{ss}}/c=0.65$, the maxima of $\overline{v_n^2}^{+}$ and $\overline{w^2}^{+}$ are approximately $2.2$ and $3.5$ on the wing, compared with $1.0$ and $1.7$ in the corresponding channel. Thus, the agreement in the location of the near-wall maximum does not extend to the outer-region stress distributions.

With uniform suction, the inner-scaled mean velocity in TCF-BDF exceeds that on the wing beyond the buffer layer (Supplementary Figure~\ref{fig:u_mean}). Wall suction and body-force damping enter the mean momentum balance differently~\citep{atzori_uniform_2021}, and the local friction velocity also differs between the cases. The near-wall maxima of $\overline{u_t^2}^{+}$ are approximately $5.8$--$6.5$ on the wing and $8.1$--$9.3$ in TCF-BDF, occurring at $y_n^+=13$--$14$ and $16$--$21$, respectively. Despite its lower near-wall maximum, the downstream wing exhibits a larger outer-region contribution to the tangential variance, as well as larger wall-normal and spanwise variances. At $x_{\mathrm{ss}}/c=0.75$, the latter maxima are approximately $1.5$ and $2.5$, compared with $0.9$ and $1.8$ in TCF-BDF-4.

OC and DRL produce an upward displacement of the inner-scaled mean-velocity profiles outside the immediate vicinity of the wall. The near-wall maximum of $\overline{u_t^2}^{+}$ also moves to larger $y_n^+$, with a generally larger displacement for DRL than for OC. In TCF without body-force damping, the maximum moves from approximately $y_n^+=14$ without control to $19$--$20$ with OC and $21$--$24$ with DRL. The corresponding locations on the wing without background suction are $12$--$13$, $18$--$19$, and $19$--$21$. Both geometries therefore exhibit a similar wall-normal displacement under control, although the peak magnitudes and outer-region profiles remain different. 

\paragraph{Spectra analysis}
To examine the scales associated with these changes in the turbulence statistics, we compare premultiplied spanwise spectra, spanwise two-point correlations, and temporal cross-correlations between wall shear and tangential velocity. The stations and configurations are those used in the profile comparisons. Each time series is analysed using its recorded sampling interval and duration. Correlations at the prescribed $y_n^+$ are evaluated on the nearest available wall-normal plane, using the local viscous scales of each case.

\begin{figure}[ht]
    \centering
    \includegraphics[width=0.85\textwidth]{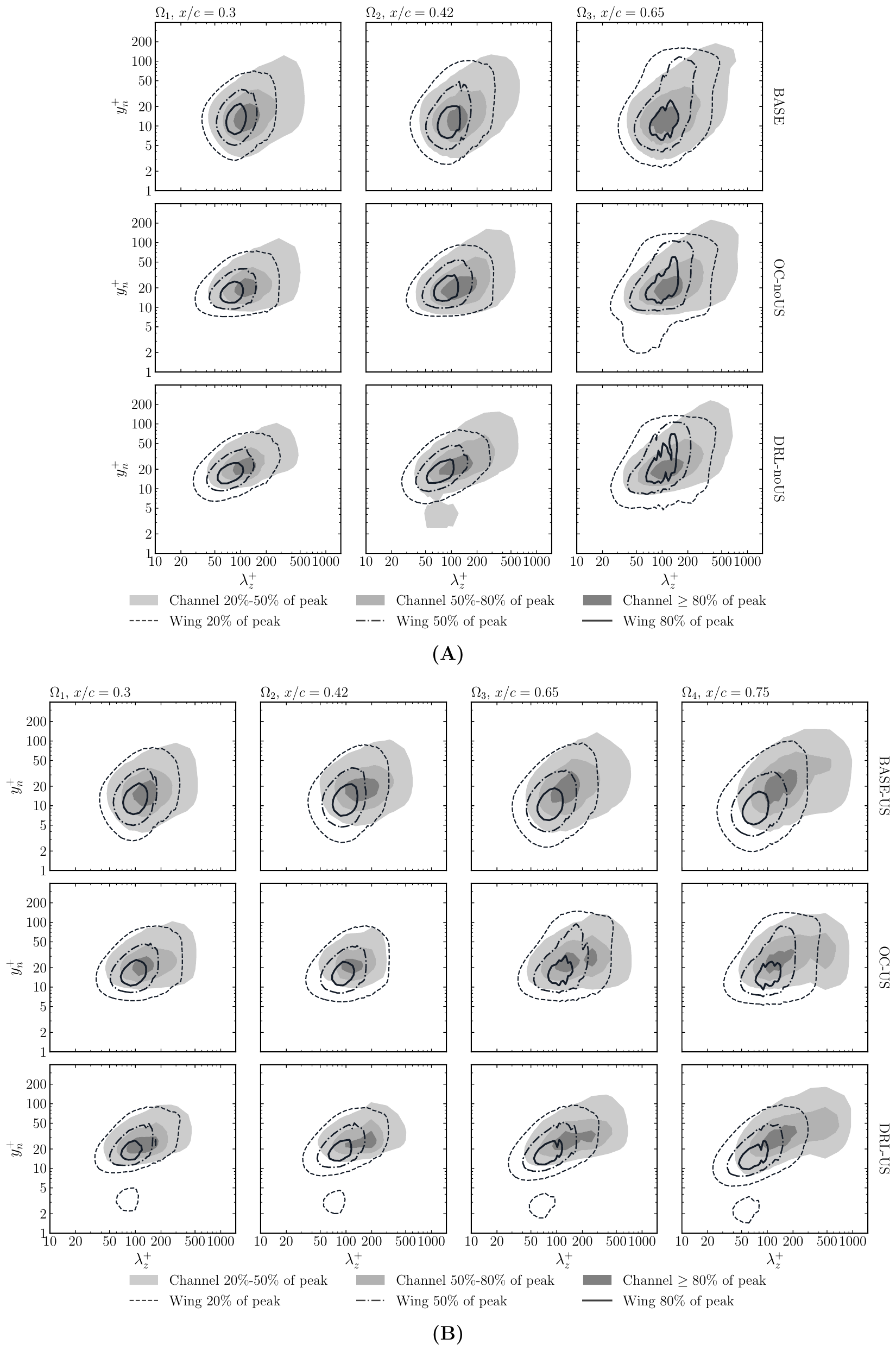}
    \caption{\textbf{Premultiplied spanwise spectra of tangential-velocity fluctuations.} (A) Wing-section cases without background suction (noUS). (B) Wing-section cases with uniform suction at $0.2\%\,U_\infty$ (US). Rows show the background without feedback, OC, and DRL, from top to bottom. Columns compare the indicated nominal wing stations with their channel surrogates, covering $\Omega_1$--$\Omega_3$ in (A) and $\Omega_1$--$\Omega_4$ in (B). Spanwise wavelength and wall-normal distance are expressed in local viscous units for each case.}
    \label{fig:psd}
\end{figure}

Supplementary Figure~\ref{fig:psd} shows the premultiplied spectrum $k_z\phi_{u_tu_t}/u_\tau^2$ as a function of spanwise wavelength and wall-normal distance. Here $\phi_{u_tu_t}$ is the one-dimensional spectrum of tangential-velocity fluctuations and $k_z=2\pi/\lambda_z$ is the spanwise wavenumber. Without background suction, the uncontrolled wing and channel spectra have near-wall maxima at $y_n^+\simeq10$--$12$ and spanwise wavelengths of order $10^2$ wall units, consistent with near-wall streak organization. The sampled maxima occur at $\lambda_z^+\simeq85$--$97$ on the wing and $98$--$125$ in TCF. The contours near the spectral maximum show the closest agreement at the upstream stations. At lower energy levels, the spectra of the first two channels extend to larger spanwise wavelengths, whereas the downstream wing spectrum extends farther from the wall. This greater wall-normal extent is consistent with the outer-region contribution observed in adverse-pressure-gradient boundary layers~\citep{harun_PGEffect_2013}.

OC and DRL generally displace the near-wall spectral maxima away from the wall, consistent with the variance profiles. The spanwise wavelength at the maximum responds differently in the wing and channel flows. At $x_{\mathrm{ss}}/c=0.42$ with background suction, the wing maximum moves from approximately $(\lambda_z^+,y_n^+)=(101,12)$ without feedback control to $(92,15)$ under OC and $(85,17)$ under DRL. The corresponding channel maxima are approximately $(162,20)$, $(121,22)$, and $(197,28)$. The maximum thus moves away from the wall in both geometries, whereas DRL reduces its spanwise wavelength on the wing and increases it in the channel. 

Next, we access the two-point correlation of the streamwise-velocity fluctuations. We define the normalized spanwise two-point correlation as
\begin{equation}
 R_{u_tu_t}(\Delta z;x,y_n)=
 \frac{\left\langle\overline{
 u_t(x,y_n,z,t)\,u_t(x,y_n,z+\Delta z,t)}\right\rangle_h}
 {\left\langle\overline{u_t^2(x,y_n,z,t)}\right\rangle_h},
 \label{eq:spanwise-two-point-correlation}
\end{equation}
\noindent where the overbar denotes time averaging and $\langle\cdot\rangle_h$ averages over the homogeneous directions: $z$ on the wing and $(x,z)$ in the channel. The channel correlation is therefore independent of $x$. For the wing, the spanwise--time mean at each station and height is subtracted to obtain the fluctuations. For the channel, the time mean at each spatial point is subtracted. Periodicity is imposed over $L_z$ when evaluating the spanwise shifts, with the duplicated endpoint excluded. For nonzero variance, $R_{u_tu_t}(0)=1$. Differences in mean removal can affect the wing--channel comparison if a stationary spatial pattern remains in the finite-record mean.

\begin{figure}[ht]
    \centering
    \includegraphics[width=0.85\textwidth]{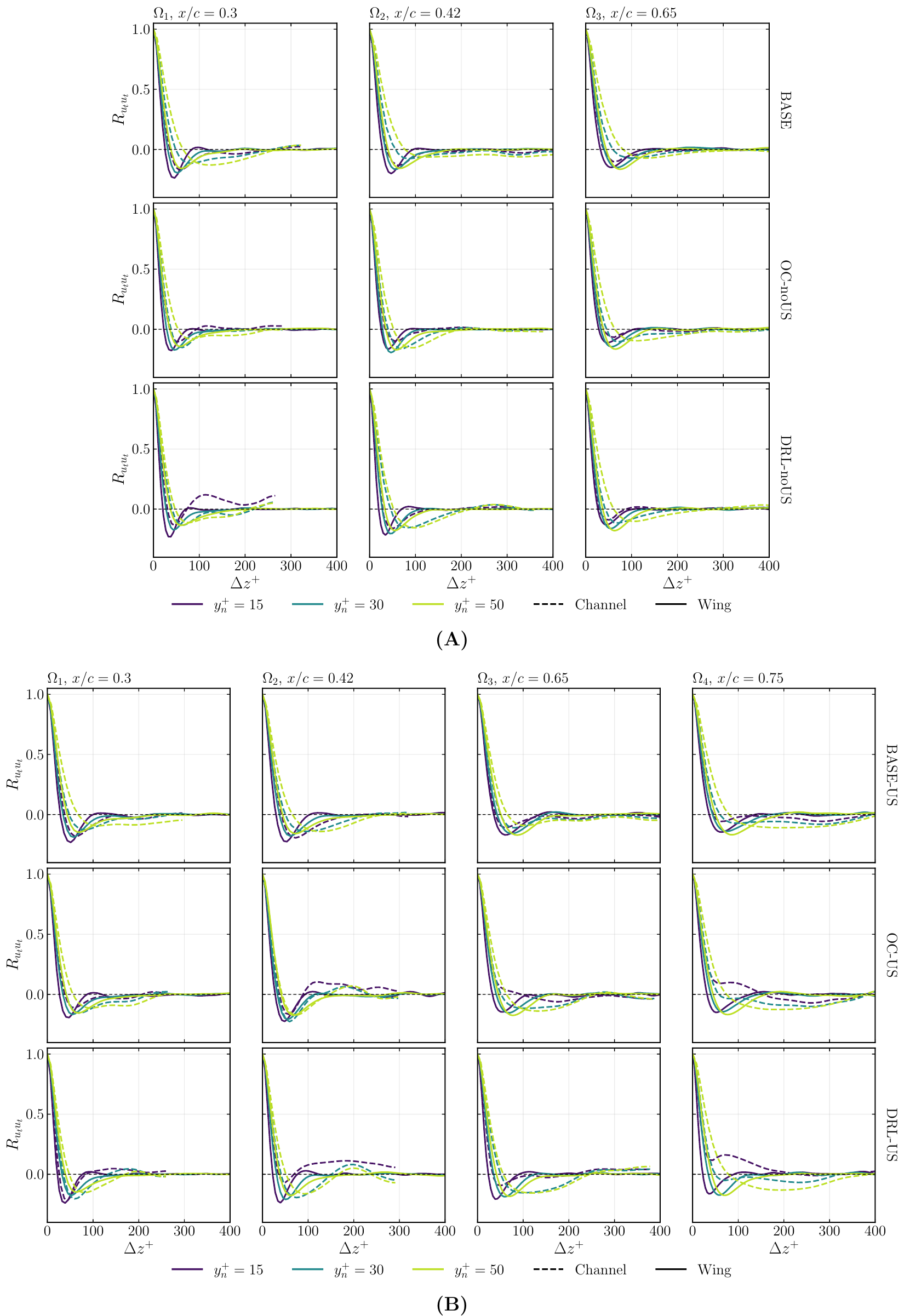}
    \caption{\textbf{Normalized spanwise two-point correlations of tangential-velocity fluctuations.} (A) Wing-section cases without background suction (noUS). (B) Wing-section cases with uniform suction at $0.2\%\,U_\infty$ (US). Rows show the background without feedback, OC, and DRL, from top to bottom. Columns compare the indicated nominal wing stations with their channel surrogates, covering $\Omega_1$--$\Omega_3$ in (A) and $\Omega_1$--$\Omega_4$ in (B).}
    \label{fig:corr_z}
\end{figure}

Supplementary Figure~\ref{fig:corr_z} compares the correlations at nominal heights $y_n^+=15$, $30$, and $50$. Without feedback control, the correlation decreases rapidly with separation and becomes negative in both geometries, consistent with the alternating low- and high-speed streaks. The first minimum generally occurs at a smaller separation and has a larger negative magnitude on the wing, particularly at $y_n^+=15$. Its position provides a characteristic spanwise separation associated with anticorrelation, rather than a unique measure of streak width.

Control modifies both the negative minimum and the correlation at larger separations. On the wing, DRL generally shifts the first minimum towards smaller separations, most consistently near $y_n^+=15$. At this height, several controlled channels, including TCF-1 and TCF-BDF-1, 2, and 3, exhibit positive correlation over a wider range of separations. The corresponding wing correlations are weaker or close to zero. The differences at larger separations indicate that the similar initial decay does not imply agreement over the full spanwise range. 

Moreover, we access the temporal correlation between wall-shear-stress fluctuations ($\tau'_w$) and wall-tangential velocity fluctuations~\citep{brown_1977_corr}.
The temporal cross-correlation pairs wall-shear fluctuations at time $t$ with tangential-velocity fluctuations at the same wall-parallel position and time $t+T$. Both signals have their time means removed at each spatial point. For a record of $N_t$ samples separated by $\delta t$, the estimator is
\begin{equation}
 R_{\tau_w,u_t}(T;y_n)=
 \frac{\displaystyle
 \left\langle\frac{1}{N_t}\sum_{n\in\mathcal I_m}
 \tau_w'(\boldsymbol{s},t_n)\,
 u_t(\boldsymbol{s},y_n,t_{n+m})\right\rangle_h}
 {\tau_{w,\mathrm{rms}}\,u_{t,\mathrm{rms}}(y_n)},
 \qquad T=m\delta t,
 \label{eq:wall-shear-velocity-temporal-correlation}
\end{equation}
\noindent where $\mathcal I_m=\{n:0\leq n<N_t,\ 0\leq n+m<N_t\}$ contains the overlapping pairs. The spatial sample $\boldsymbol{s}$ denotes $z$ at a fixed wing station and $(x,z)$ in the channel. The denominator contains the root-mean-square fluctuations over the full record and homogeneous samples. Time pairs do not wrap periodically. The numerator retains the full-record normalization $N_t$, rather than $N_t-|m|$, thus finite-record attenuation increases with absolute lag. The lag in viscous units is $T^+=T u_\tau^2/\nu$, using the local friction velocity of the case considered. Positive $T^+$ corresponds to a velocity sample taken after the wall-shear sample.

\begin{figure}[ht]
    \centering
    \includegraphics[width=0.85\textwidth]{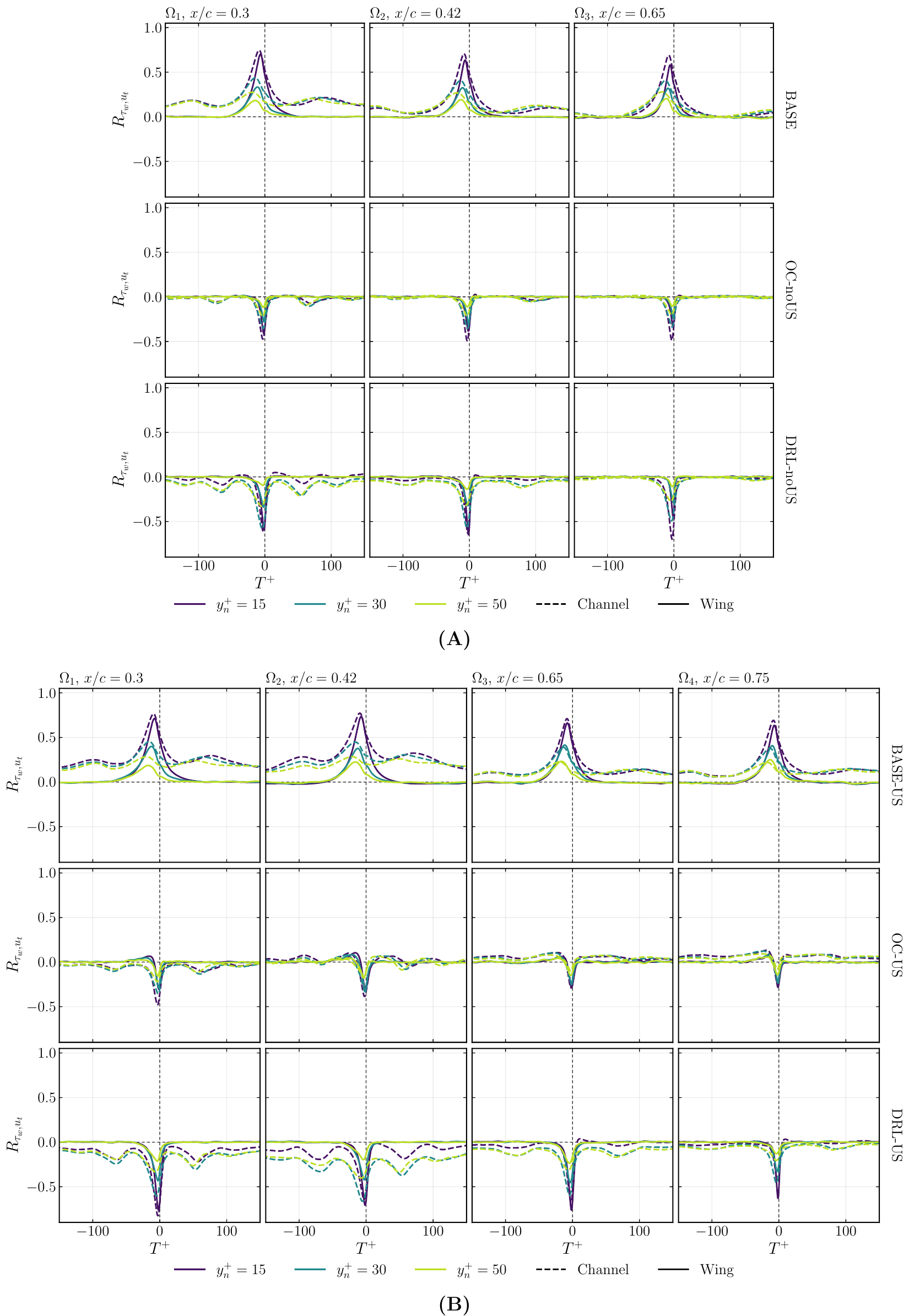}
    \caption{\textbf{Temporal cross-correlations between wall-shear and tangential-velocity fluctuations.} (A) Wing-section cases without background suction (noUS). (B) Wing-section cases with uniform suction at $0.2\%\,U_\infty$ (US). Rows show the background without feedback, OC, and DRL, from top to bottom. Columns compare the indicated nominal wing stations with their channel surrogates, covering $\Omega_1$--$\Omega_3$ in (A) and $\Omega_1$--$\Omega_4$ in (B).}
    \label{fig:corr_tau_u}
\end{figure}

Without feedback control, both geometries exhibit a positive maximum near zero lag. Its magnitude decreases and its location moves towards more negative lags as the wall-normal distance increases (Supplementary Figure~\ref{fig:corr_tau_u}). Near $y_n^+=15$ in the no-suction $\Omega_3$ pair, the peaks are approximately $0.58$ on the wing and $0.69$ in TCF, at $T^+\simeq-6$ and $-8$, respectively. The channels also retain broader correlations at larger lags. For example, in uncontrolled TCF-BDF-2, the mean correlations over $50\leq|T^+|\leq150$ are about $0.23$--$0.25$ on either side of zero, whereas the wing values are close to zero. Several DRL-controlled channels also exhibit negative correlation over a wider range of lags than on the wing. These differences can be affected by the flow configuration and the finite record lengths.

OC and DRL change the central correlation from positive to negative, with DRL usually producing stronger anticorrelation near $y_n^+=15$. The magnitude of this response differs between the two geometries at the downstream stations. In the no-suction $\Omega_3$ pair, the DRL minima are approximately $-0.45$ on the wing and $-0.70$ in TCF. In the uniform-suction $\Omega_4$ pair, they are approximately $-0.63$ and $-0.43$. Thus, both geometries exhibit a change in the sign of the central correlation, while the magnitude and lag dependence remain different.

The mean profiles, Reynolds stresses, and spectra show similar near-wall features and wall-normal displacements under control in the channel and wing flows. The temporal correlations also exhibit a common change of sign under OC and DRL. Differences remain in the outer-region stresses, the spanwise wavelength of the spectral maximum, and the correlations at larger separations and lags. The observed transfer of the policies therefore occurs despite incomplete agreement of the turbulence statistics. These comparisons characterize the shared control response without identifying a unique mechanism for transfer.

\section*{Supplementary Information 8: DRL training parameters}
Supplementary Table~\ref{tab:drl_hyper} summarizes the network architecture and hyperparameters for the MARL training described in Methods. Supplementary Table~\ref{tab:tcf_train_configs} lists the number of pseudo-environments, mini-batch size and replay-buffer capacity for each channel surrogate.
\begin{table}[ht]
\centering
\begin{tabular}{ll}
\toprule
{Parameter} & {Value} \\
\midrule

Learning algorithm & TD3 \\

Actor network architecture & Fully connected (8 neurons) \\
Critic network architecture & Fully connected (16--64--64 neurons) \\

Number of input features ($N_{\rm feat}$) & 2 (i.e.\ $u'^+,\,v'^+$) \\

Replay buffer size ($\mathcal{D}$) & Scaled with grid resolution \\
Mini-batch size ($\mathcal{B}$) & Scaled with grid resolution \\

Gradient steps per update & 64 \\
Policy update interval & $\Delta t^+=180$ \\
Actuation update interval & $\Delta t^{+}=0.6$ \\
Episode length & $t^{+}=1{,}500$ \\
Number of episodes & 100 \\
Number of initial conditions & 6 \\
Exploration noise & Gaussian ($\mu = 0$, $\sigma = 0.1\,u_\tau$) \\
Action bound & ($-u_\tau$, $u_\tau$) \\

\botrule
\end{tabular}
\caption{\textbf{Summary of the main DRL hyperparameters used in the present MARL framework.}
See Table~\ref{tab:tcf_train_configs} for the batch size and replay-buffer size employed in each configuration.}
\label{tab:drl_hyper}
\end{table}

\begin{table}[ht]
  \centering
  \begin{tabular}{cccc}
  \toprule
  Name
  & {$N_{\rm penv}$}
  & {$\mathcal{B}$}
  & {$\mathcal{D}$} \\
  \midrule
  TCF-1  & 4,410 & 2,048 & $2\times10^7$ \\
  TCF-2  & 4,410 & 2,048 & $2\times10^7$ \\
  TCF-3  & 10,890 & 4,096 & $1\times10^8$ \\
  TCF-BDF-1  & 4,410 & 2,048 & $2\times10^7$ \\
  TCF-BDF-2  & 4,410 & 2,048 & $2\times10^7$ \\
  TCF-BDF-3  & 10,890 & 4,096 & $1\times10^8$ \\
  TCF-BDF-4  & 15,972 & 6,144 & $4\times10^8$ \\
  \botrule
  \end{tabular}
  \caption{\textbf{Training configurations in the channel flows.} $\mathcal{B}$ and $\mathcal{D}$ denote the mini-batch size and replay-buffer capacity, respectively.}
  \label{tab:tcf_train_configs}
\end{table}

\clearpage

\bibliography{sn-bibliography}%